\documentclass[aps,pra,twocolumn,showpacs,floatfix,superscriptaddress]{revtex4}

\usepackage{amsmath}
\usepackage{amsfonts}
\usepackage{amssymb}
\usepackage{graphicx}
\usepackage{epsfig}
\usepackage[hang,nooneline]{subfigure}
\usepackage[normalem]{ulem}
\usepackage{color}
\usepackage{braket}
\usepackage{hyperref}
\usepackage{todonotes}
\usepackage{algorithm}
\usepackage{multirow}
\usepackage{hhline}
\usepackage{shortvrb}
\usepackage{cprotect}
\usepackage[noend]{algpseudocode}
\usepackage{overpic}
\usepackage[capitalise]{cleveref}

\definecolor{darkblue}{rgb}{0.00,0.00,0.55}
\definecolor{black}{rgb}{0.00,0.00,0.00}
\hypersetup{
    linkcolor = darkblue,
    anchorcolor = darkblue,
    citecolor = darkblue,
    filecolor = darkblue,
    urlcolor = darkblue,
    colorlinks  = true,
}
\definecolor{brightcerulean}{rgb}{0.11, 0.67, 0.84}
\DeclareMathAlphabet\mathbfcal{OMS}{cmsy}{b}{n}

\newcounter{fig}

\newcommand{\linfigwidth}{2.6cm}
\begin{document}

\title{Deflation-based Identification of Nonlinear Excitations of the 3D Gross--Pitaevskii equation}

\author{N. Boull\'e}
\email[Email: ]{nicolas.boulle@maths.ox.ac.uk}
\affiliation{Mathematical Institute, University of Oxford, Oxford, UK}
\author{E. G. Charalampidis}
\email[Email: ]{echarala@calpoly.edu}
\affiliation{Mathematics Department, California Polytechnic State University, 
San Luis Obispo, CA 93407-0403, USA}
\author{P. E. Farrell}
\email[Email: ]{patrick.farrell@maths.ox.ac.uk}
\affiliation{Mathematical Institute, University of Oxford, Oxford, UK}
\author{P. G. Kevrekidis}
\email[Email: ]{kevrekid@math.umass.edu}
\affiliation{Department of Mathematics and Statistics, University of Massachusetts
  Amherst, Amherst, MA 01003-4515, USA}
\affiliation{Mathematical Institute, University of Oxford, Oxford, UK}

\date{\today}

\begin{abstract}
We present previously unknown solutions to the 3D Gross--Pitaevskii equation
describing atomic Bose-Einstein condensates. This model supports elaborate
patterns, including excited states bearing vorticity. The discovered coherent
structures exhibit striking topological features, involving combinations of
vortex rings and multiple, possibly bent vortex lines. Although unstable, many of them
persist for long times in dynamical simulations. These solutions were identified
by a state-of-the-art numerical technique called deflation, which is
expected to be applicable
to many problems from other areas of physics.
\end{abstract}


\maketitle


\section{Introduction} The nonlinear Schr\"odinger (NLS) or Gross--Pitaevskii (GP) equation~\cite{ablowitz,ablowitz1,sulem,pethick,stringari,siambook}
is a fundamental partial differential equation that combines dispersion 
and nonlinearity. It has been central to a variety of areas of mathematical 
physics for several decades. The NLS/GP model has facilitated a universal 
description of a wide range of phenomena, including electric fields in 
optical fibers~\cite{hasegawa}, Langmuir waves in plasmas~\cite{zakh1}, 
freak waves in the ocean~\cite{slunaev}, and Bose-Einstein condensates (BECs). 
In the past 25 years since the experimental realization of atomic BECs, 
the NLS/GP model has enabled the theoretical identification and experimental
observation of a wide range of coherent structures, including (but not limited to)
dark~\cite{djf} and bright~\cite{tomio} solitary waves, two-dimensional vortical 
patterns and lattices~\cite{fetter1,fetter2} as well as vortex lines and 
rings~\cite{komineas}.

The examination of three-dimensional (3D) systems has been a key frontier
of recent studies in BECs. Recent theoretical advances have enabled the capturing of 
a number of such states~\cite{mateo2014chladni}. Some, especially topological ones 
such as skyrmions, monopoles and Alice rings~\cite{ruost1,ruost2} have been of 
particular interest since the early exploration of BECs, while others such as 
knots~\cite{irvine} have been studied more recently. 
In this manuscript, we apply a powerful numerical technique called deflation~\cite{farrell2015deflation,charalampidis2018computing,charalampidis2019bifurcation} 
to identify multiple solutions of the 3D NLS/GP equation.

Many of the solutions obtained by this process are identifiable as nonlinear 
extensions of solutions of the linear limit of the problem, or as bifurcations 
therefrom. Yet other solutions are highly unexpected and are not previously known, 
to the best of our knowledge. Without deflation, it would be very difficult to 
identify these complex (literally and figuratively) topological stationary points 
of the infinite-dimensional energy landscape. In fact, as we increase the atom number 
of the system, we observe this complexity to be substantially enhanced and lead to
states which, while stationary, are not straightforwardly decomposable into simpler 
linear or nonlinear building blocks. To further investigate the nature of the identified
solutions, we compute their spectral linearization (so-called Bogoliubov-de Gennes)
modes, and conduct transient simulations of prototypical unstable states to
explore the dynamical behavior of their instabilities. The present work showcases,
in our view, the utility and potential impact of the deflation method to complex 3D 
physical problems well beyond atomic BECs. 

The structure of this paper is as follows. In Section II, we present the model and
computational techniques employed in this work. Our numerical results on the existence,
stability and selected transient simulations of nonlinear excitations are demonstrated
in Section III. Section IV summarizes our findings and presents directions for future 
study.

\section{Theoretical and Numerical Setup}
The 3D NLS/GP model of interest is of the form~\cite{pethick,stringari,siambook}:
\begin{eqnarray} \label{eq:transient}
  i \psi_t =- \frac{1}{2} \nabla^2 \psi + |\psi|^2 \psi +V(\mathbf{r})\psi,
  \label{eq0}
\end{eqnarray}
subject to homogeneous Dirichlet conditions on the boundary of the domain 
$D = [-6, 6]^3$. Here, $\psi=\psi(\mathbf{r},t)$ plays the role of the suitably 
normalized (see~\cite{siambook} for details) wavefunction, while $V$ is the external 
confining potential of the form $V(\mathbf{r}) = \frac{1}{2}\Omega^2|\mathbf{r}|^2$,
a spherically symmetric trap of 
strength $\Omega$, which we fix to $\Omega = 1$.
The boundary conditions do not affect the solutions for this choice of
the trap strength since the domain is chosen large enough so that the
solutions vanish well before reaching the boundary. 
Using the standard standing wave decomposition $\psi(\mathbf{r},t)=e^{-i \mu t}
\phi(\mathbf{r})$ (where $\mu > 0$ is the chemical potential), we obtain 
the stationary NLS/GP elliptic problem of the form:
\begin{equation} \label{eq:stationary}
F(\phi) \doteq -\frac{1}{2}\nabla^2\phi+|\phi|^2\phi+V(\mathbf{r})\phi-\mu\phi=0.
\end{equation}
This equation is discretized using piecewise cubic Lagrange finite 
elements on a structured hexahedral grid using the Firedrake finite 
element library~\cite{rathgeber2016}. Multiple solutions to the discretized
problem are sought using deflation, which we briefly describe here.

Suppose that Newton's method has discovered an isolated root $\phi_1$ 
of $F$. Deflation constructs a new problem $G$ via
\begin{equation}
G(\phi)\doteq\left(\frac{1}{\|\phi - \phi_1\|^2} + 1   \right) F(\phi),
\label{eq_deflation}
\end{equation}
where $\|\cdot\|$ is a suitable norm, in this case the $H^1$ norm. The 
essential idea is that $\|\phi - \phi_1\|^{2}$ approaches $0$
faster than $F(\phi)$ does as $\phi \to \phi_1$, hence avoiding 
the convergence to $\phi_1$ of a Newton iteration applied to $G$. The 
addition of $1$ ensures that $G(\phi) \approx F(\phi)$ far from $\phi_1$. 
By applying Newton's method to $G$, an additional root $\phi_2 \neq \phi_1$ 
can be found, and the process repeated (by premultiplying with additional
factors) until Newton's method fails to converge from the available initial
guesses.

Previous applications of deflation to the study of BECs in 2D interleaved 
with continuation in $\mu$, capturing solutions as they bifurcate from known
ones~\cite{charalampidis2018computing,charalampidis2019bifurcation}. This
strategy is too expensive in 3D and so a different approach is taken here. 
We fix $\mu = 6$ and exploit the linear (low-density, i.e. $|\phi|^2\to 0$) 
limit states to furnish a large number of initial guesses for Newton's method.
The algorithm proceeds as follows. Given an initial guess, the inner
loop applies Newton's method and deflation until no more solutions are found.
The outer loop iterates over the available initial guesses, and terminates 
when no guess yields any solutions. We emphasize that at each application 
of Newton's method, \emph{all} previously computed solutions are deflated, to 
avoid their rediscovery.

The initial guesses used were the eigenstates of the linear limit in Cartesian,
cylindrical and spherical coordinates. The Cartesian eigenstates are given by
\begin{equation} \label{extra1}
\ket{k,m,n} \doteq H_k(\sqrt{\Omega} x) H_m(\sqrt{\Omega} y) %
H_{n}(\sqrt{\Omega} z) e^{-\Omega r^2/2},
\end{equation}
with associated energy (i.e.~chemical potential) $E_{k,m,n}\doteq (k+ m + n +3/2)\Omega$.
The $H_{k,m,n}$ in~\eqref{extra1} stand for the Hermite polynomials and $k,m$ and 
$n$ are nonnegative integers.
The cylindrical eigenstates are given by
\begin{equation} \label{extra2}
\ket{K,l,n}_{\textrm{cyl}}\doteq  q_{K,l}(R) e^{i l \theta} H_{n}(\sqrt{\Omega} z) e^{-\Omega (R^2+z^2)/2},
\end{equation}
with $E_{K,l,n} \doteq (2K + |l| +n + 3/2)\Omega$ where
$K$, $n$ are nonnegative integers, and $l=0,\pm1,\pm2,\dotsc$.
The radial profile $q_{K,l}$ in~\eqref{extra2} is given by
$q_{K,l} \sim r^l L_K^l(\Omega R^2) e^{-\Omega R^2/2} \Omega$ 
where $L_K^l$ are the associated Laguerre polynomials in $R=\sqrt{x^2+y^2}$.

Finally, the spherical eigenstates are given by $\ket{K,l,m}_{\textrm{sph}}$, 
where the radial part is similar but now in the spherical variable 
$r = \sqrt{x^2 + y^2 + z^2}$, and the angular part is described by
the spherical harmonics $Y_{lm}(\theta,\phi)$ with $E_{K,l,m}=(2 K + l
+ 3/2) \Omega$. The quantum numbers $K$ and $l$ are nonnegative integers 
and $m=0,\pm 1, \dotsc, \pm l$. 
All these states with $E \le \mu = 6$ were used in the process described 
above.

Once a solution has been discovered, the next step is the consideration
of the spectral stability of the solutions via the well-established~\cite{pethick,stringari,siambook}
Bogoliubov-de Gennes (BdG) analysis. More specifically, we assume the 
following perturbation ansatz around a stationary solution $\phi^0$:
\begin{equation}\label{eq:tildephi}
\tilde{\psi}(\mathbf{r},t)=e^{-i\mu t}\left\{\phi^0(\mathbf{r})+%
\epsilon[a(\mathbf{r}) e^{i\omega t}+b^{\ast}(\mathbf{r})e^{-i\omega^{\ast}t}]\right\},
\end{equation}
where $\epsilon$ is a (formal) small perturbation parameter, $\omega$ is the 
eigenfrequency, and $(a,b)^\top$ the corresponding eigenvector. After 
substituting Eq.~\eqref{eq:tildephi} into the time-dependent NLS equation 
[cf. Eq.~\eqref{eq:transient}] we obtain the following complex eigenvalue 
problem
\begin{equation} \label{eq_eig}
\begin{pmatrix}
A_{11} & A_{12} \\
-A_{12}^* & -A_{11}
\end{pmatrix}
\begin{pmatrix}
a\\
b
\end{pmatrix} = 
\rho
\begin{pmatrix}
a\\
b
\end{pmatrix},
\end{equation}
where $\rho = -\omega$ and the matrix elements are given by
\begin{subequations}
\begin{align}
A_{11} & = -\frac{1}{2}\nabla^2+2|\phi^0|^2+V(\mathbf{r})-\mu,\\
A_{12} &= \left(\phi^0\right)^2.
\end{align}
\end{subequations}
We solve the above eigenvalue problem for the eigenfrequencies $\omega$ and 
eigenvectors $(a,b)^\top$ by using a Krylov--Schur algorithm with a shift-and-invert 
spectral transformation~\cite{stewart2002}, implemented in the SLEPc 
library~\cite{hernandez2005slepc} (details about the decomposition of 
Eq.~\eqref{eq_eig} into real and imaginary parts are presented in Appendix~\ref{app_spec}).
Upon convergence of the eigenvalue solver, we draw conclusions on the stability 
characteristics of the stationary state $\phi^0$ i.e., real $\omega$ implies 
stability (vibrations), while complex $\omega$ is associated with instability.

Finally, we explore the dynamical evolution of unstable solutions via transient
numerical simulations of Eq.~\eqref{eq:transient}. To that end, let $\phi^{0}$ 
be an unstable stationary solution discovered by deflation, and $(a,b)^\top$ be 
its most unstable eigendirection normalized according to
\begin{equation}
  \int_{D}\left(\,|a|^2+|b|^2\,\right)\,dx=1.
\end{equation}
We integrate Eq.~\eqref{eq:transient} forward in time until $t=50$ using
the following perturbed solution as initial state
\begin{equation}\label{eq:init_state}
\psi(x,y,z,t=0)=\phi^0+\epsilon[a+b^*],
\end{equation}
thus perturbing $\phi^0$ along its most unstable eigendirection with 
perturbation parameter $\epsilon$ chosen to be $0.1$. Next, let $\Delta t$ 
be the time step-size ($\Delta t=5\times 10^{-2}$ in this work) such 
that $t_{n}=n\Delta t$ and $\psi^{(n)}\doteq\psi(\mathbf{r},t_{n})$ 
with $n\geq 0$. Then, for given $\psi^{(n)}$ at the $n$th time step, $\psi^{(n+1)}$ 
is obtained (implicitly) by a modified Crank-Nicolson method~\cite{delfour1981finite}:
\begin{equation}\label{eq:CN}
\begin{aligned}
& i\frac{\psi^{(n+1)}-\psi^{(n)}}{\Delta t}=\\
& \left(-\frac{1}{2}\nabla^2+V(\mathbf{r})+\frac{1}{2}(|\psi^{(n+1)}|^2+|\psi^{(n)}|^2)\right)%
\frac{\psi^{(n+1)}+\psi^{(n)}}{2},
\end{aligned}
\end{equation}
where cubic finite elements are employed for the spatial discretization as before. 
At each time step $n$, a nonlinear problem is solved by using Newton's method.
It should be pointed out in passing that the time-marching scheme employed in this 
work [cf.~Eq.~\eqref{eq:CN}] preserves both the squared $L^2$ norm (i.e., atom 
number)
\begin{equation}\label{eq:noa}
N(\psi)\doteq\int_{D}|\psi|^2\,dx,
\end{equation}
and the energy of the solutions
\begin{equation}
\mathcal{E}(\psi)=\int_{D}\bigg\lbrace{\frac{1}{4}|\nabla\psi|^2+%
\frac{1}{2}V(\mathbf{r})|\psi|^2+\frac{1}{4}|\psi|^4\bigg\rbrace} dx,
\end{equation}
to machine precision. We now turn to discussing the solutions obtained 
through the application of these numerical methods for the 3D NLS/GP problem.

\section{Numerical Results}
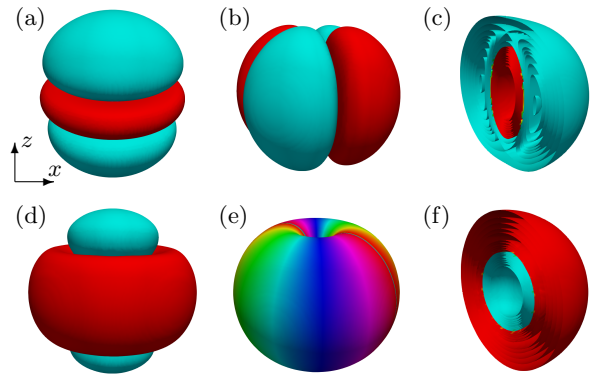
\begin{figure}[tbp]
\begin{center}
\begin{overpic}[width=\linfigwidth]{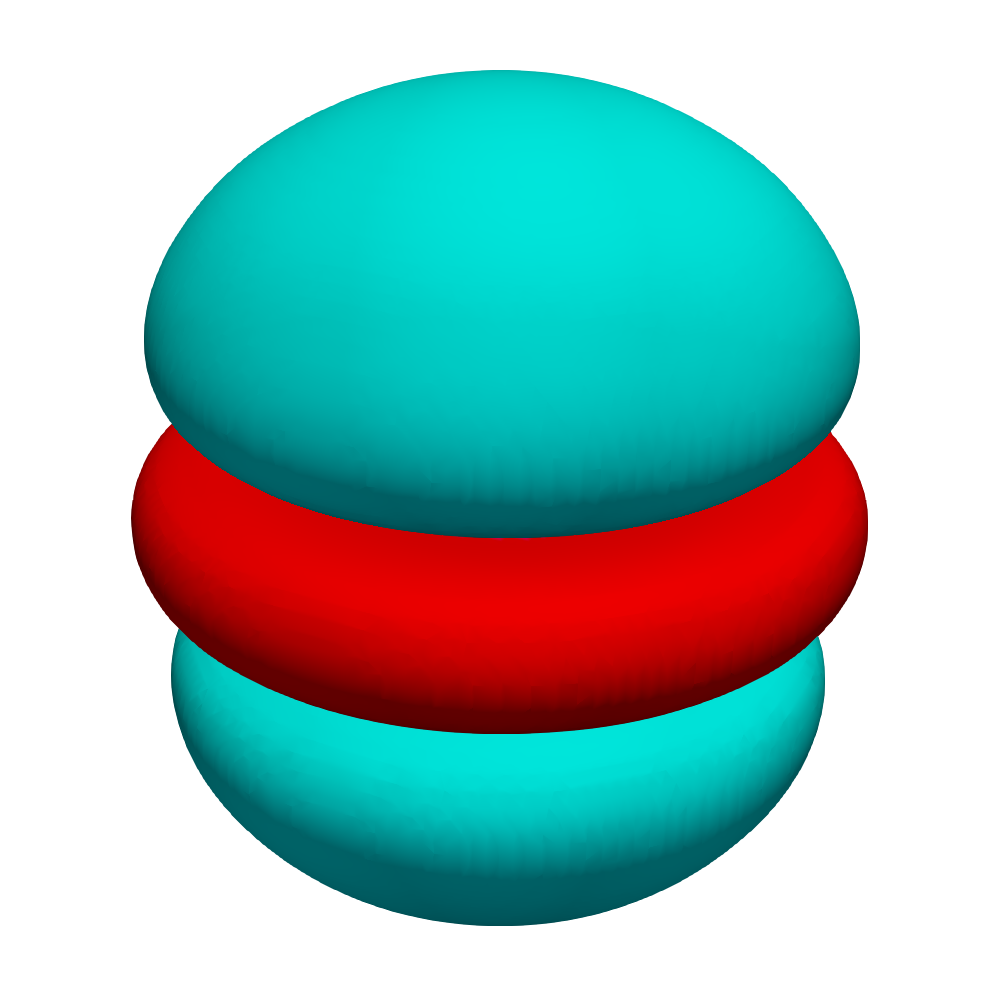}
\put(0,85){(a)}
\put(0,5){\color{black}\vector(1,0){20}}
\put(17,8){$x$}
\put(0,5){\color{black}\vector(0,1){20}}
\put(3,23){$z$}
\end{overpic}
\begin{overpic}[width=\linfigwidth]{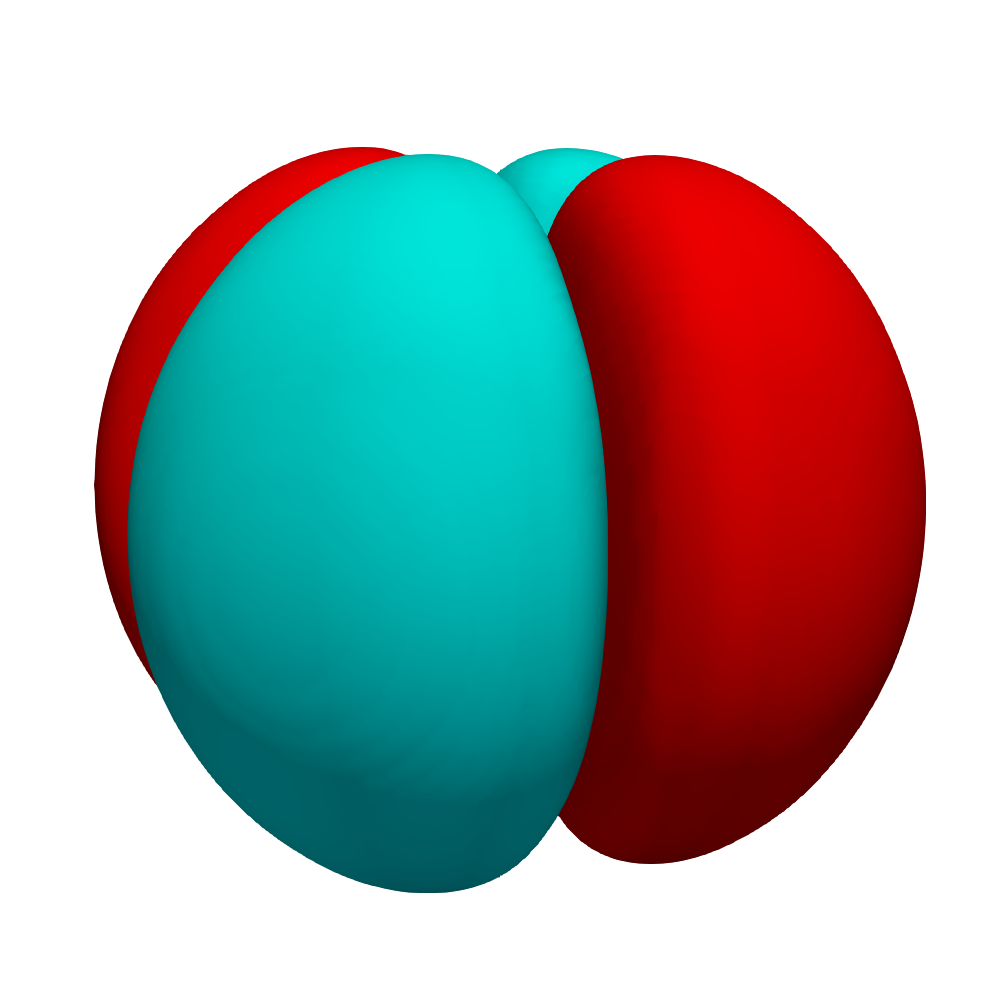}
\put(0,85){(b)}
\end{overpic}
\begin{overpic}[width=\linfigwidth,trim={200 150 100 150},clip]{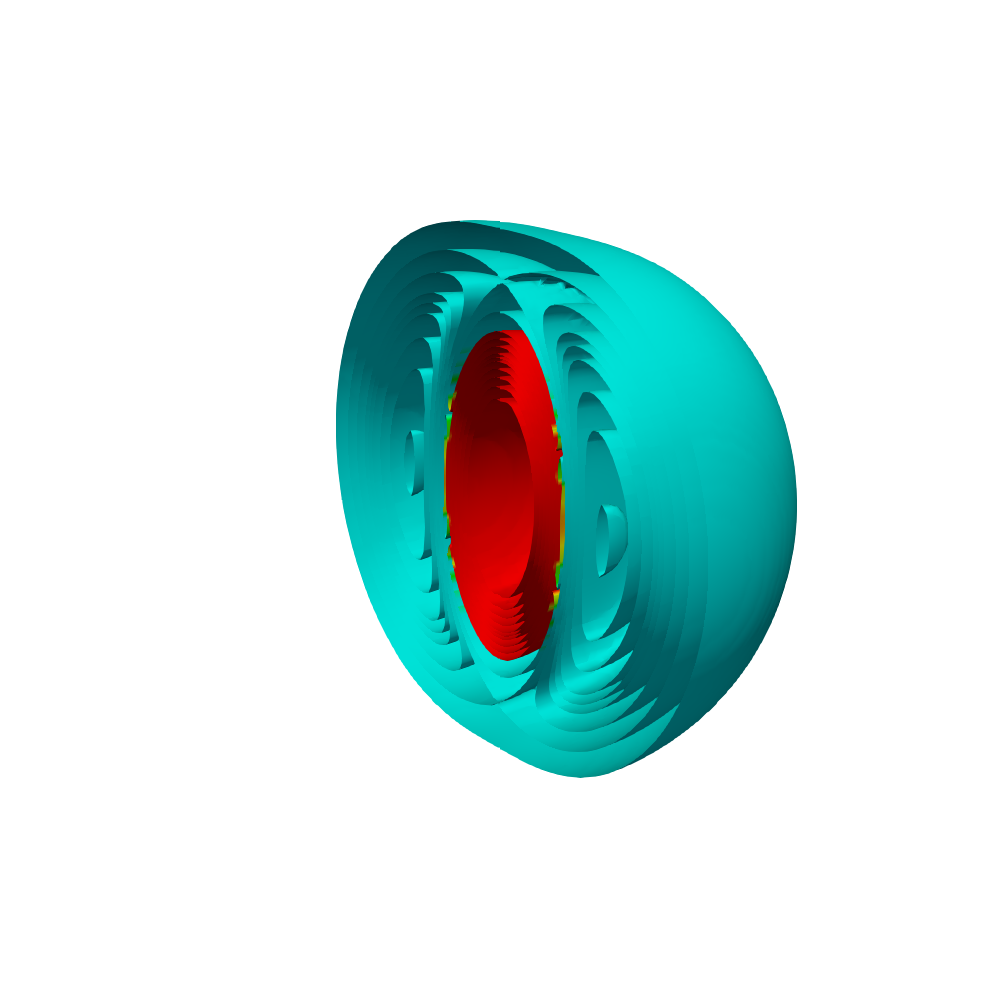}
\put(0,85){(c)}
\end{overpic}\\
\begin{overpic}[width=\linfigwidth]{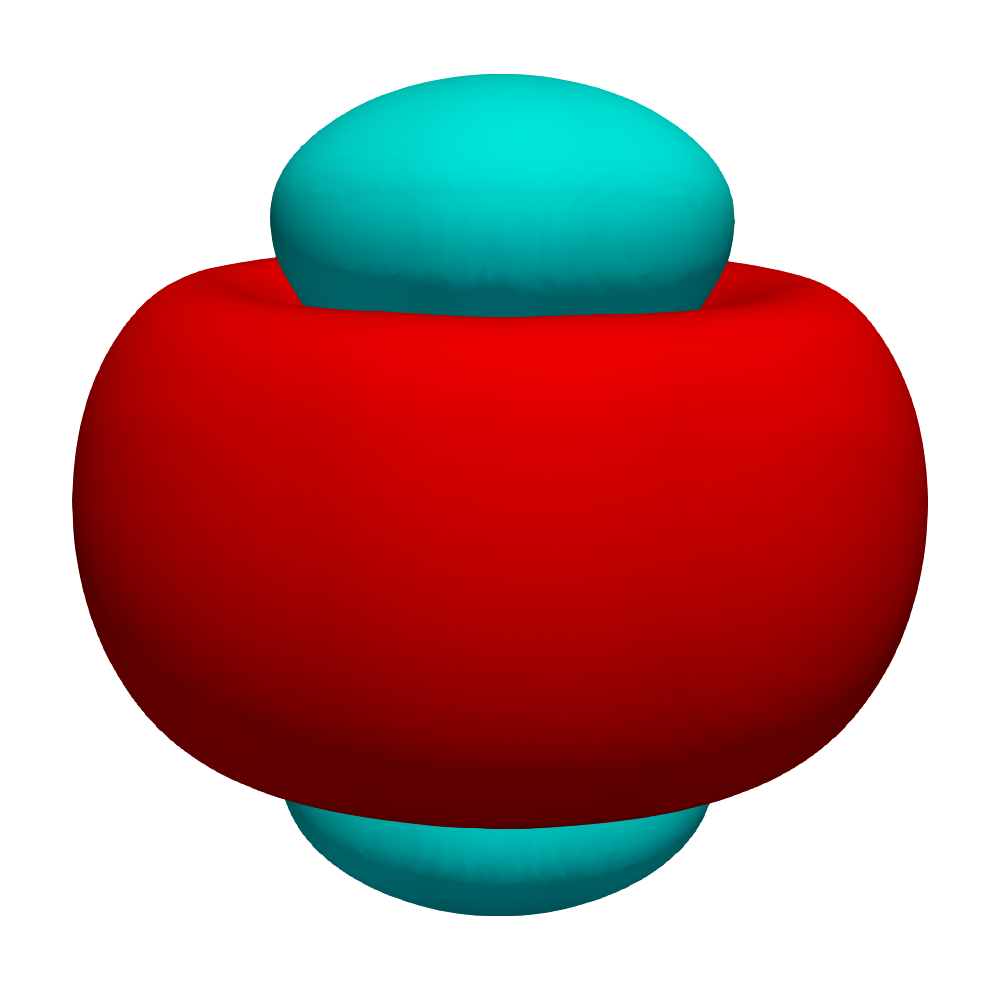}
\put(0,85){(d)}
\end{overpic}
\begin{overpic}[width=\linfigwidth]{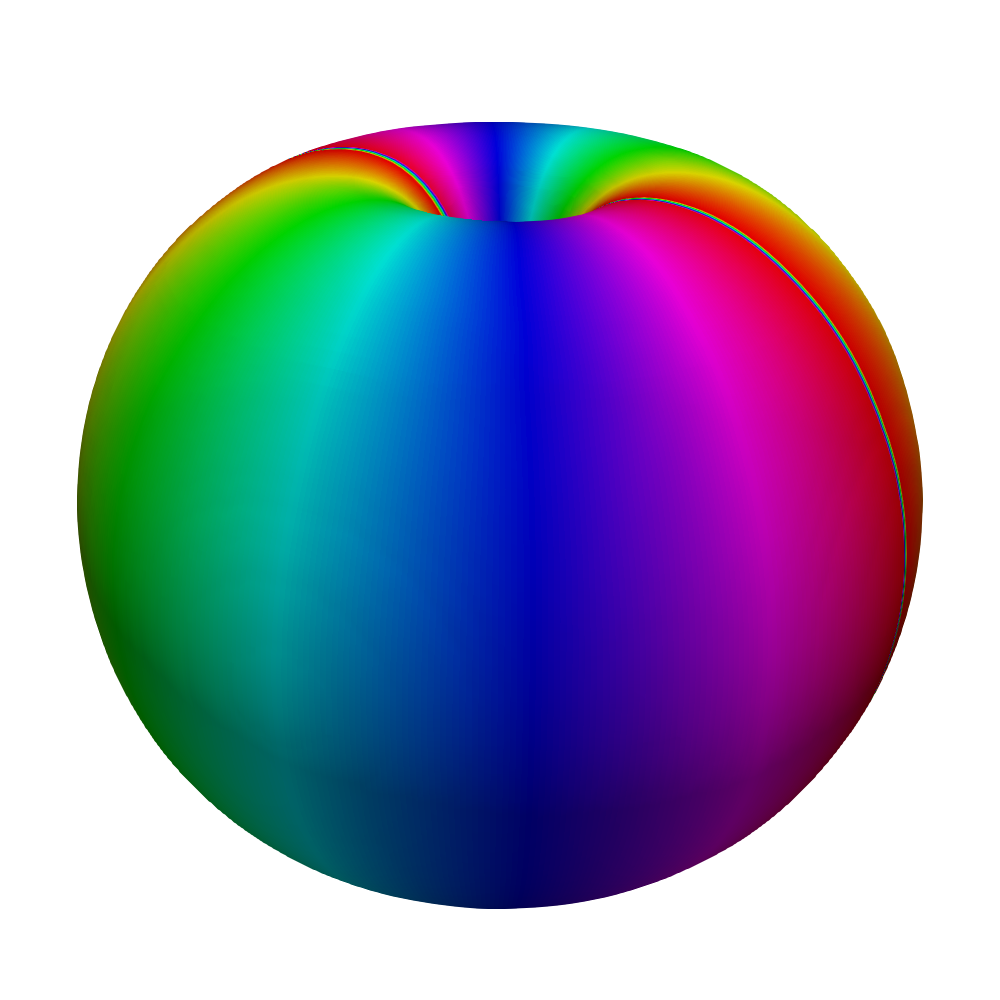}
\put(0,85){(e)}
\end{overpic}
\begin{overpic}[width=\linfigwidth,trim={200 150 100 150},clip]{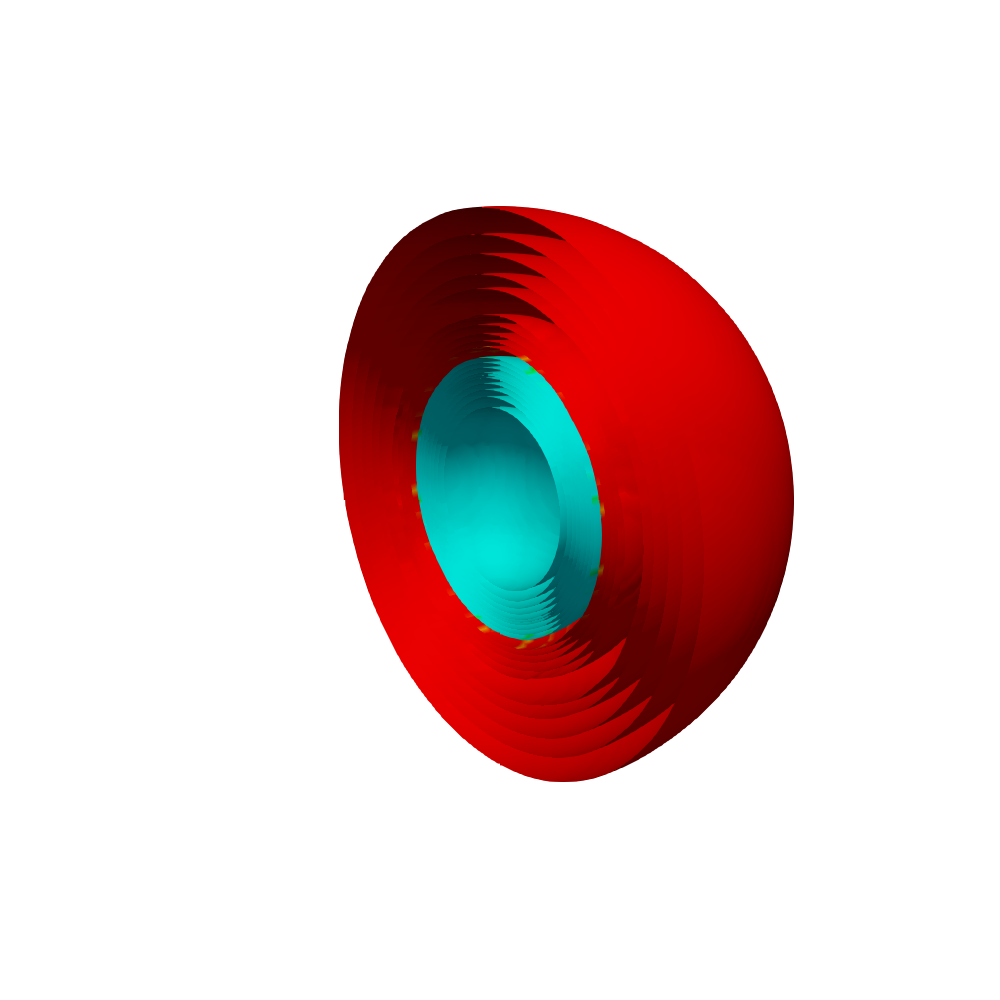}
\put(0,85){(f)}
\end{overpic}
\end{center}
\caption{Some solutions obtained by deflation that emanate from the 
second eigenvalue of the linear spectrum at $\mu=7/2$.
The colors represent the argument of the solutions, ranging from $-\pi$ 
to $\pi$ (blue and red represent a phase of $0$ and $\pm\pi$, respectively). 
The states in panels (a)-(d) and (f) are real, while (e) is complex.}
\label{fig_linear_states}
\end{figure}

We briefly describe the physical meaning of the quantum numbers
for the Cartesian, cylindrical and spherical states, as they are
useful in what follows. In the case of the Cartesian eigenfunctions, 
the quantum numbers $k, m$, and $n$ are simply the numbers of cuts along 
the $x$-, $y$-, and $z$-axes respectively. For instance, in~\cref{fig_linear_states}, 
panel (a) represents a $\ket{0,0,2}$ Cartesian state with 2 cuts along the 
$z$-axis (and $\pi$ phase differences across them), while panel (b)
is $\ket{1,1,0}$, bearing one planar cut along the $x$-axis, and 
one along the $y$-axis. Combinations of states are also possible, 
such as the one in panel (c) of $\ket{2,0,0} + r \ket{0,2,0} +
\ket{0,0,2}$
(in the particular example of this panel $r \approx 3.39$), 
which forms a 2D ring along the $y$- and $z$-axes embedded in 3D space.

Vortical structures and rings can be identified in the cylindrical
system of coordinates. Here, $K$ denotes the number of cylindrical 
(nodal) surfaces, $l$ the topological charge of the configuration and 
$n$ the number of planar cuts along the $z$-axis. For example, panel (d) 
of Fig.~\eqref{fig_linear_states} is a so-called ring dark soliton state 
(extended in 3D) $\ket{1,0,0}_{\rm cyl}$ that has been recently considered 
in~\cite{wenl}, and panel (e) is the $\ket{0,2,0}_{\rm cyl}$ state, i.e., 
a vortex line, piercing through the BEC with topological charge $l=2$.

Finally, in the spherical representation, $K$ denotes the number of
spherical (nodal) shells within the solution, $l-m$ denotes the number 
of planar cuts along the $z$-axis, and $m$ denotes the topological charge 
of vortical lines. Panel (f) is the $\ket{1,0,0}_{\rm sph}$ state corresponding
to a spherical shell dark solitary wave, which is also connected with recent 
work~\cite{wenl1}.

\begin{figure}[htbp]
\begin{center}
\begin{overpic}[width=\linfigwidth]{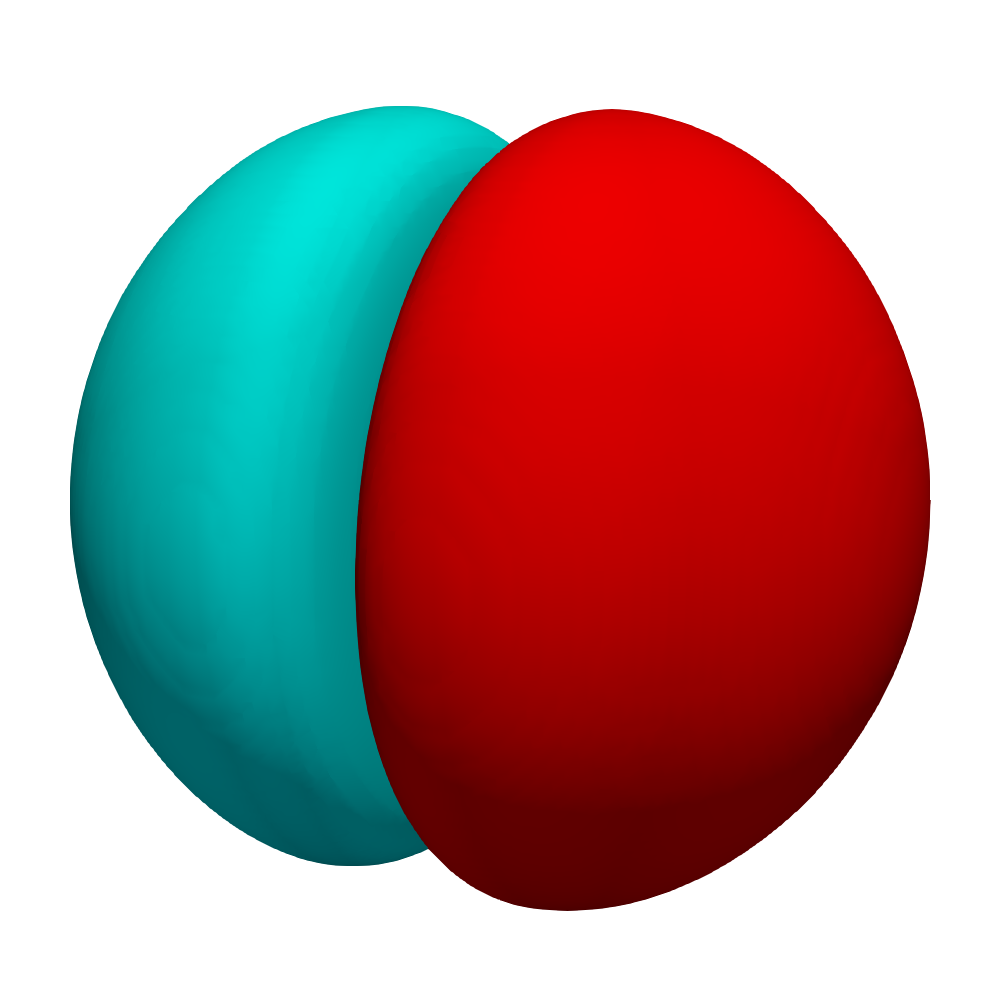}
\put(0,85){(a)}
\end{overpic}
\begin{overpic}[width=\linfigwidth]{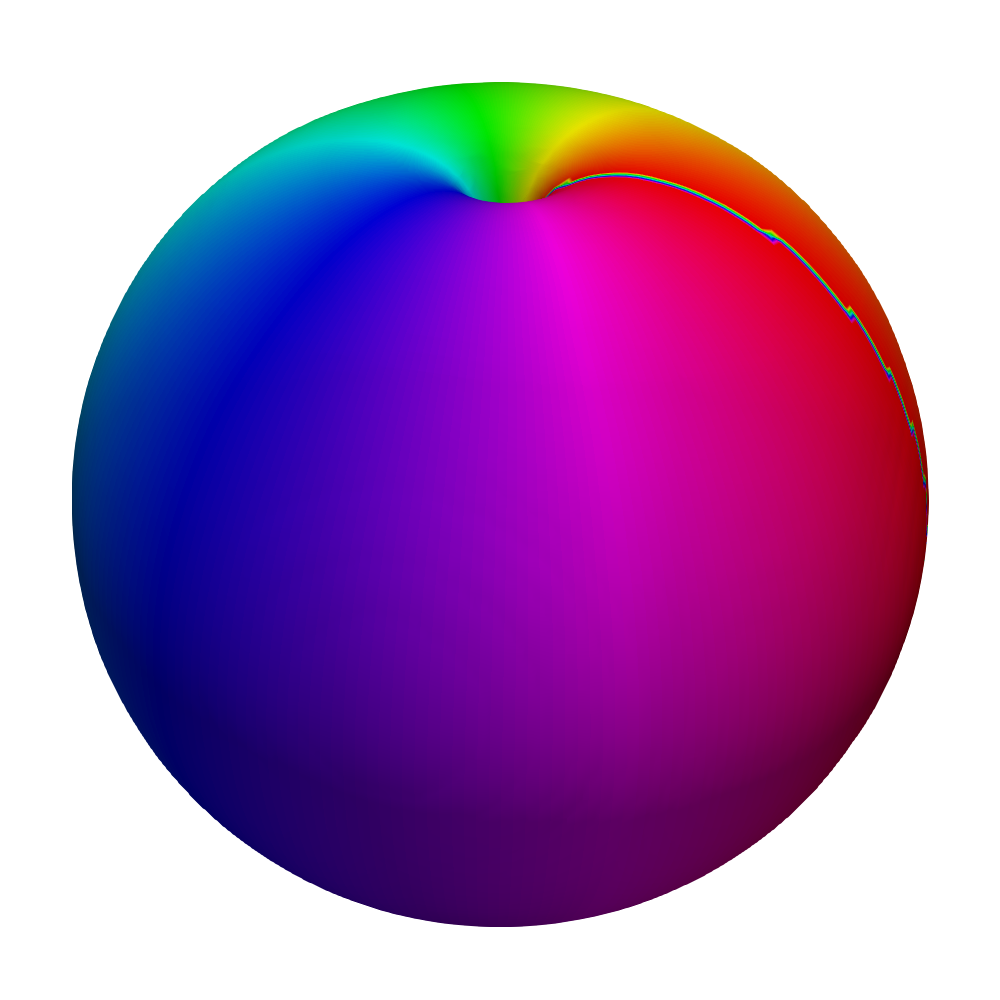}
\put(0,85){(b)}
\end{overpic}
\begin{overpic}[width=\linfigwidth]{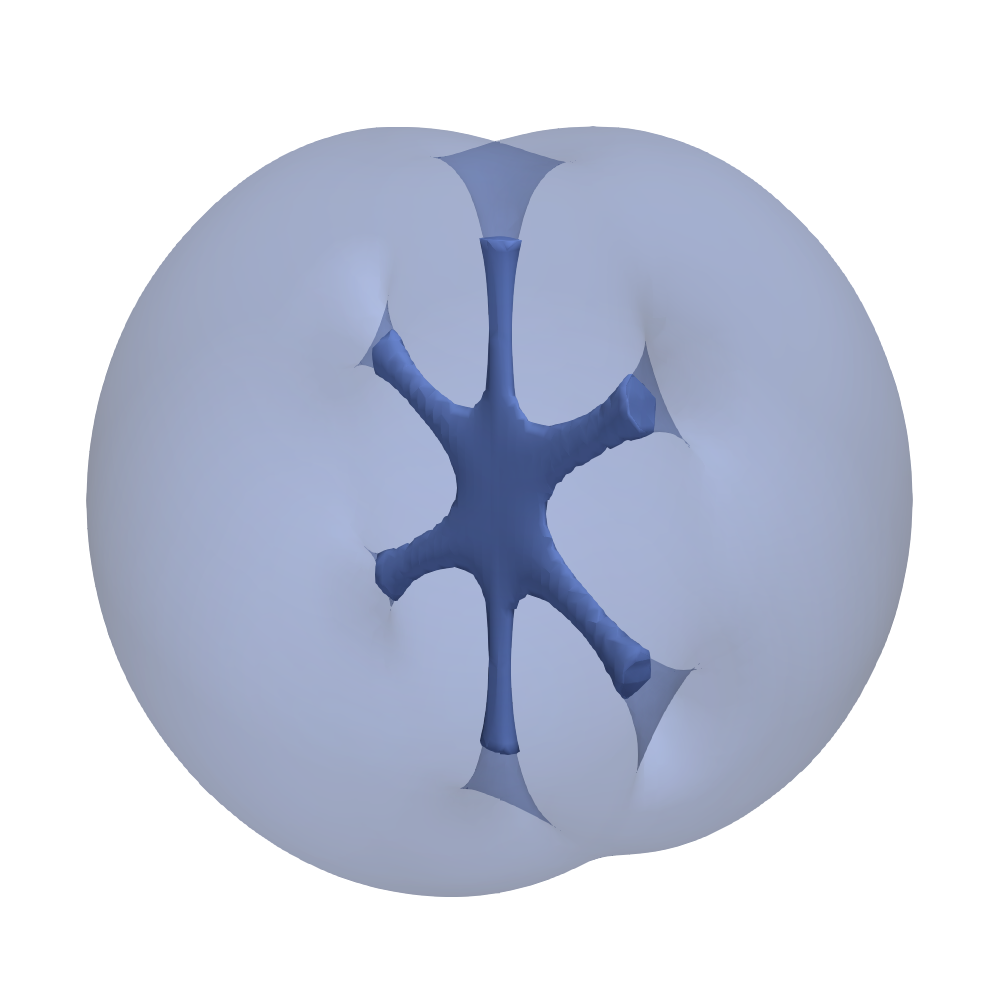}
\put(0,85){(c)}
\end{overpic}
\end{center}
\caption{Solutions emanating from the 1st excited state $\mu=5/2$
are shown in panels (a) and (b). In particular, panels (a) and (b) 
show a dipole and single vortex line solution. The colors represent 
the argument of the solutions, ranging from $-\pi$ to $\pi$ (blue and 
red represent a phase of $0$ and $\pm\pi$, respectively). Panel (c) 
also emanates from the 1st excited state and corresponds to 
the density isosurfaces of the Chladni soliton at densities $0.30$ 
and $0.35$.}
\label{fig_2_5}
\end{figure}

The ground state of the system (starting
at $\mu=3/2$) is known to always be spectrally and nonlinearly stable~\cite{pethick,stringari}.
The case of the 1st excited states (e.g.~dipolar states and single
vortex lines) emanating from $\mu=5/2$ is interesting but reasonably well
understood on the basis of corresponding 2D studies~\cite{siambook},
since no fundamentally novel states appear to emerge in 3D. Indicatively,~\cref{fig_2_5}(a) shows a 
$\ket{1,0,0}$ Cartesian state with one cut along the $x$-axis whereas~\cref{fig_2_5}(b) 
presents the $\ket{0,1,0}_{\textrm{cyl}}$ state corresponding to a single 
vortex line with topological charge $l=1$. The rotations of these solutions 
along the $x$, $y$, and $z$ axes such as the $\ket{0,1,0}$ and $\ket{0,0,1}$ 
Cartesian states are also obtained by deflation but are not reported. 
A typical example of a relevant solution is shown in \cref{fig_2_5}(c) 
and represents a so-called Chladni soliton, previously identified in cylindrical 
geometry in~\cite{mateo2014chladni, mateo2015stability}. However, the 
states that follow next are sufficiently complex to feature the emergence 
of unexpected patterns, yet it will still be possible to connect them to 
fundamental building blocks of topological patterns such as vortex lines 
and rings~\cite{komineas}.

We thus focus our discussion on states emanating from the 2nd excited 
state of the linear problem at $\mu = 7/2$. To that end, steady-state 
solutions to the NLS equation are identified by the deflation method at
$\mu=6$ (all the solutions presented in this paper are displayed at $\mu=6$). The branches are then continued backward in $\mu$ down to the linear limit by a standard zero-order continuation method~\cite[\S4.4.2]{seydel2010}. 

\begin{figure}[htbp]
\begin{center}
\begin{overpic}[width=\linfigwidth]{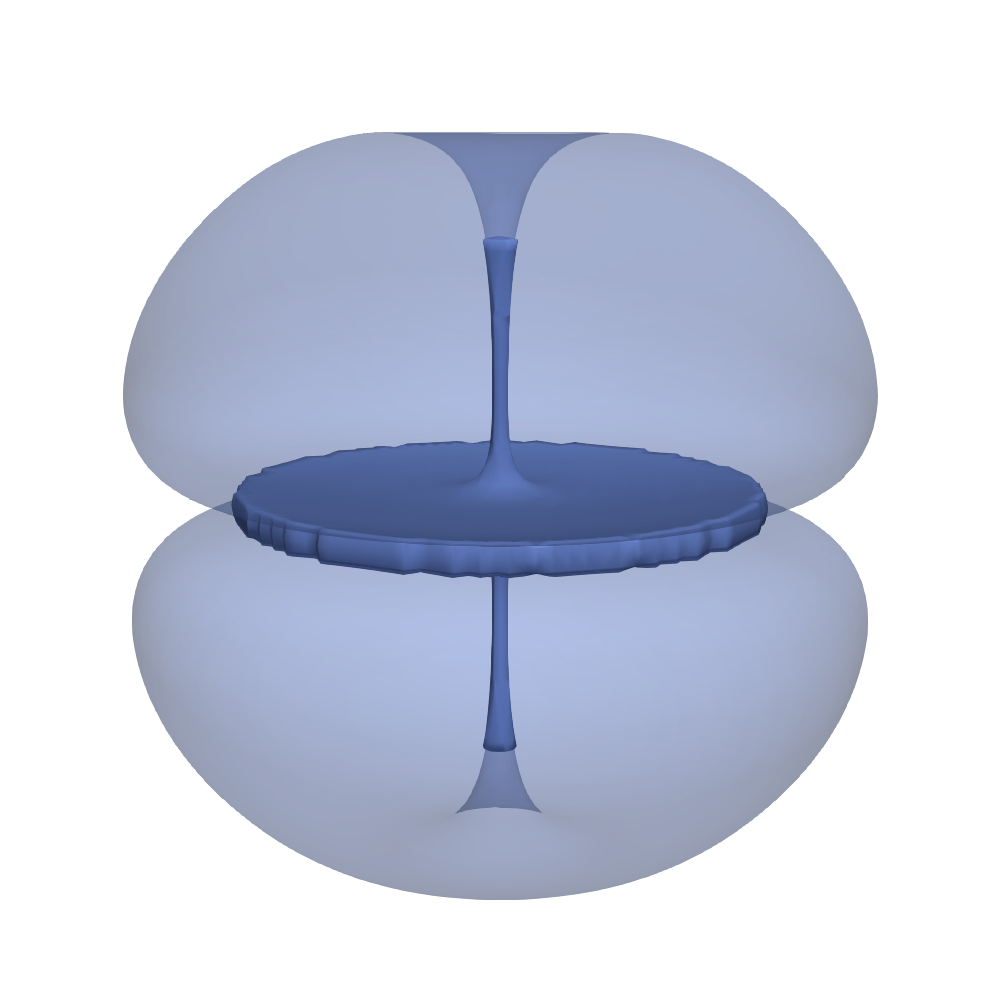}
\put(0,85){(a)}
\end{overpic}
\begin{overpic}[width=\linfigwidth]{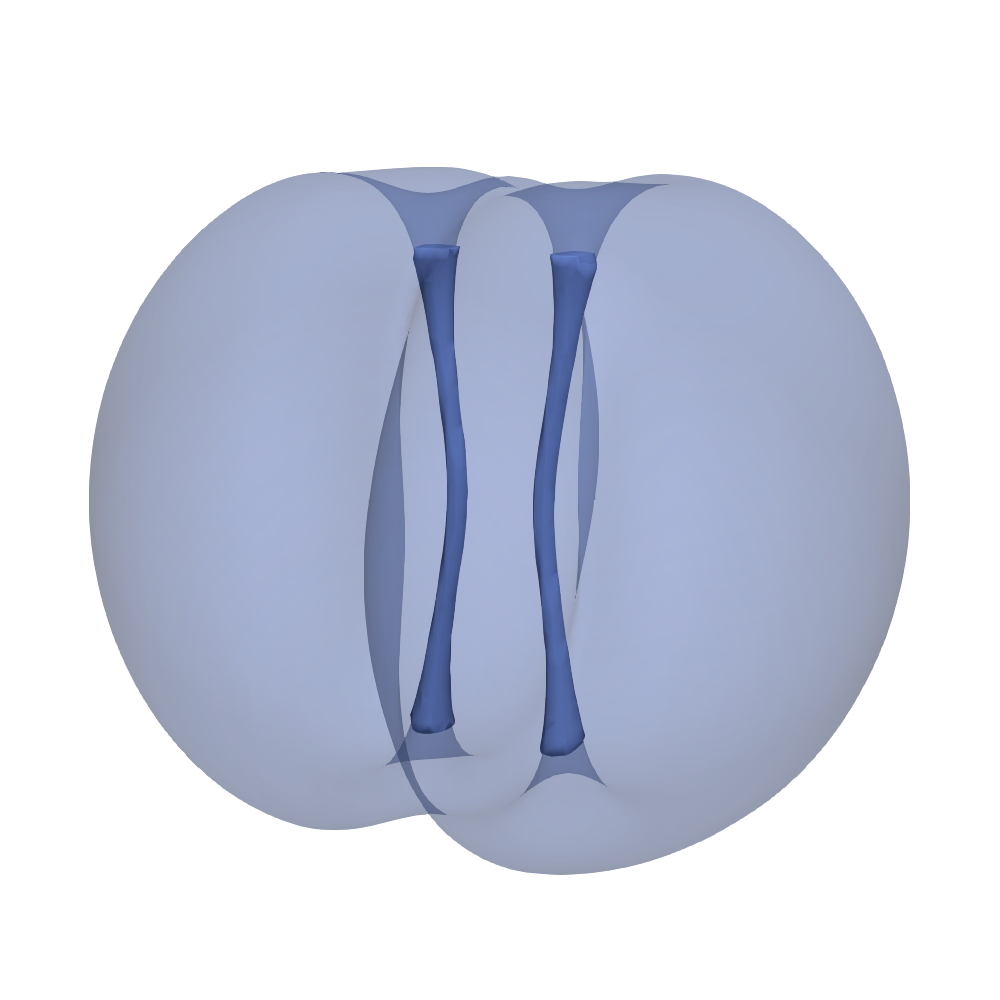}
\put(0,85){(b)}
\end{overpic}
\begin{overpic}[width=\linfigwidth]{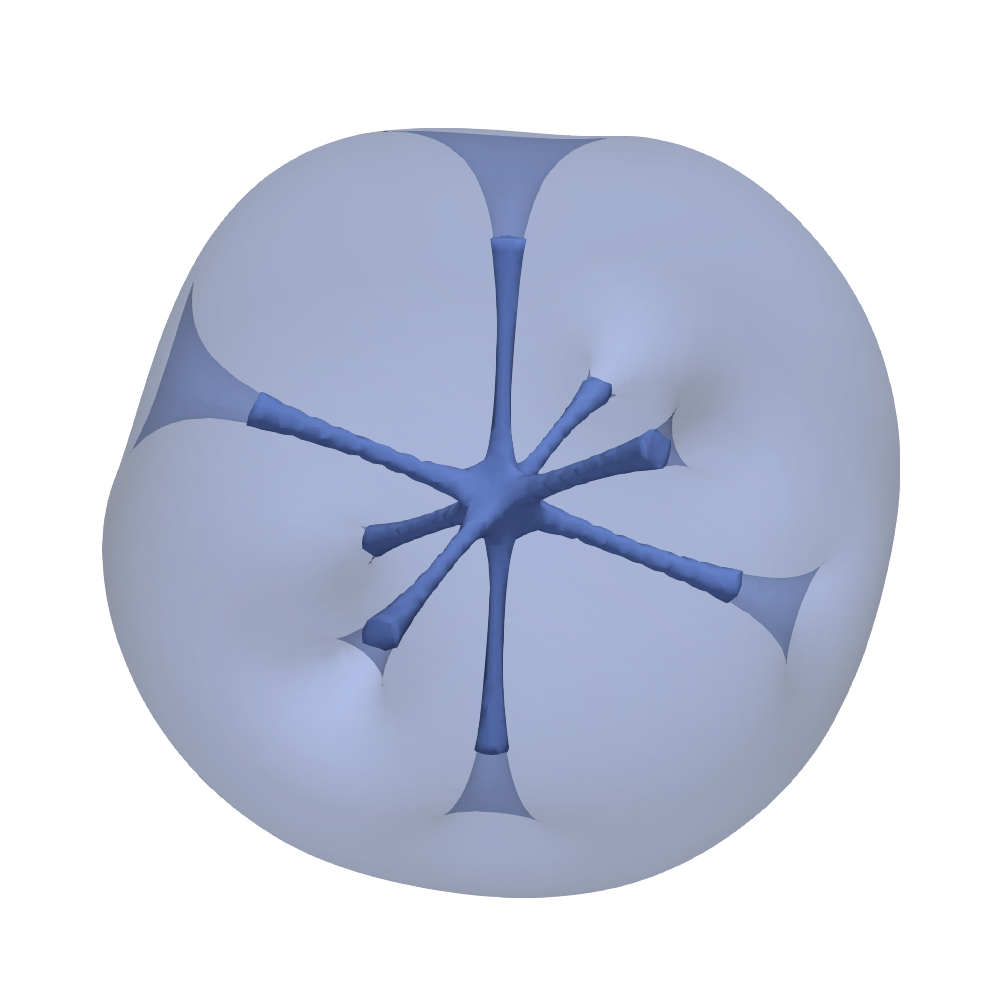}
\put(0,85){(c)}
\end{overpic}\\
\begin{overpic}[width=\linfigwidth]{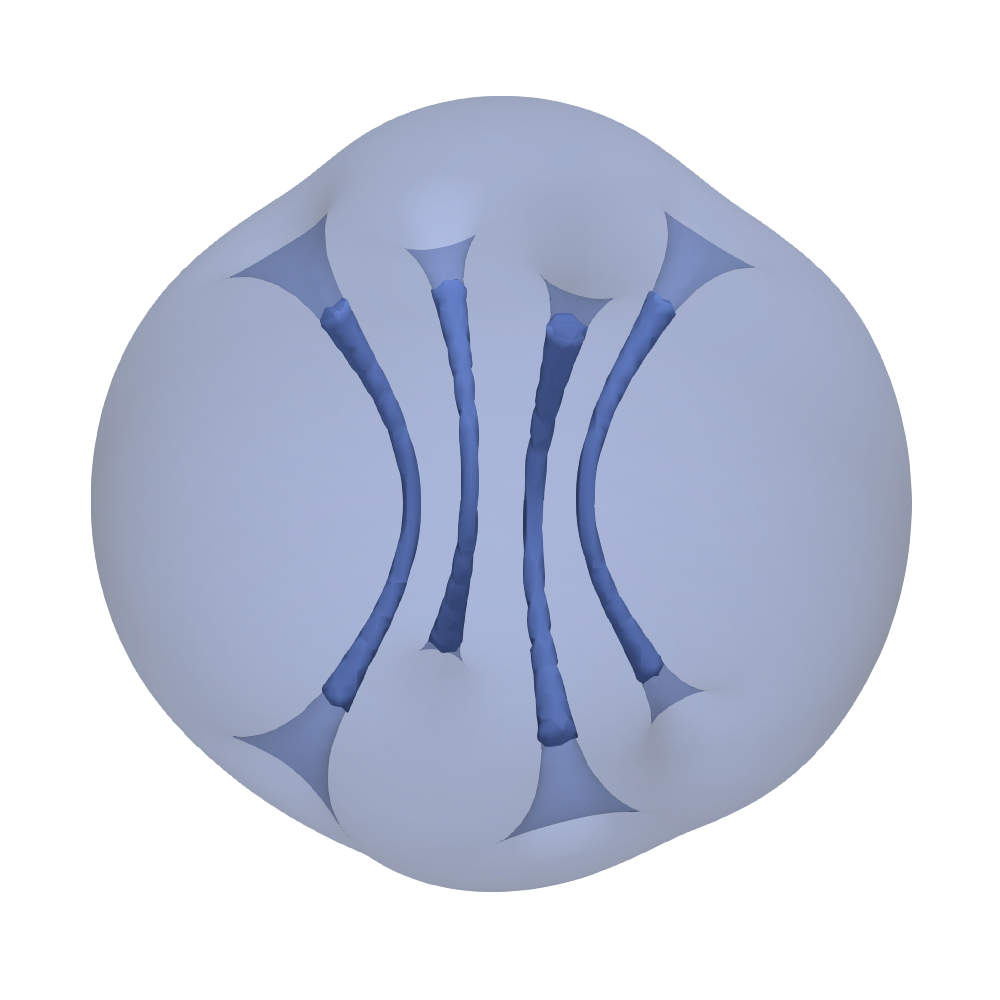}
\put(0,85){(d)}
\end{overpic}
\begin{overpic}[width=\linfigwidth]{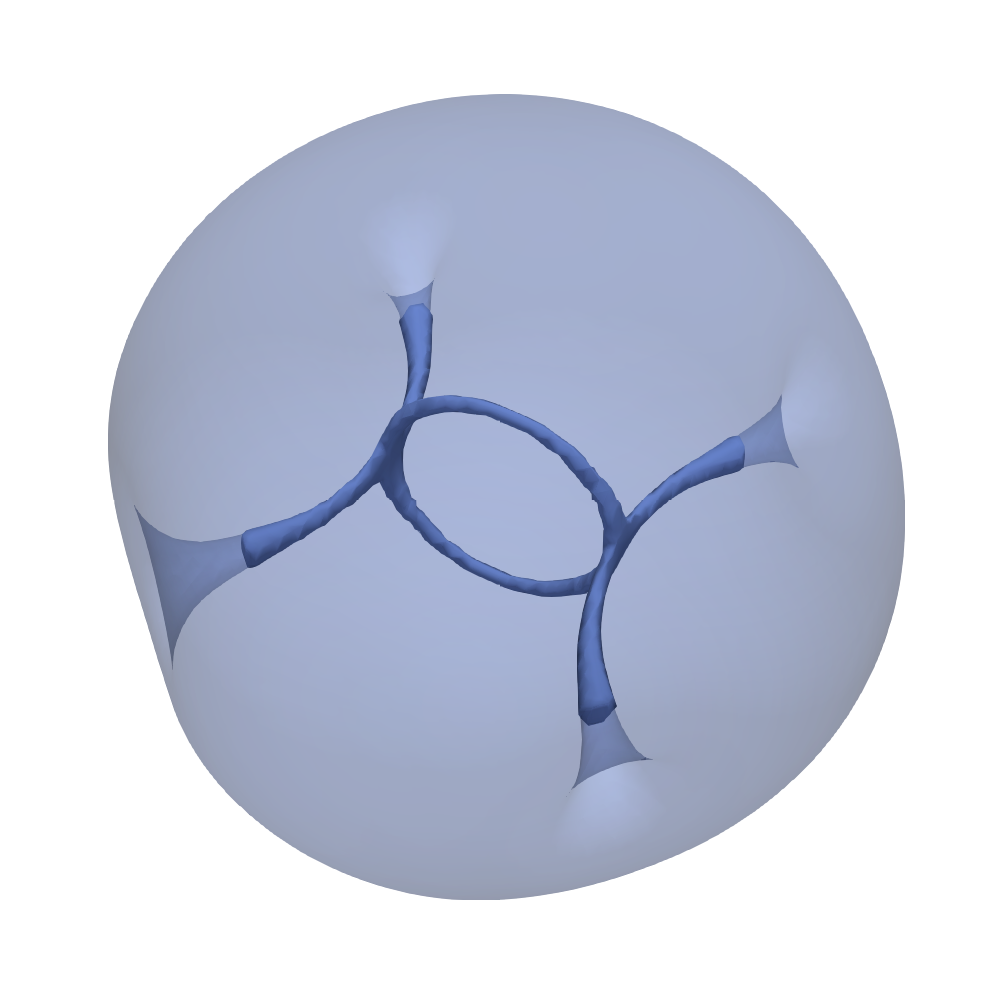}
\put(0,85){(e)}
\end{overpic}
\begin{overpic}[width=\linfigwidth]{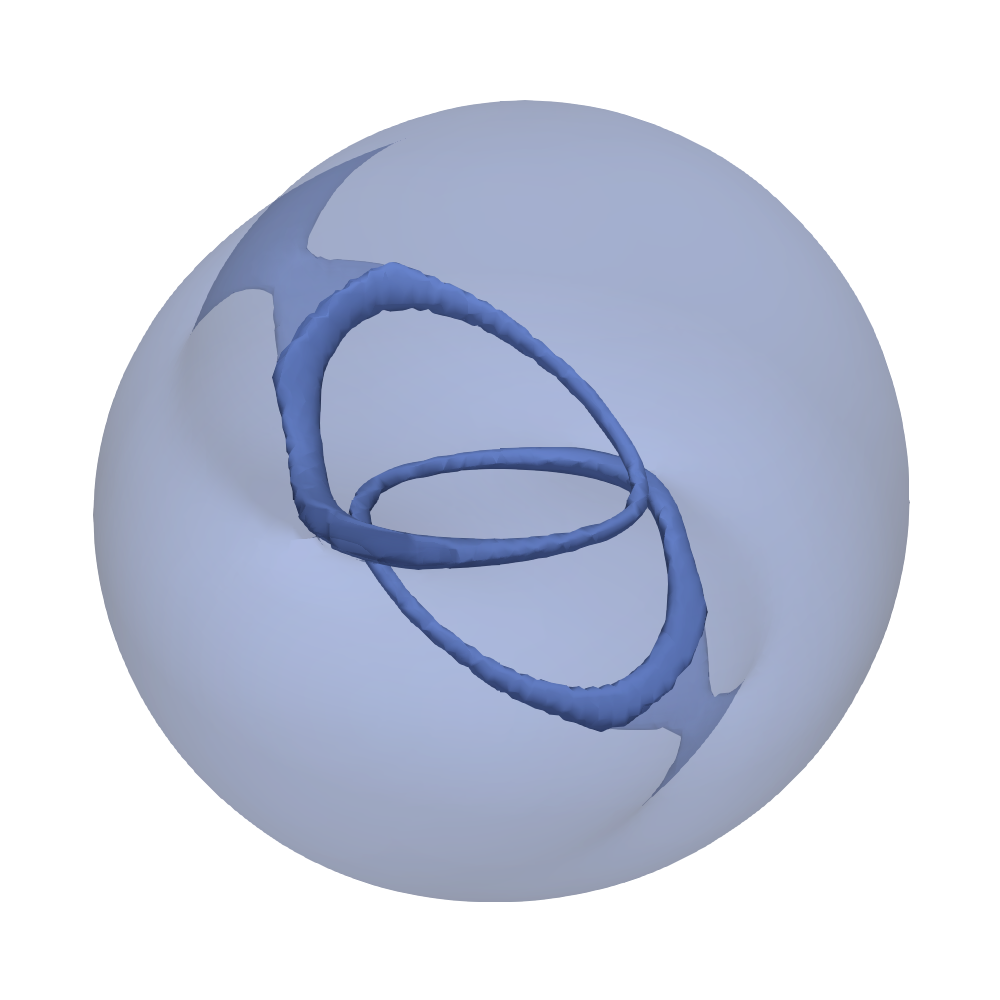}
\put(0,85){(f)}
\end{overpic}
\end{center}
\caption{More exotic solutions discovered by deflation that emanate 
from the 2nd excited state. The panels show density isosurfaces
of the states at densities $0.30$ and $0.35$.}
\label{fig_new_states}
\end{figure}

\begin{figure}[htbp]
\begin{center}
\begin{overpic}[width=\linfigwidth]{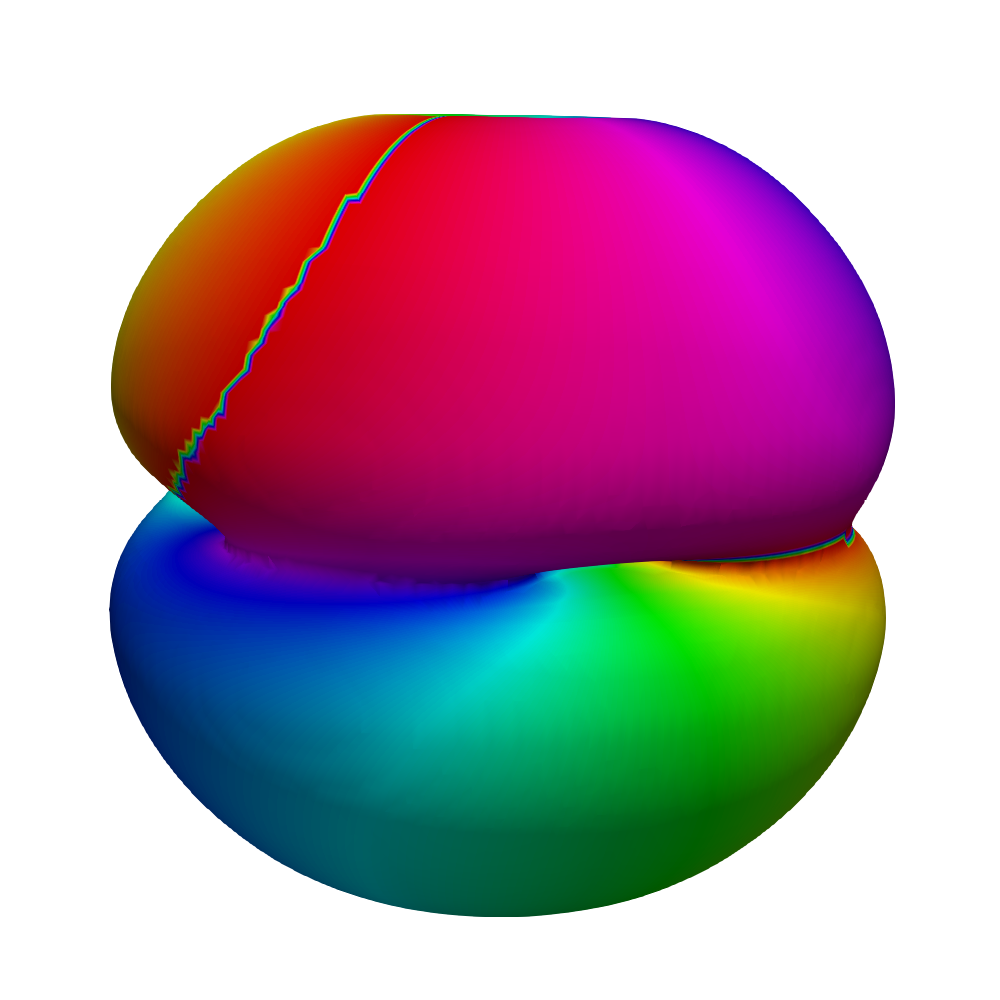}
\end{overpic}
\hspace{0.5cm}
\begin{overpic}[width=\linfigwidth]{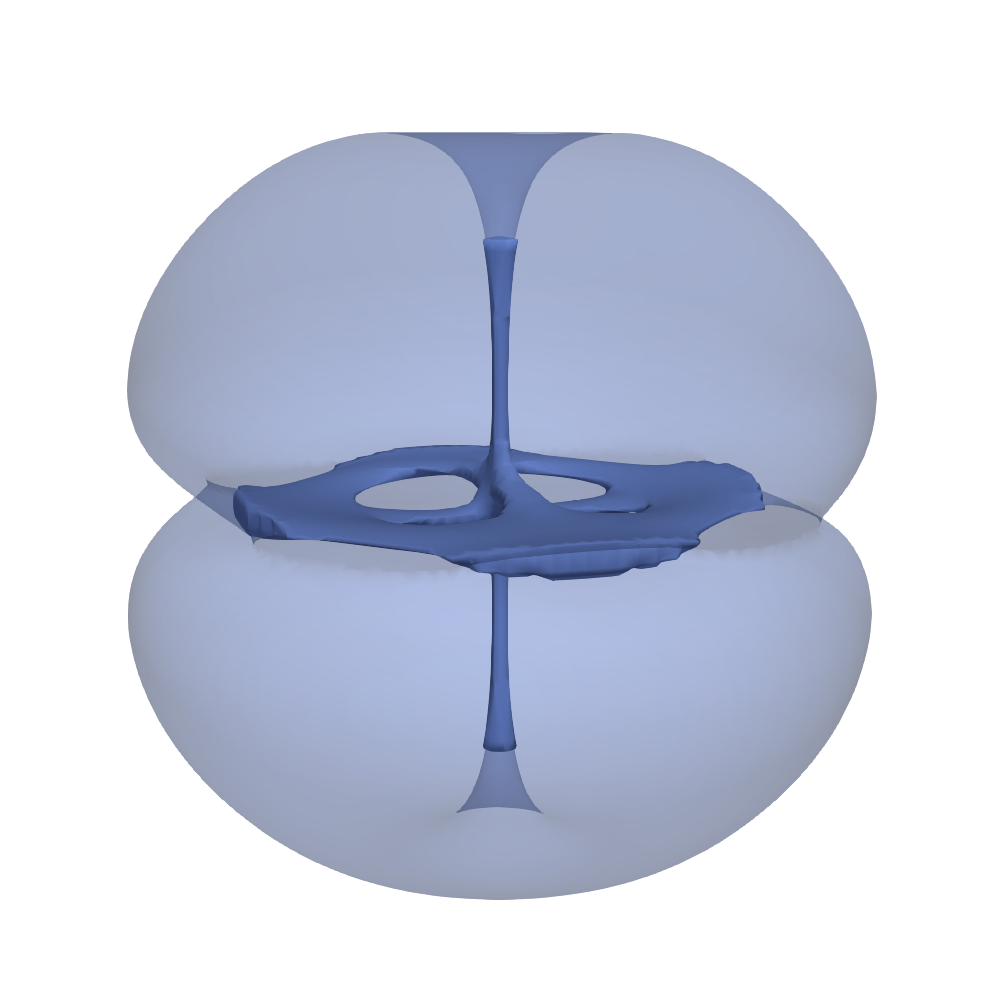}
\end{overpic}
\end{center}
\caption{Left: argument of the solution at $\mu = 6$. This branch 
bifurcates from the state of~\cref{fig_new_states}(a) at $\mu=5.84$ 
(see~\cref{fig_bifurcation_diagram}). Right: density isosurfaces 
of the state at densities $0.30$ and $0.35$.}
\label{fig_sol_2}
\end{figure}

Examples of these solutions are shown in \cref{fig_new_states}. In this 
figure, we observe that deflation enables us to converge to states with 
multiple coherent structures such as the one of panel (a) consisting 
of a vortex line and a planar dark soliton. The linear state corresponding 
to such a nonlinear waveform is $\ket{0,1,1}_{\rm cyl}$. This nonlinear state 
undertakes a symmetry--breaking bifurcation at $\mu=5.84$ and gives birth 
to the waveform of \cref{fig_sol_2}. However, more complex multi-vortex topological
states can progressively be identified as well. Panel (b) of~\cref{fig_new_states}
represents a pair of vortex lines: at the linear limit such a state can be 
formulated as the linear combination $\ket{1,1,0} + i \ket{0,2,0}$, in line with
what is known about vortex dipole bifurcations~\cite{siambook}. Panel (c) represents
what was termed a vortex star in~\cite{crasovan2004three}, arising at the linear 
limit via the linear combination $\ket{2,0,0}-\ket{0,2,0}+i[\ket{2,0,0}-\ket{0,0,2}]$. 
Panel (d) shows a generalization of the well-known 2D vortex quadrupole~\cite{mikko} 
consisting of 4 bent yet alternating charged vortex lines.

\begin{figure*}[htbp]
\begin{center}
\begin{overpic}[width=0.7\textwidth]{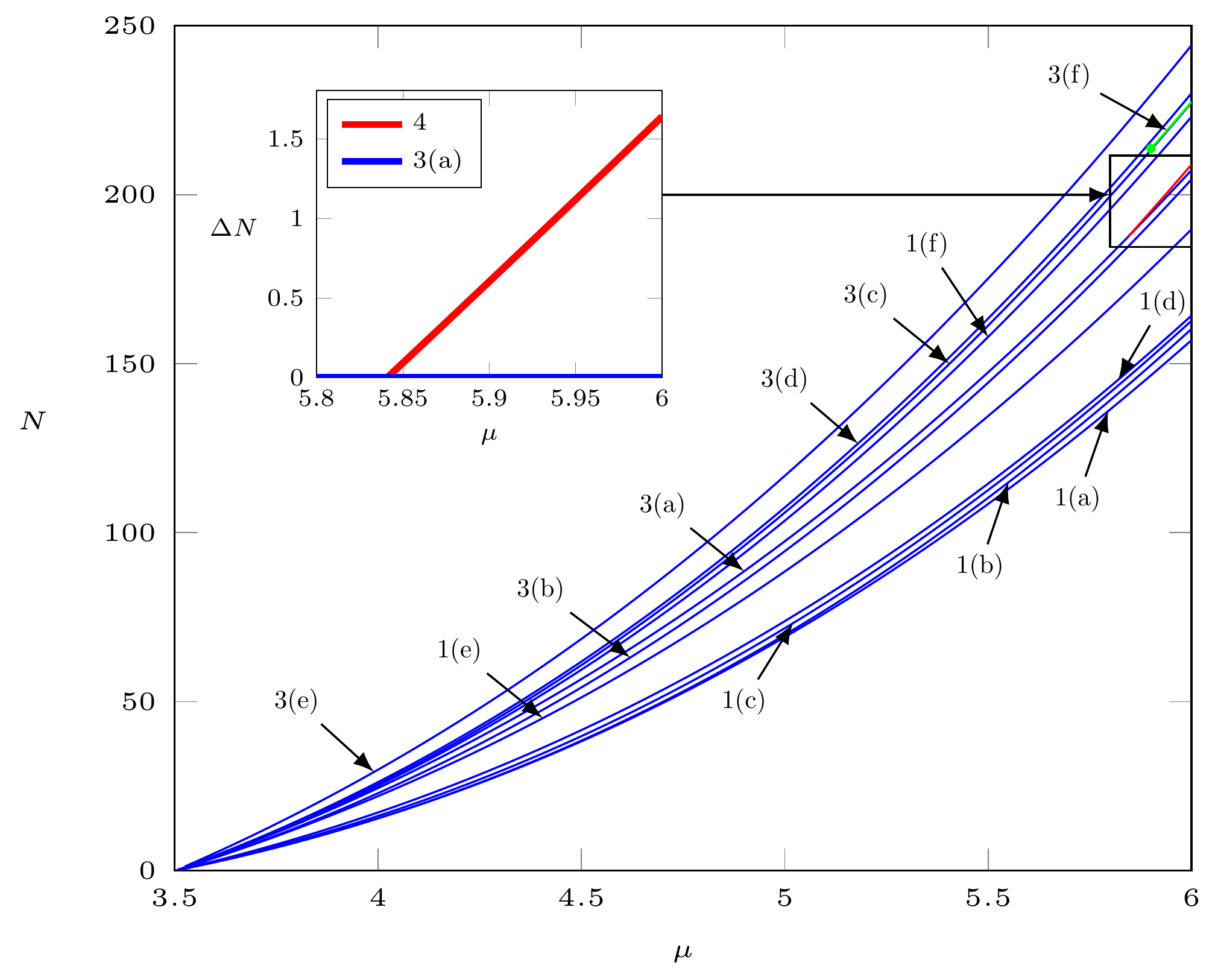}
\end{overpic}
\end{center}
\caption{Bifurcation diagram of the solutions emanating from the 
2nd excited state at $\mu=7/2$. The labels indicate the solutions 
represented in the different panels of \cref{fig_linear_states,fig_new_states}. 
The main panel corresponds to the total number of atoms $N$ as a 
function of $\mu$, while the top-left inset shows the atom number 
difference $\Delta N$ between the branch \cref{fig_new_states}(a) 
and \cref{fig_sol_2}, colored in red. The green branch illustrates the solution displayed in \cref{fig_new_states}(f) and its terminal point at $\mu=5.9$.}
\label{fig_bifurcation_diagram}
\end{figure*}

The solutions in \cref{fig_new_states}(a)-(d) either allow for a direct tracing 
of their linear limit or have been previously identified. However, deflation allows 
us to go well beyond these. Important examples of this arise in panels (e) and (f) 
of \cref{fig_new_states}. Panel (e) consists of a vortex ring combined with 2 
(oppositely charged) vortex line ``handles''. This state, too, can be identified at 
the linear limit through a more complex topologically charged combination, as 
$\ket{2,0,0} + \ket{0,2,0} + i \ket{1,0,1}$. Such a state exhibiting a vortex ring 
with multiple vortex lines attached to it has not been previously reported, to the 
best of our knowledge. Even more complex is the state in panel (f), which does not 
bear a linear analogue. This state involves 2 vortex rings, both of which are bent; 
i.e.~instead of having two ``perpendicular'' vortex rings (e.g., in the $xy$- and 
$yz$-planes), it is as if the top half of the one has connected itself with the right 
half of the other and the bottom half of one with the left half of the other. This 
configuration was discovered by deflation at $\mu = 6$ but the branch terminates
by $\mu = 5.9$ without ever reaching the linear limit of $\mu=7/2$. In other words, 
this appears to be a purely nonlinear state not derivable by some suitable combination 
of linear eigenstates. We conclude the presentation of our numerical results 
for the 2nd exciting states by presenting the associated bifurcation diagram in 
\cref{fig_bifurcation_diagram} where the total number of atoms [cf. Eq.~\eqref{eq:noa}]
is used as the diagnostic functional. The inset panel in the top-left corner 
therein uses the atom--number--difference $\Delta N$ between the branches \cref{fig_new_states}(a) 
and~\cref{fig_sol_2} to illustrate a bifurcation in the diagram. 

\begin{figure}[tbp]
\begin{center}
\hspace{0.1cm}
\begin{overpic}[width=\linfigwidth,trim={50 100 50 40}, clip]{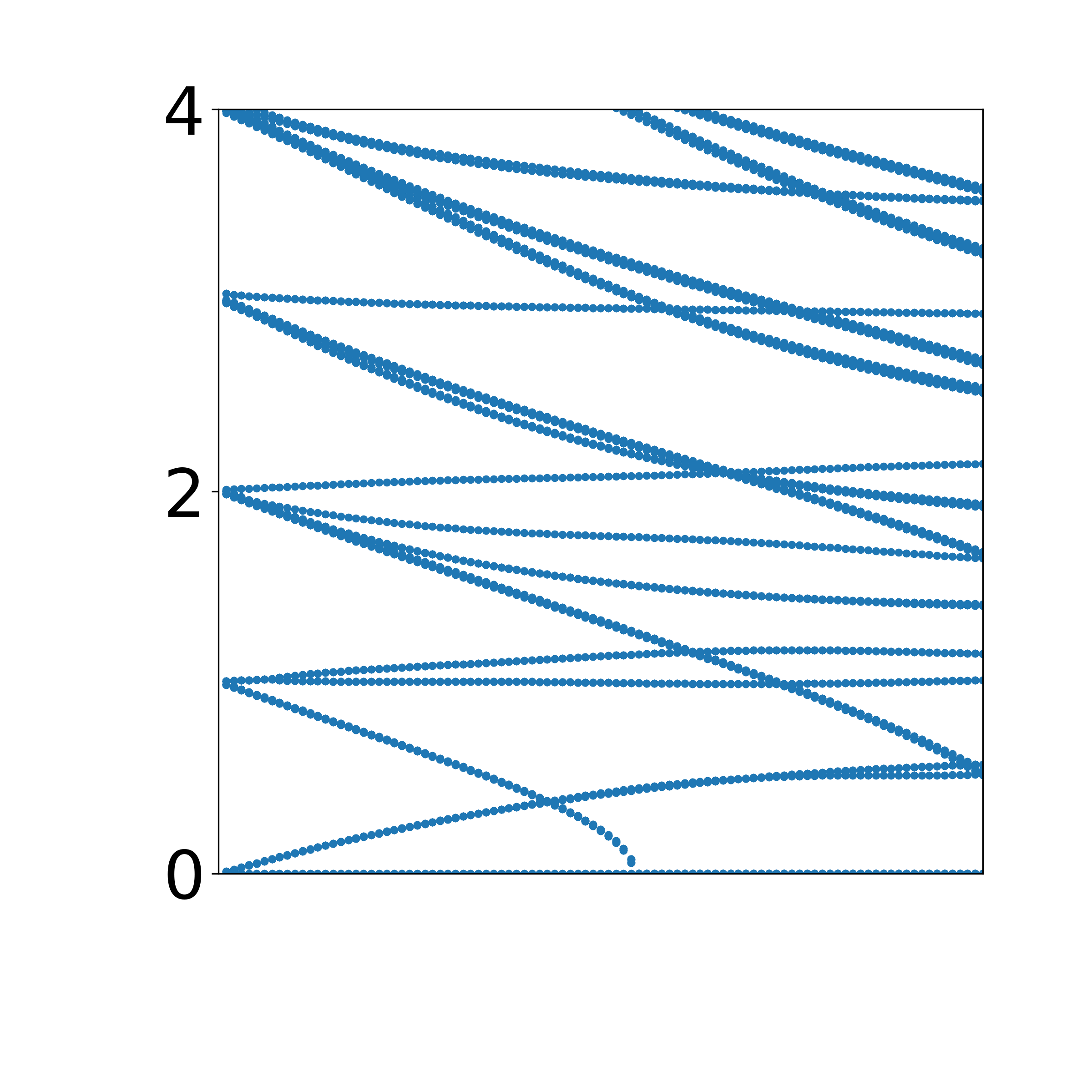}
\put(-14,32){\rotatebox{90}{$\mathcal{R}(\omega)$}}
\end{overpic}
\begin{overpic}[width=\linfigwidth,trim={50 100 50 40}, clip]{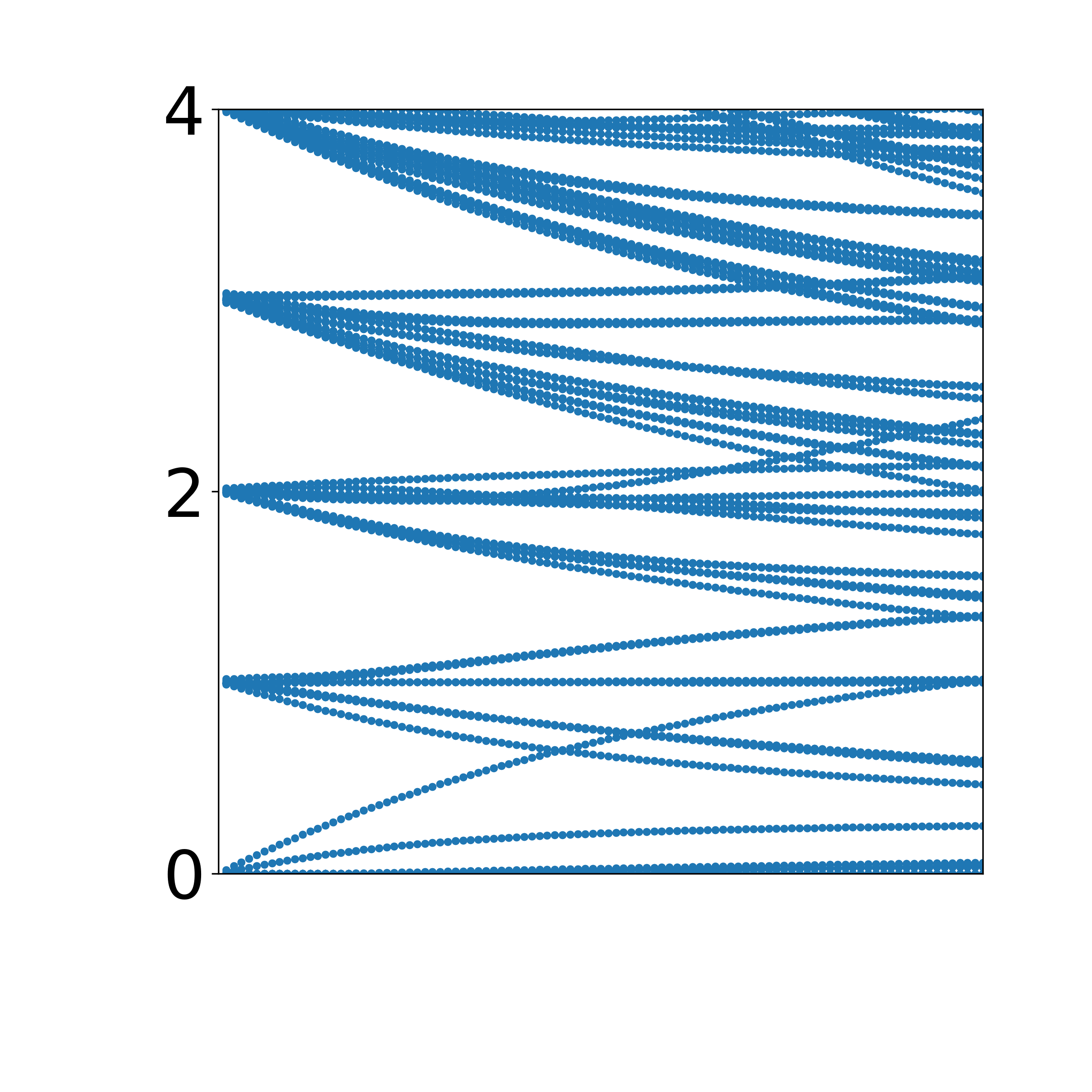}
\end{overpic}
\begin{overpic}[width=\linfigwidth,trim={50 100 50 40}, clip]{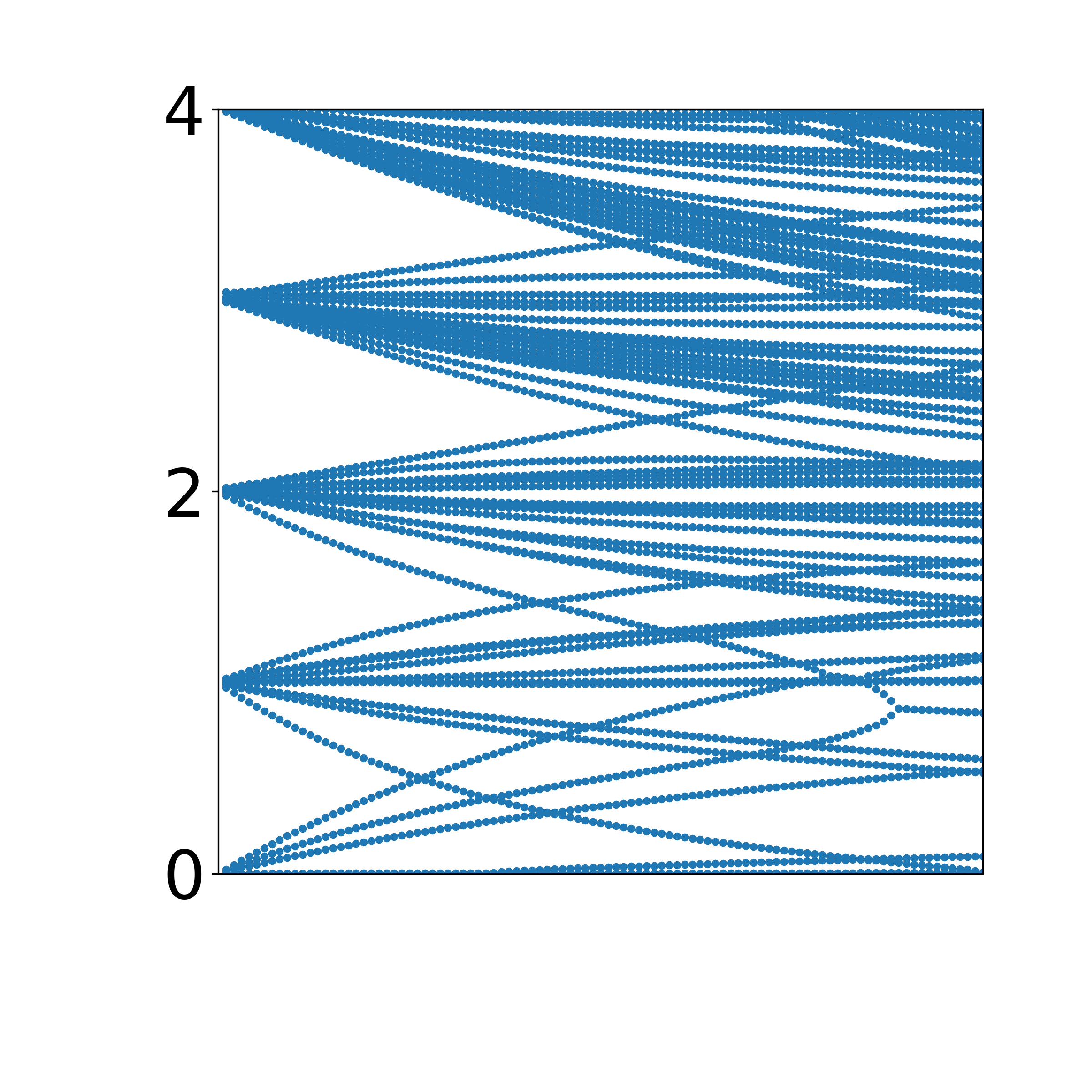}
\end{overpic}
\\
\vspace{0.1cm}
\hspace{0.1cm}
\begin{overpic}[width=\linfigwidth,trim={50 75 50 40}, clip]{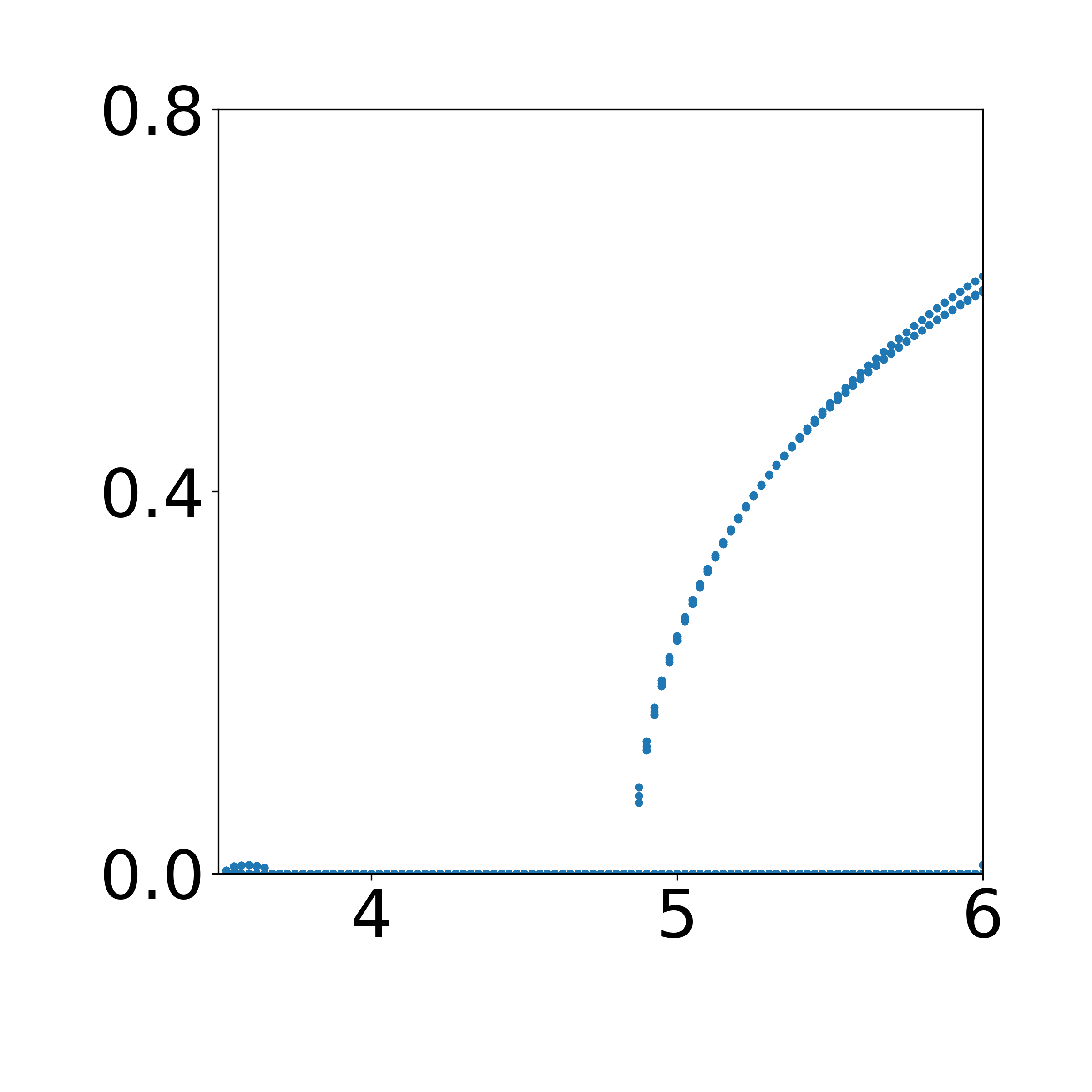}
\put(55,-6){$\mu$}
\put(-14,38){\rotatebox{90}{$\mathcal{I}(\omega)$}}
\put(15,82){(a)}
\end{overpic}
\begin{overpic}[width=\linfigwidth,trim={50 75 50 40}, clip]{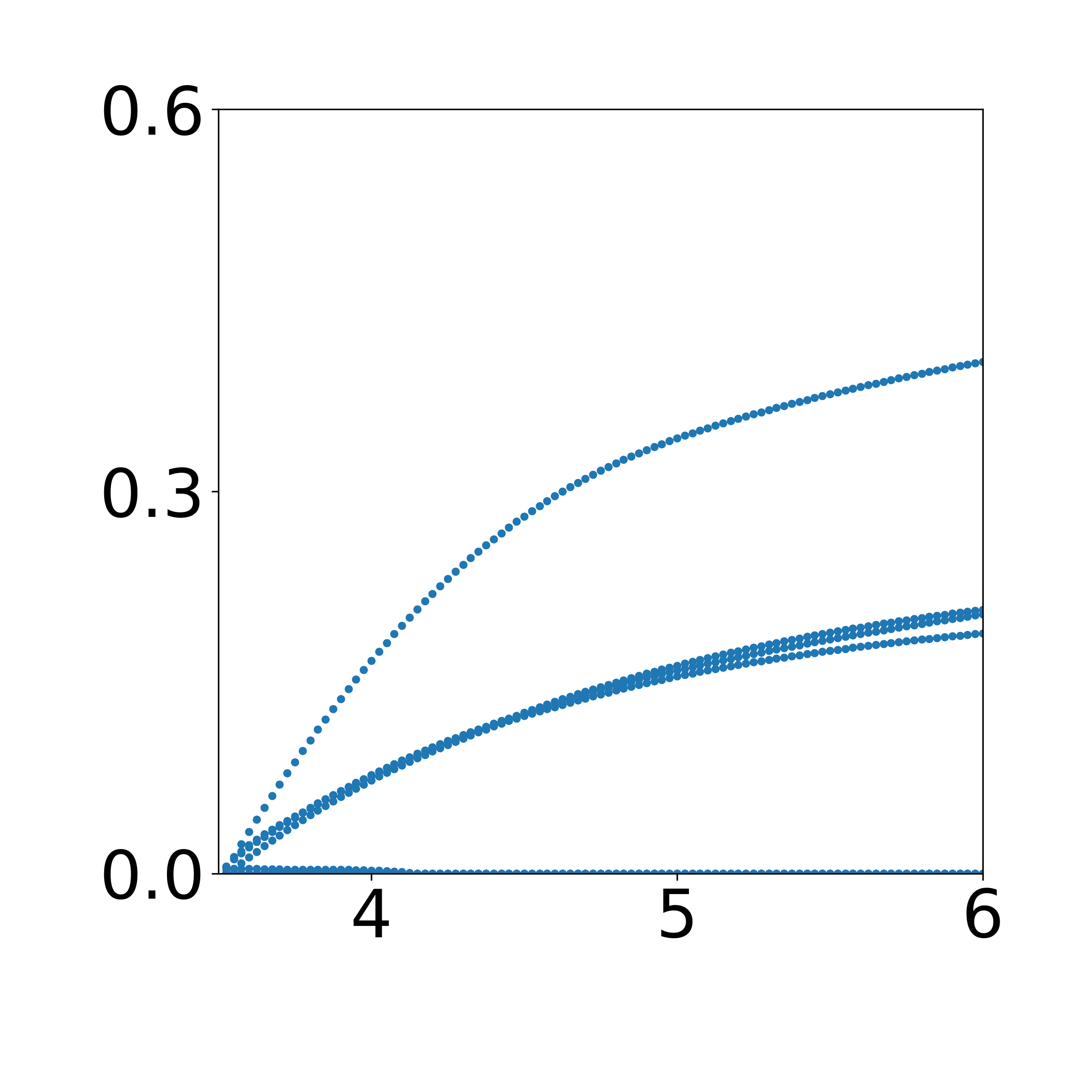}
\put(55,-6){$\mu$}
\put(15,82){(b)}
\end{overpic}
\begin{overpic}[width=\linfigwidth,trim={50 75 50 40}, clip]{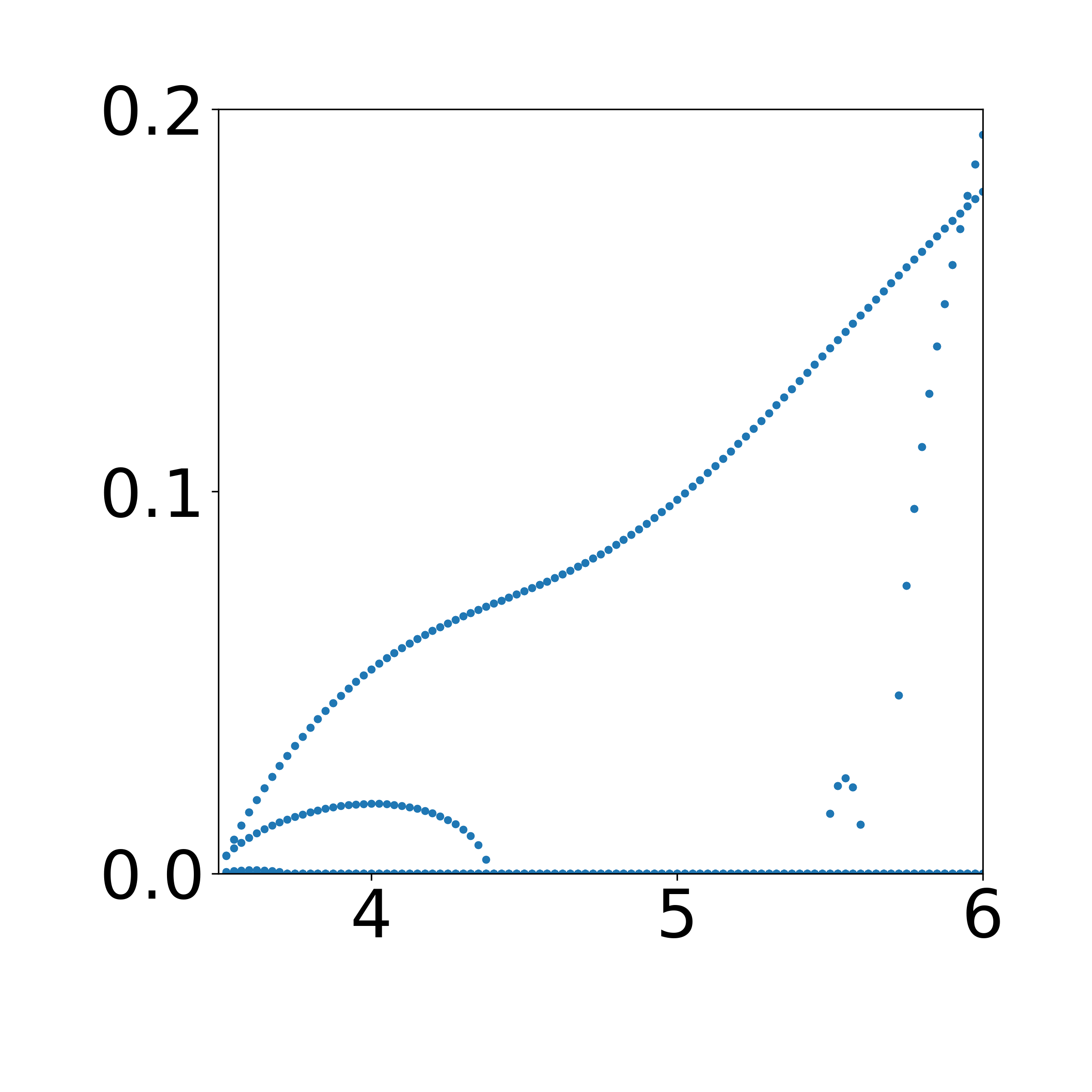}
\put(55,-6){$\mu$}
\put(15,82){(c)}
\end{overpic}
\end{center}
\caption{Spectra of (a) the spherical shell dark solitary wave presented in
\cref{fig_linear_states}(f); (b) the vortex star presented in
\cref{fig_new_states}(c); (c) the vortex ring with two vortex lines
presented in \cref{fig_new_states}(e). The real and imaginary parts of the
corresponding eigenfrequencies $\omega$ are depicted in the top and bottom
panels respectively.}
\label{fig_stab}
\end{figure}

We now explore the BdG spectral stability of selected solutions (the 
spectra of all states shown in Figs.~\ref{fig_linear_states} and~\ref{fig_new_states} 
are presented in Figs.~\ref{fig_stab_1} and~\ref{fig_stab_2} in Appendix~\ref{app_spec}, 
respectively). In fact, some of the identified waveforms are dynamically robust 
for an interval within their existence range. An example of this form is the 
spherical shell dark soliton of \cref{fig_linear_states}(f) with its spectrum 
presented in \cref{fig_stab}(a). However, most are indeed dynamically unstable, 
as expected; see, e.g., the case of the vortex star in \cref{fig_stab}(b). 
Interestingly, our BdG computations reveal that it is not the case that the most 
complex states are also the most unstable ones (see \cref{fig_stab}). An example 
of this type can be found in the vortex ring-double vortex line state of \cref{fig_new_states}(e) 
with spectrum presented in \cref{fig_stab}(c). While the solution is highly complex, 
it only bears a single unstable mode for a wide parametric interval, and at most bears 
two over the interval studied. Even more importantly, in our dimensionless units 
(scaled by the harmonic trapping frequency), the relevant growth is typically 
of the order of 0.1-0.2. This means that the characteristic dimensionless growth time 
is about 5-10 oscillation times, and if the perturbation is sufficiently small the 
instability manifestation time can be expected to be 2-3 times larger, in line with
the dynamical observations to which we now turn.

\begin{figure}[tbp]
\begin{center}
\begin{overpic}[width=\linfigwidth]{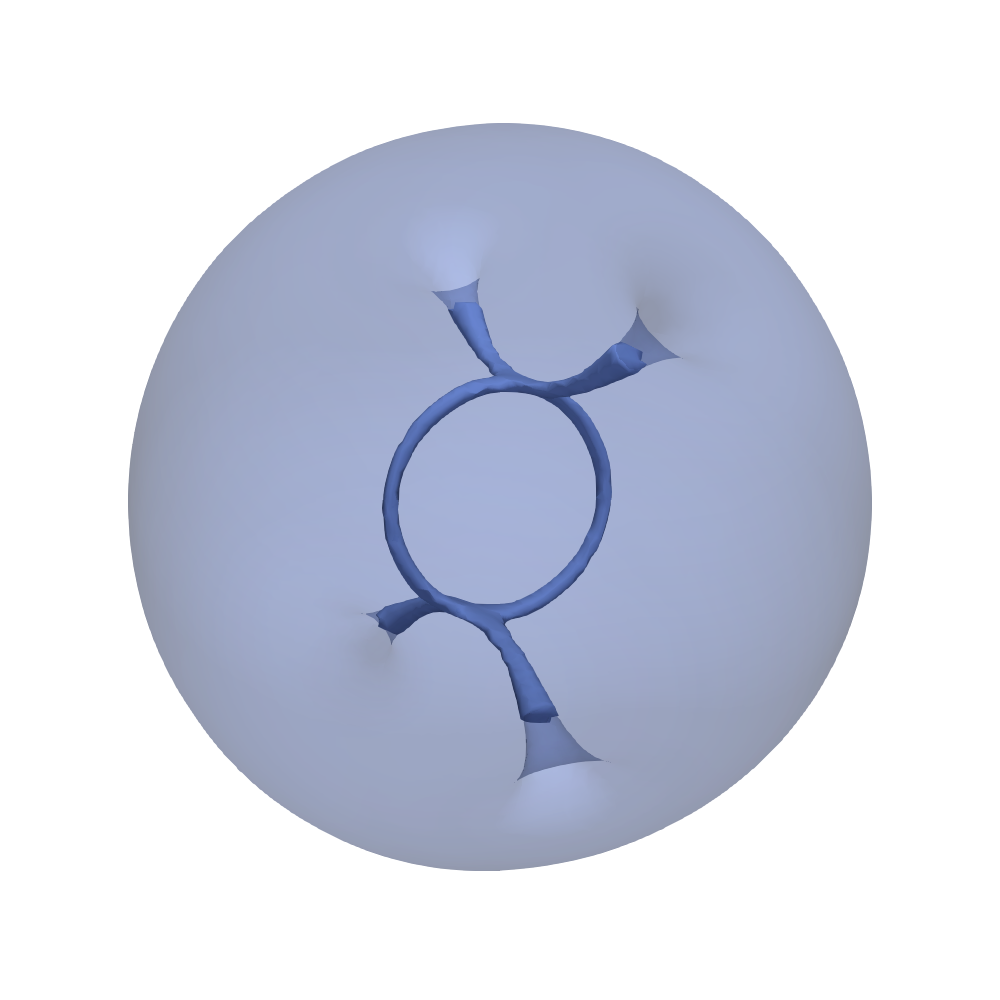}
\put(35,0){$t=15$}
\end{overpic}
\begin{overpic}[width=\linfigwidth]{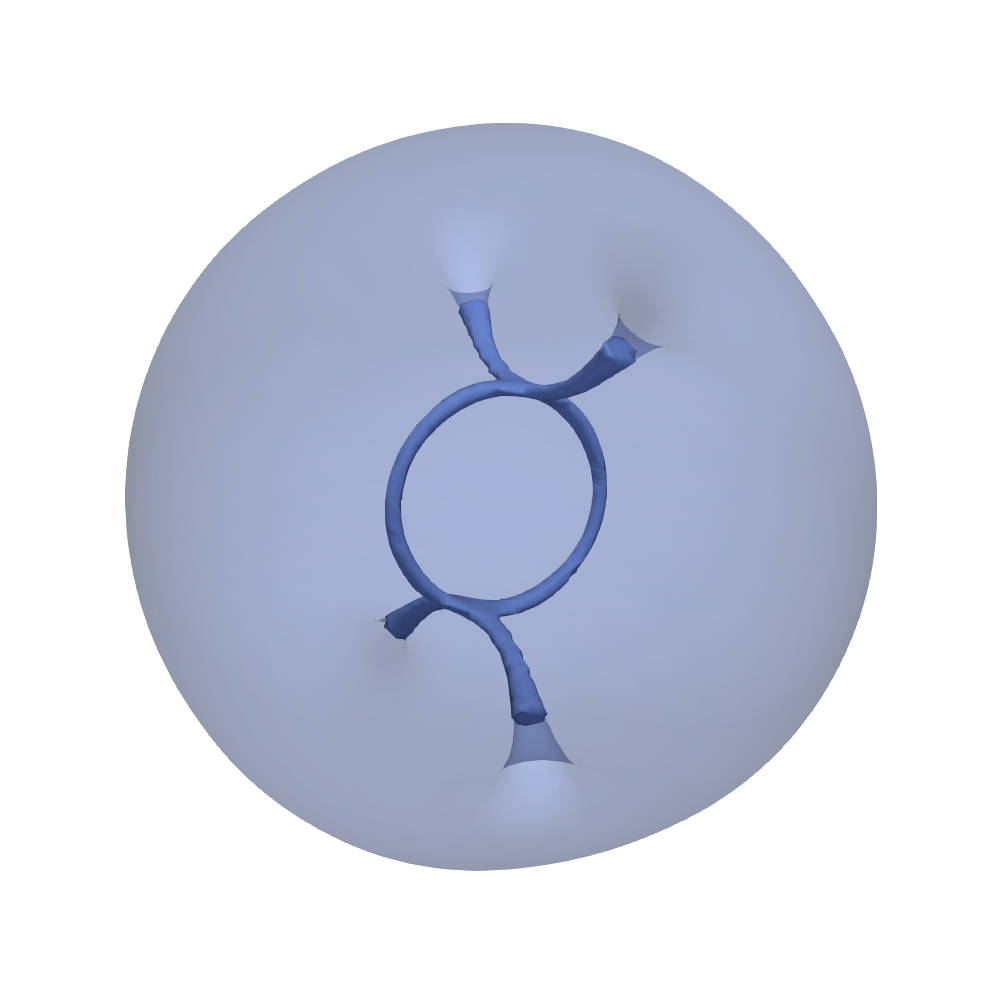}
\put(35,0){$t=19$}
\end{overpic}
\begin{overpic}[width=\linfigwidth]{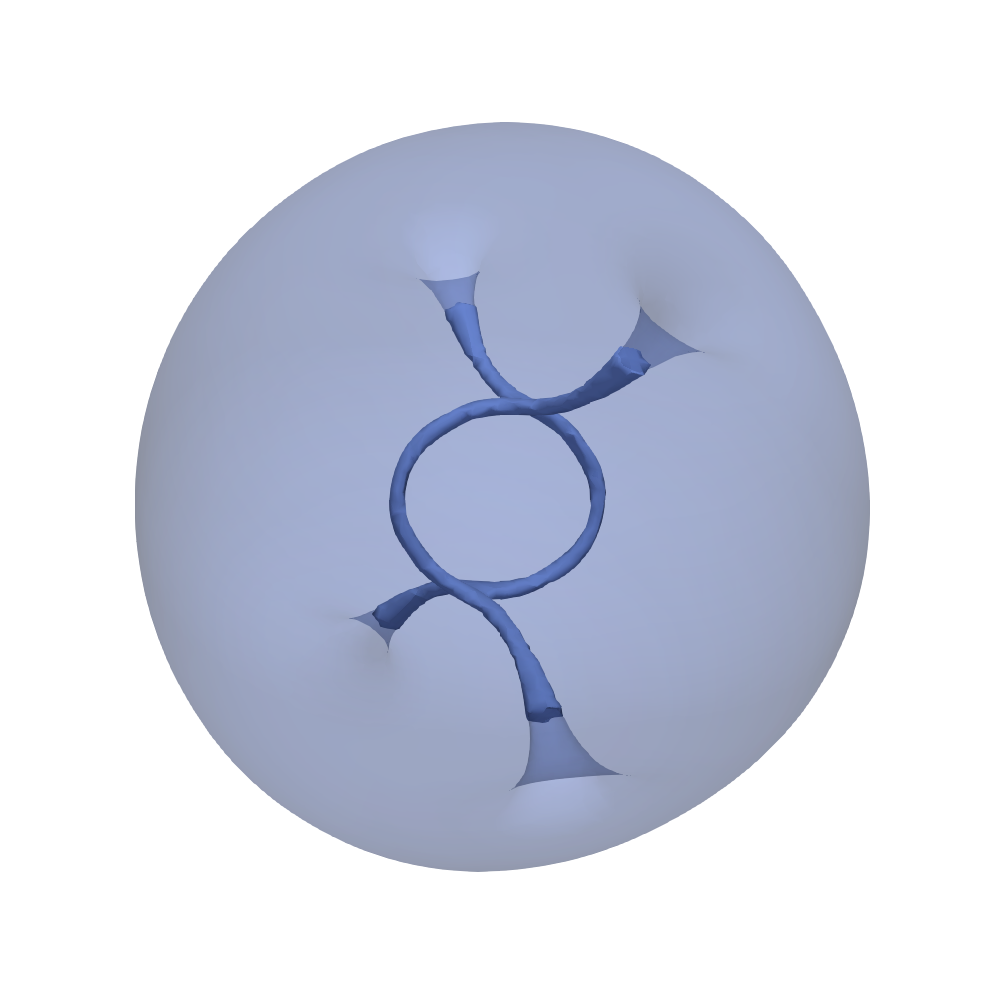}
\put(35,0){$t=23$}
\end{overpic}\\
\begin{overpic}[width=\linfigwidth]{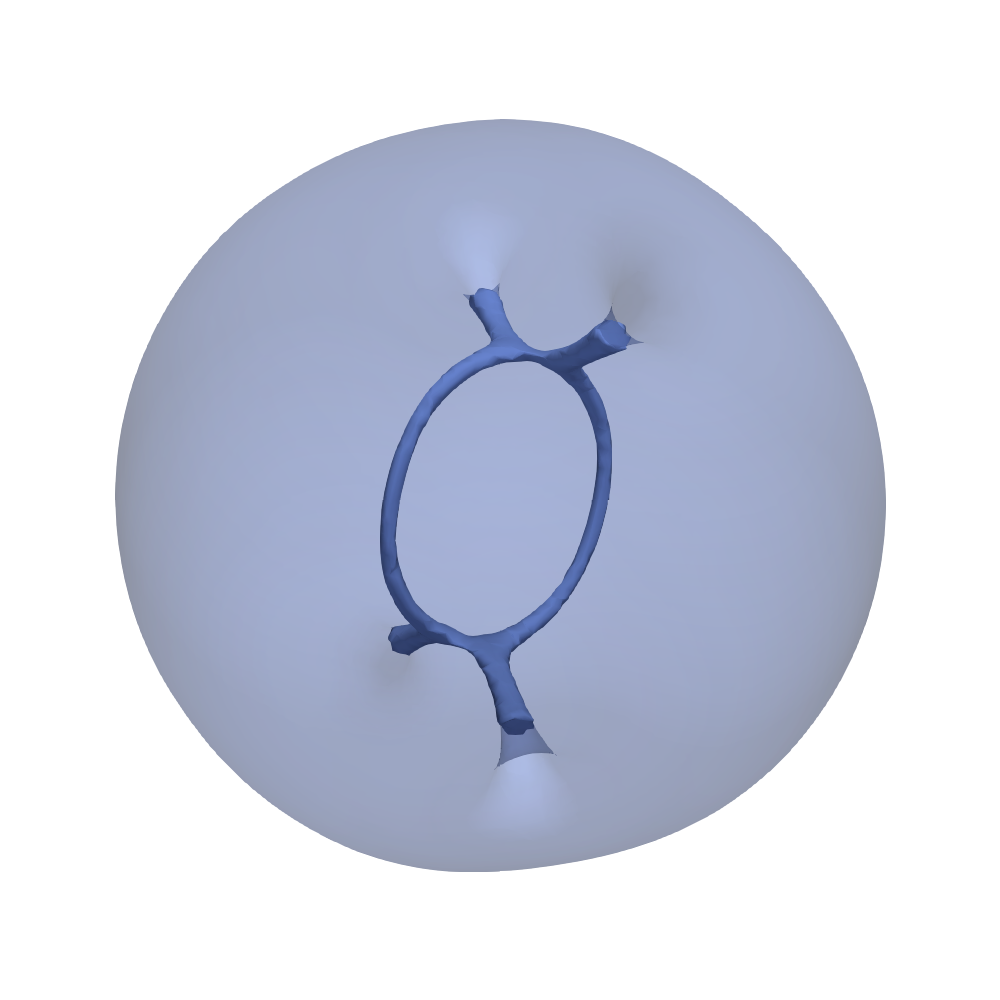}
\put(35,0){$t=27$}
\end{overpic}
\begin{overpic}[width=\linfigwidth]{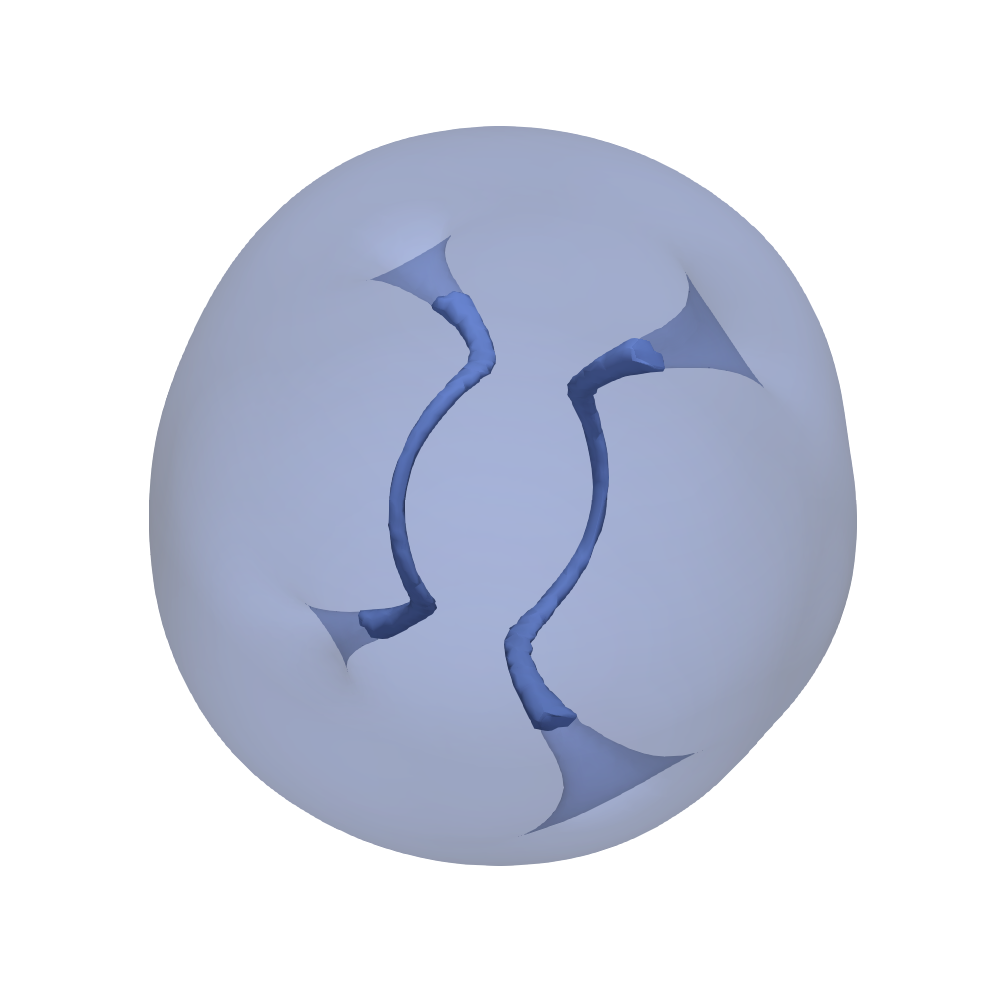}
\put(35,0){$t=31$}
\end{overpic}
\begin{overpic}[width=\linfigwidth]{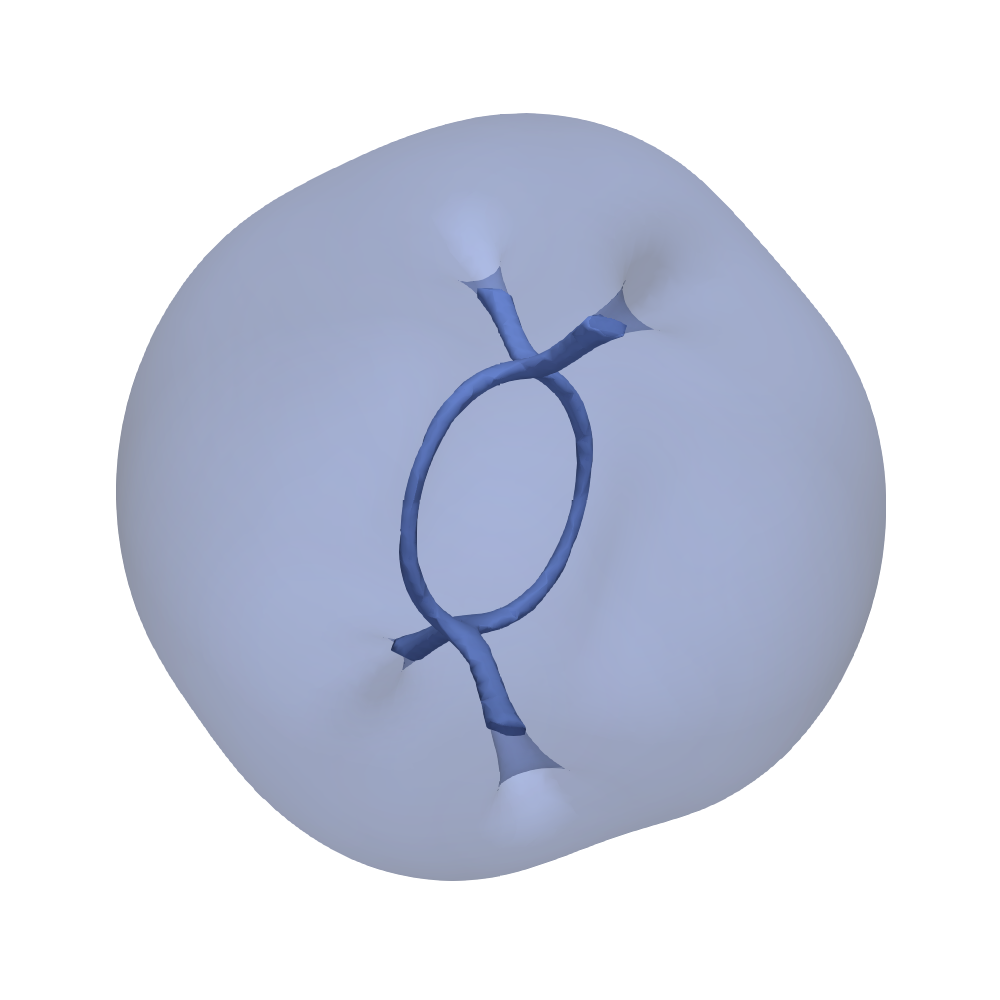}
\put(35,0){$t=35$}
\end{overpic}
\end{center}
\caption{Snapshots of the vortex ring-double
vortex line state obtained by solving the time-dependent NLS equation. The steady-state solution 
of panel (e) in \cref{fig_new_states} is initially perturbed along its
dominant unstable eigenmode.
}
\label{fig_dynamics}
\end{figure}

We employ the Crank--Nicolson method to explore the dynamical implications
of these instabilities. The initial states are prepared according to Eq.~\eqref{eq:init_state}.
In the case of \cref{fig_new_states}(e), we observe in the snapshots of the 
evolution of \cref{fig_dynamics} that the vortex ring and two vortex lines 
break up into two vortex lines which are strongly bent (in fact, they are 
somewhat reminiscent of the U-shaped vortex lines of~\cite{danaila3}). After 
about 20-30 dimensionless time units --- in line with our eigenvalue predictions
 --- the configuration is characterized by splittings and reconnections 
(including ones re-formulating the original configuration). We have performed 
similar computations for other complex, topological states such as that shown 
in \cref{fig_new_states}(f), showcasing in some such cases more radical dynamical
breakups, i.e., the eventual persistence of a single, strongly excited vortex 
line; see relevant snapshots in \cref{fig_dynamics_supplemental} and the movies 
in the Supplemental Material at [URL will be inserted by publisher]. Importantly, we note that the relevant time scales 
for standard choices of the trap strength are on the order of hundreds of milliseconds, 
and hence the configurations are expected to be well within windows of experimental 
accessibility. Furthermore, even when the configurations become unstable, as in 
the dynamics of \cref{fig_dynamics}, they appear to result in oscillatory dynamics
reconstructing the relevant states in a nearly periodic fashion for far longer 
times (rather than dispersing or yielding chaotic dynamics). This further enhances 
the potential observability window of the states of interest.

\begin{figure}[tbp]
\begin{center}
\begin{overpic}[width=\linfigwidth]{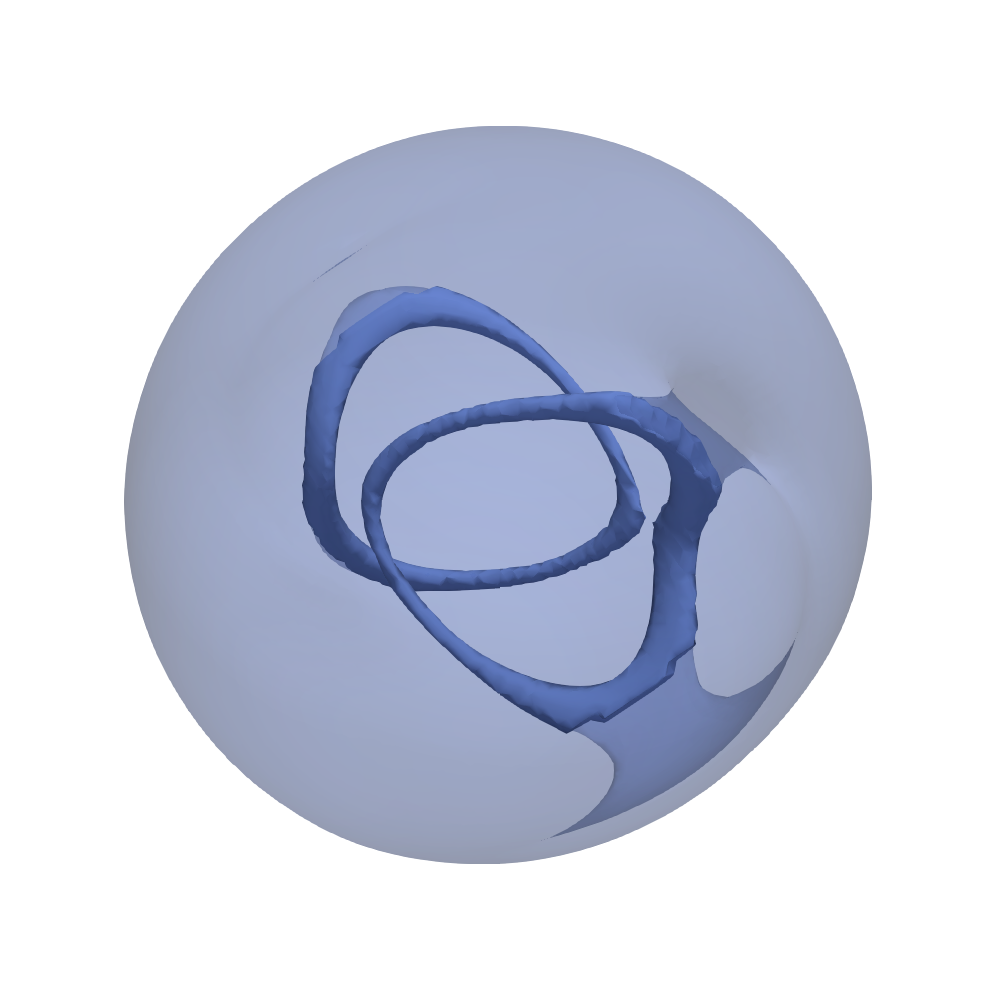}
\put(35,0){$t=10$}
\end{overpic}
\begin{overpic}[width=\linfigwidth]{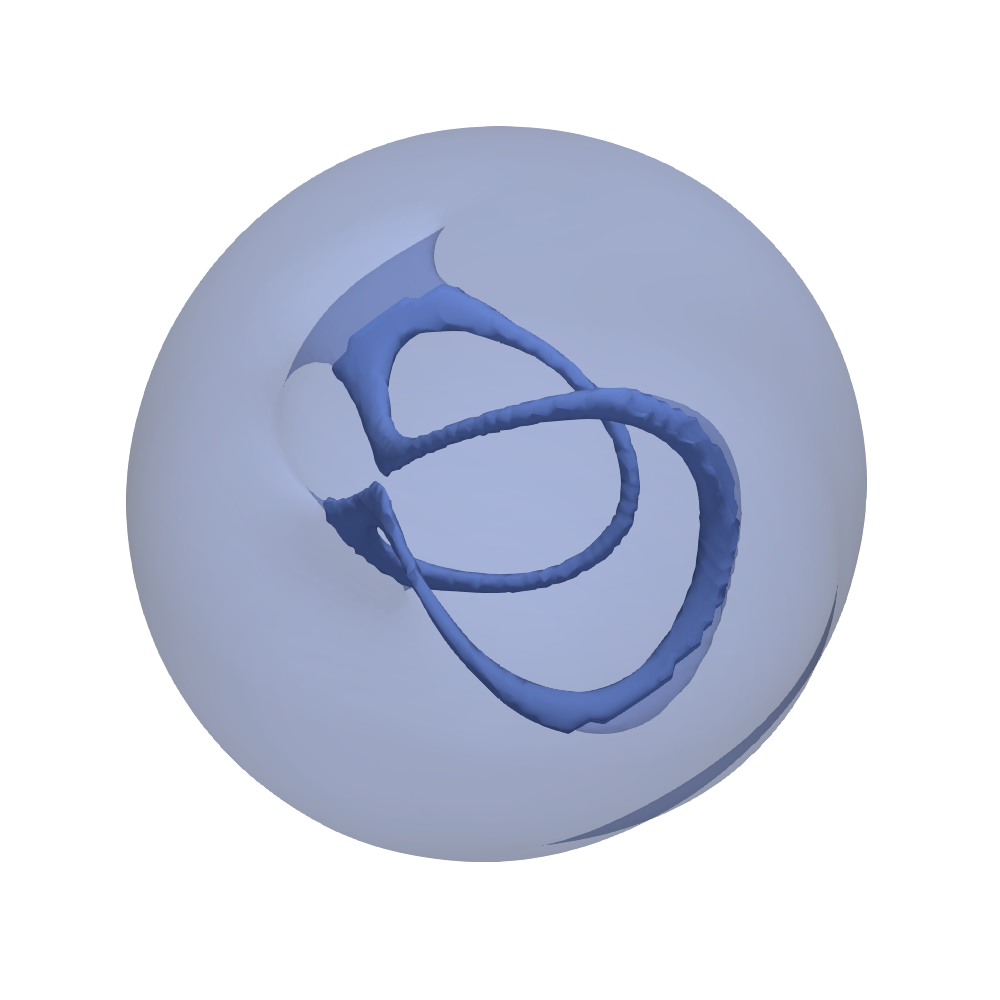}
\put(35,0){$t=15$}
\end{overpic}
\begin{overpic}[width=\linfigwidth]{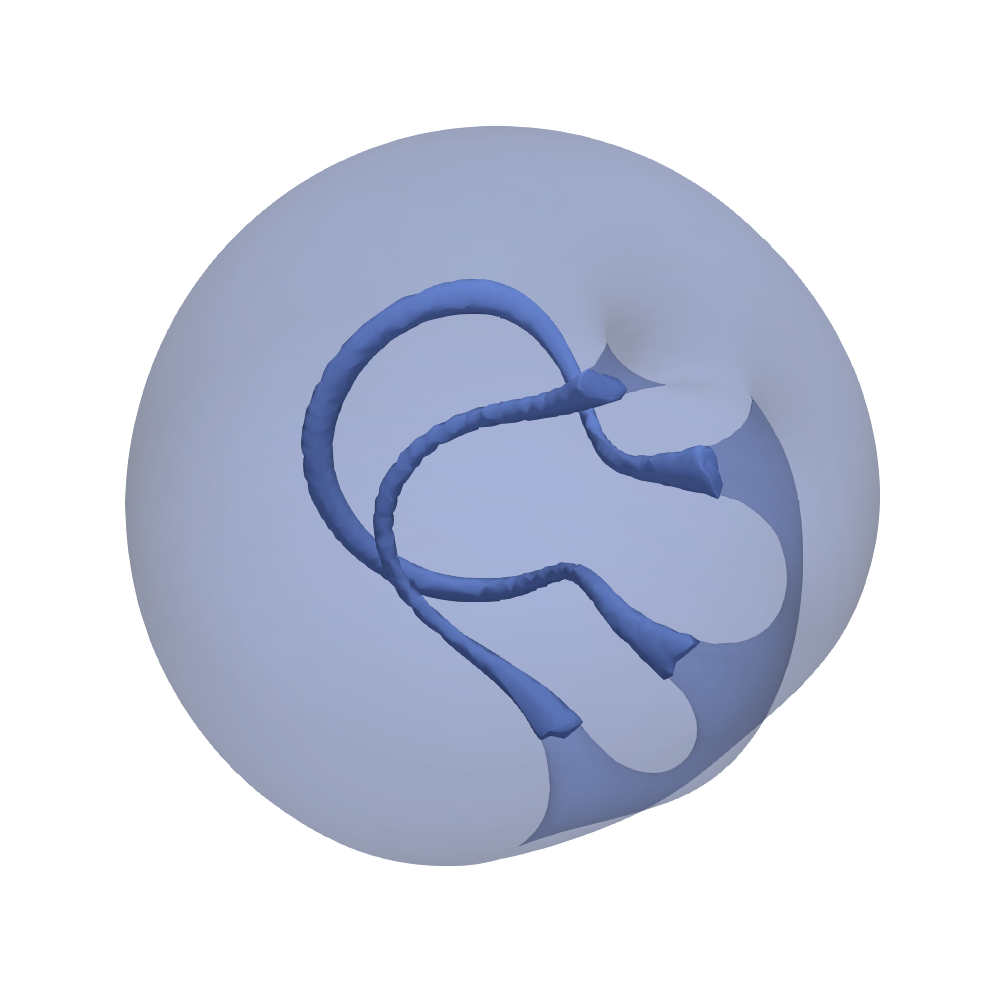}
\put(35,0){$t=18$}
\end{overpic}\\
\vspace{0.2cm}
\begin{overpic}[width=\linfigwidth]{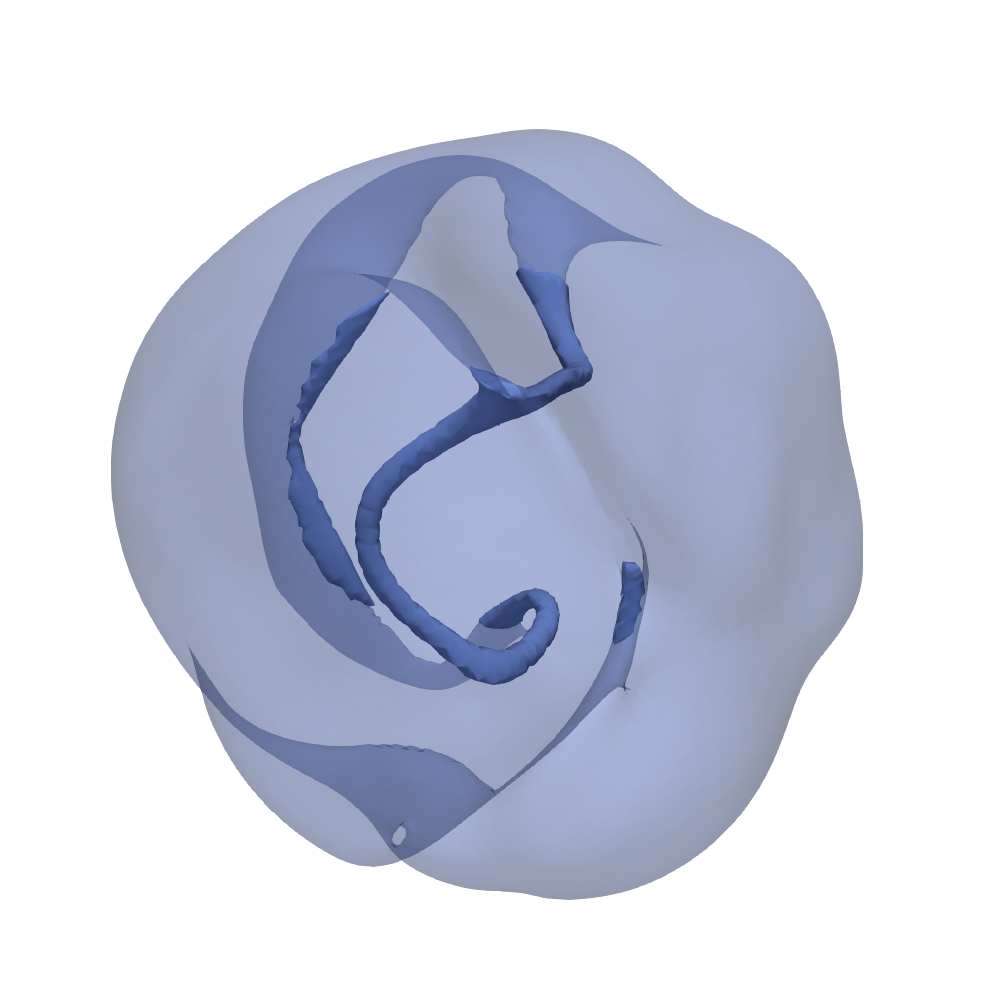}
\put(35,0){$t=28$}
\end{overpic}
\begin{overpic}[width=\linfigwidth]{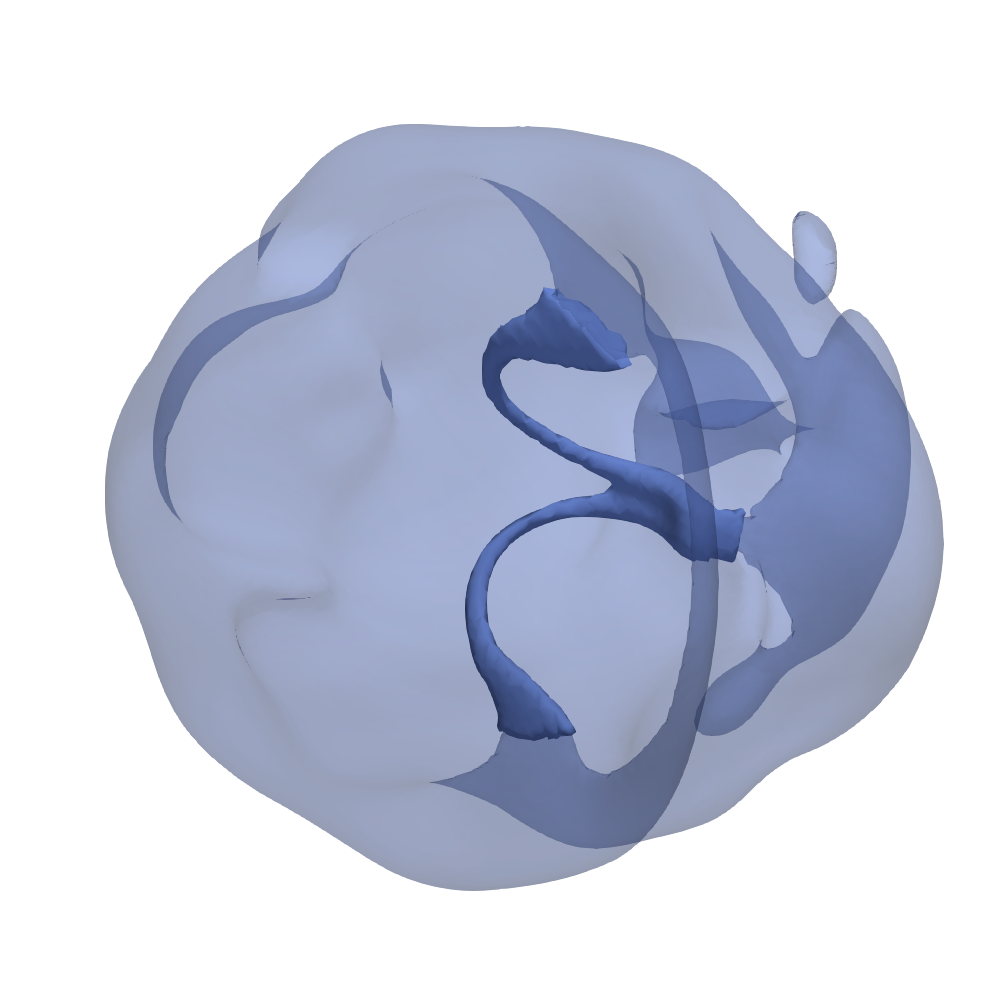}
\put(35,0){$t=34$}
\end{overpic}
\begin{overpic}[width=\linfigwidth]{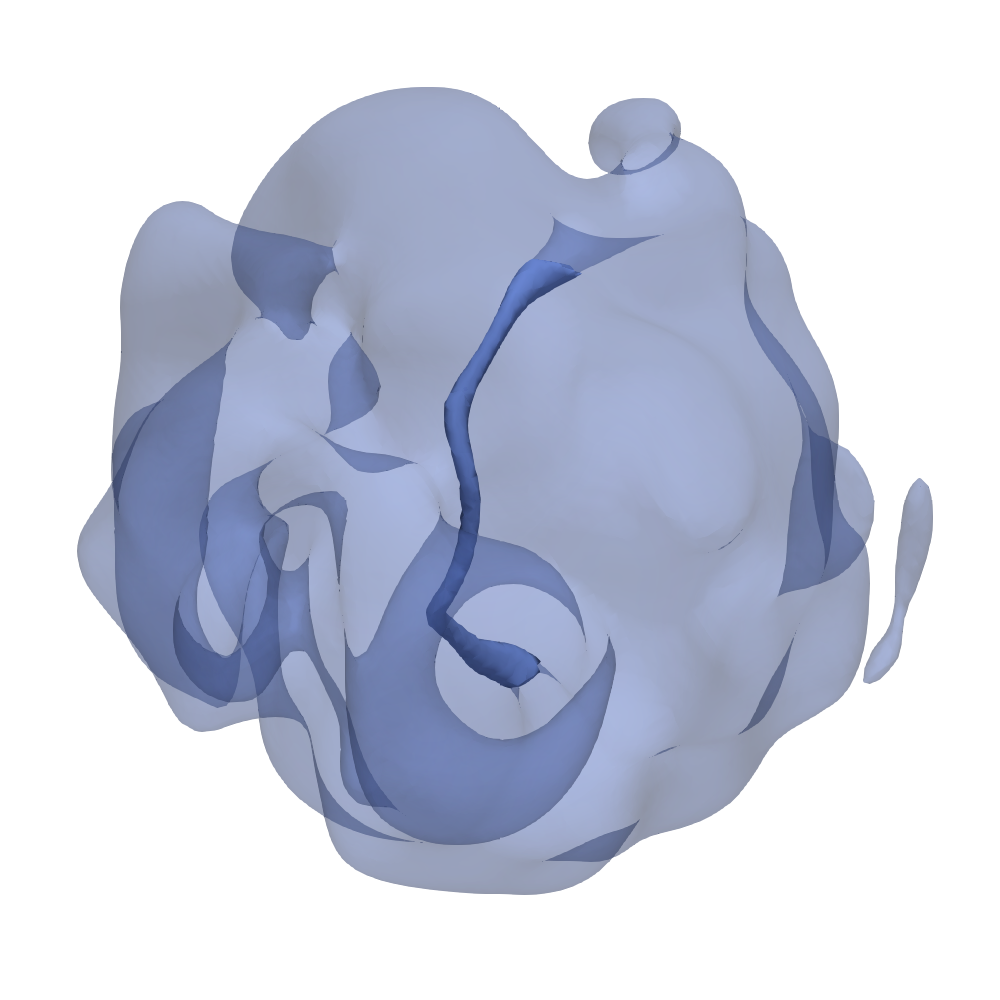}
\put(35,0){$t=40$}
\end{overpic}
\end{center}
\caption{Snapshots of the two vortex rings state of \cref{fig_new_states}(f), obtained by solving the time-dependent NLS equation with the modified Crank-Nicolson time-stepping scheme.}
\label{fig_dynamics_supplemental}
\end{figure}

\begin{figure}[tbp]
\begin{center}
\begin{overpic}[width=\linfigwidth]{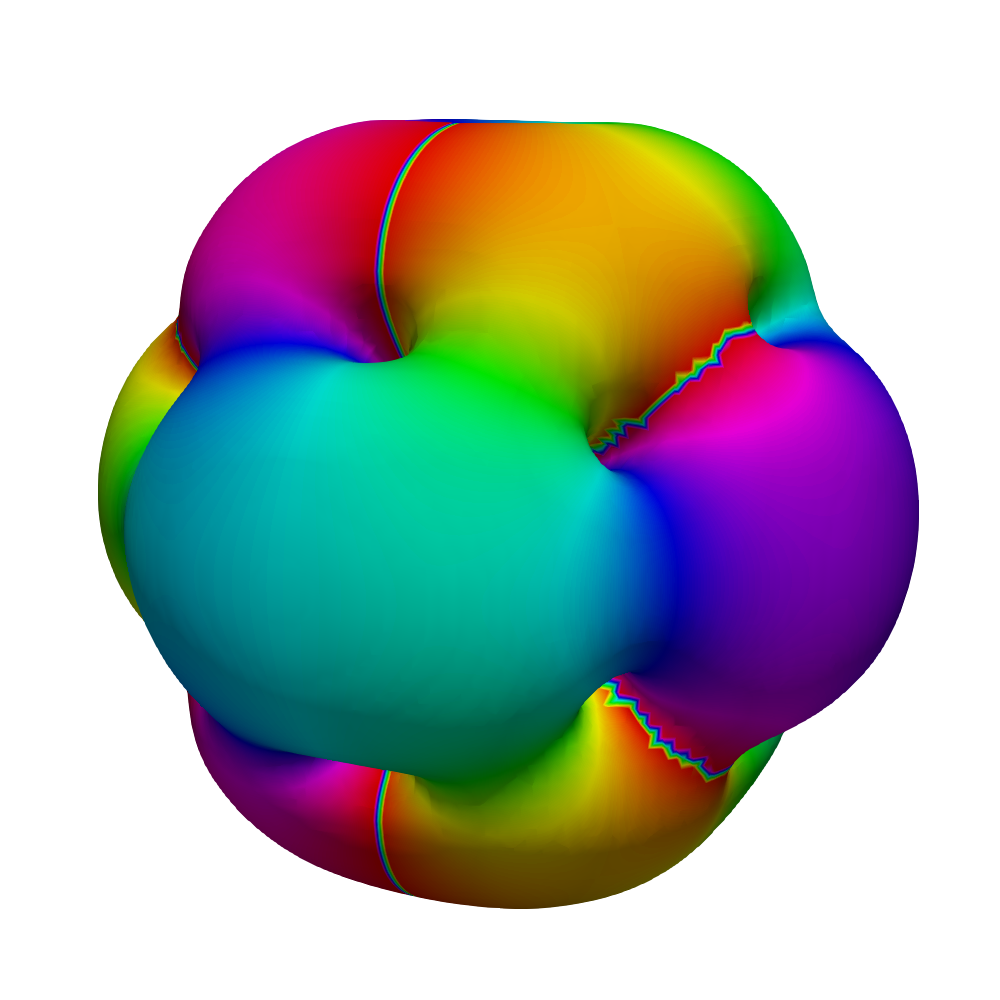}
\put(0,85){(a)}
\end{overpic}
\begin{overpic}[width=\linfigwidth]{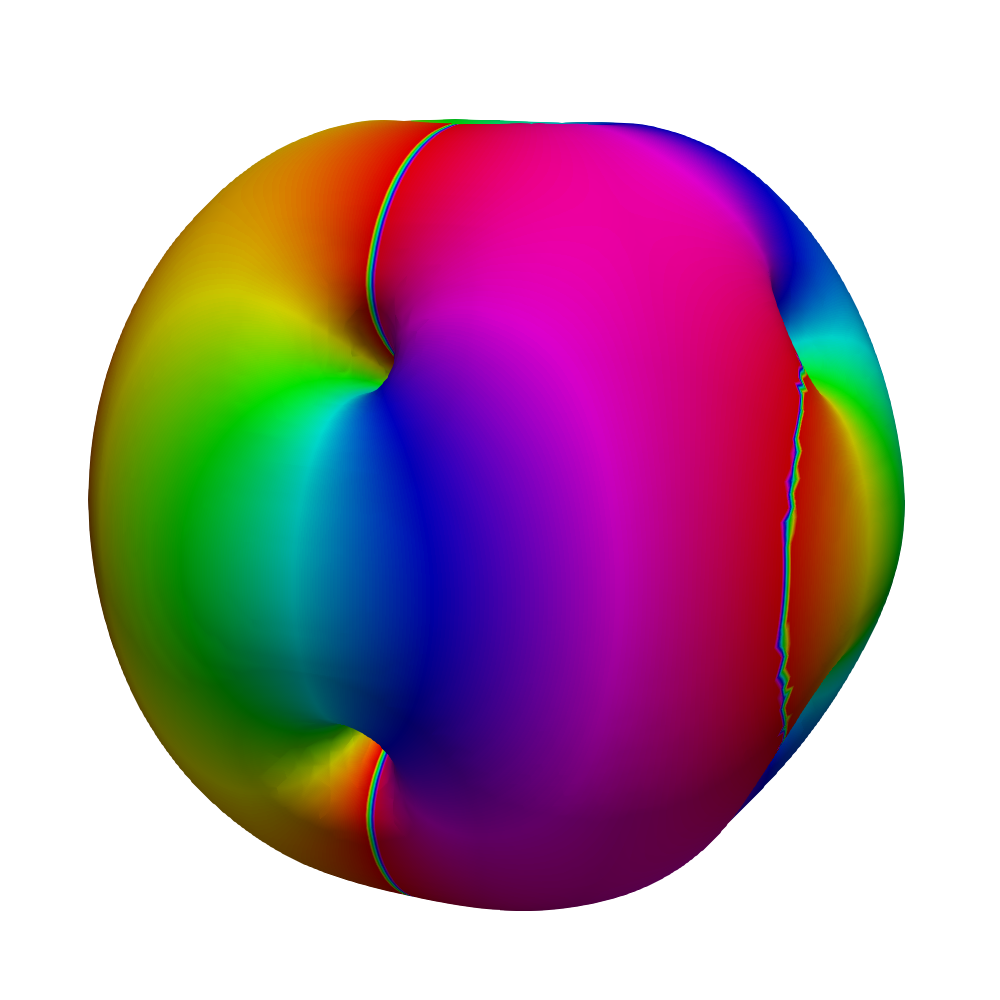}
\put(0,85){(b)}
\end{overpic}
\begin{overpic}[width=\linfigwidth]{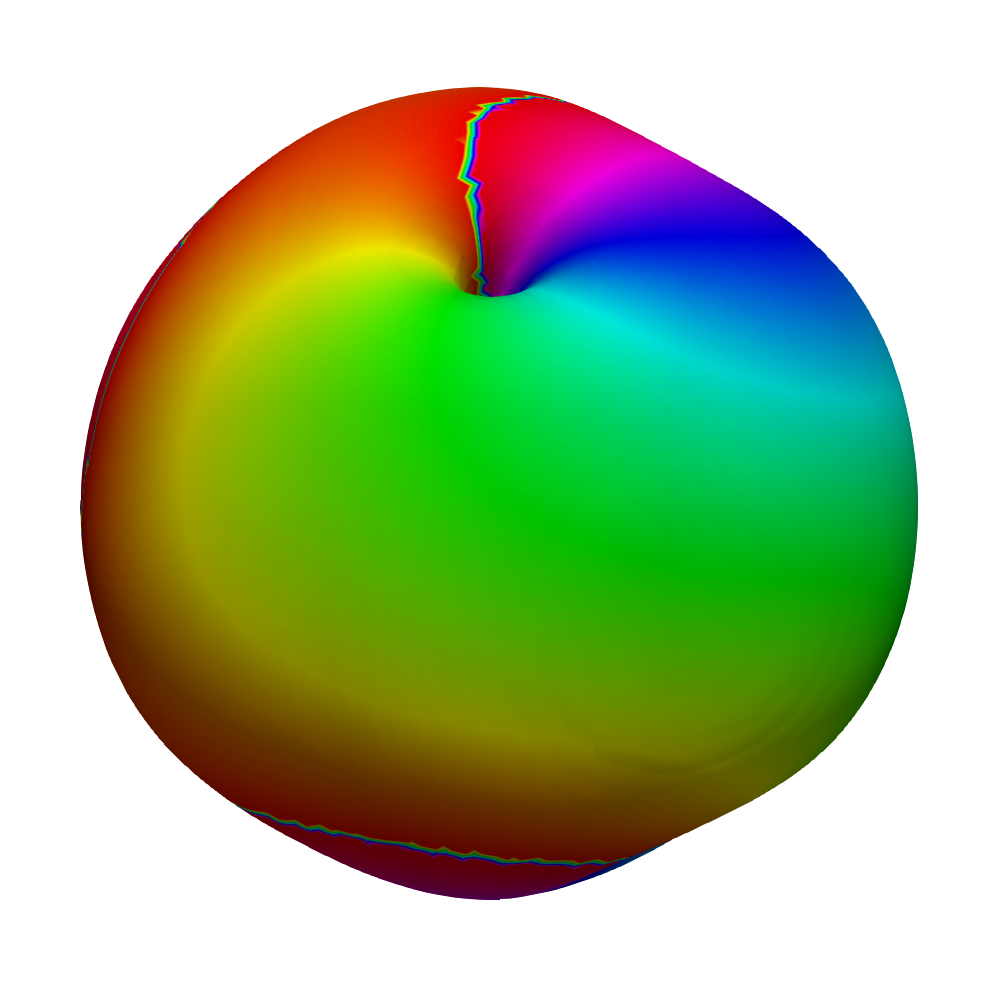}
\put(0,85){(c)}
\end{overpic}\\
\begin{overpic}[width=\linfigwidth]{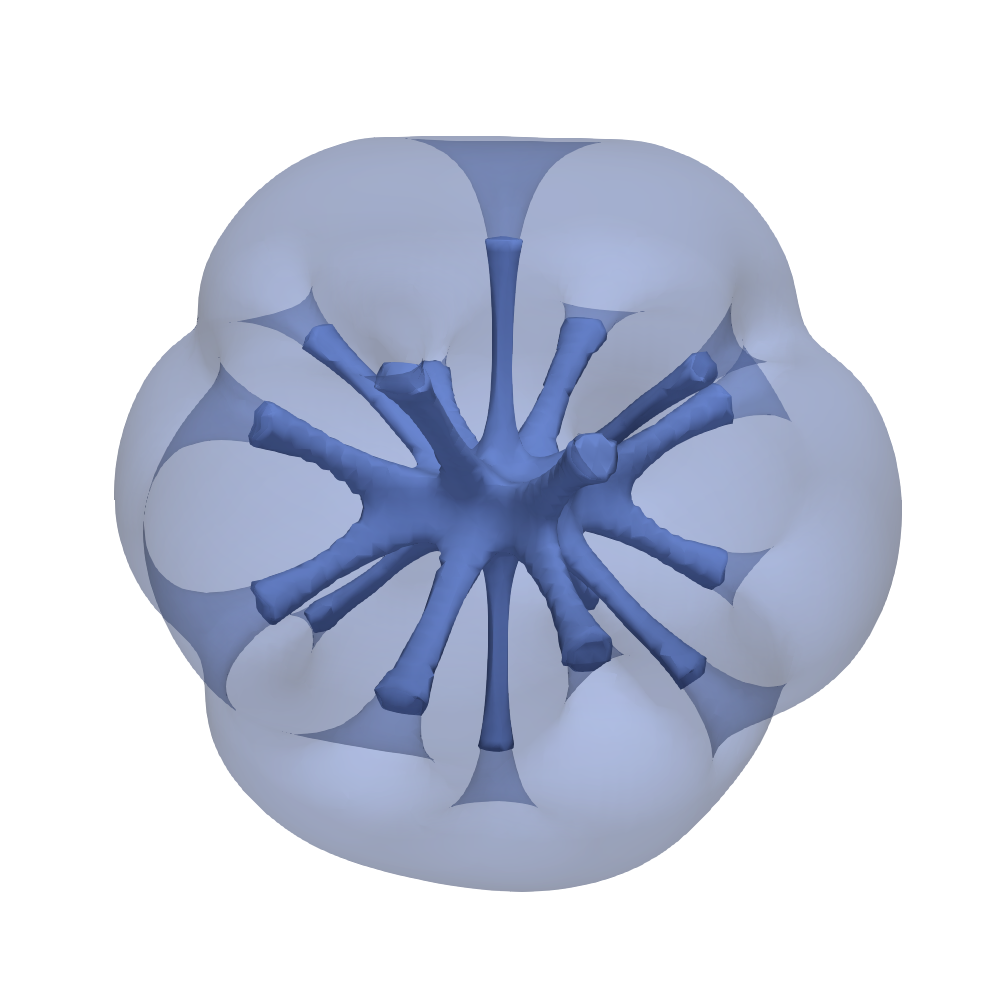}
\put(40,-13){(67)}
\end{overpic}
\begin{overpic}[width=\linfigwidth]{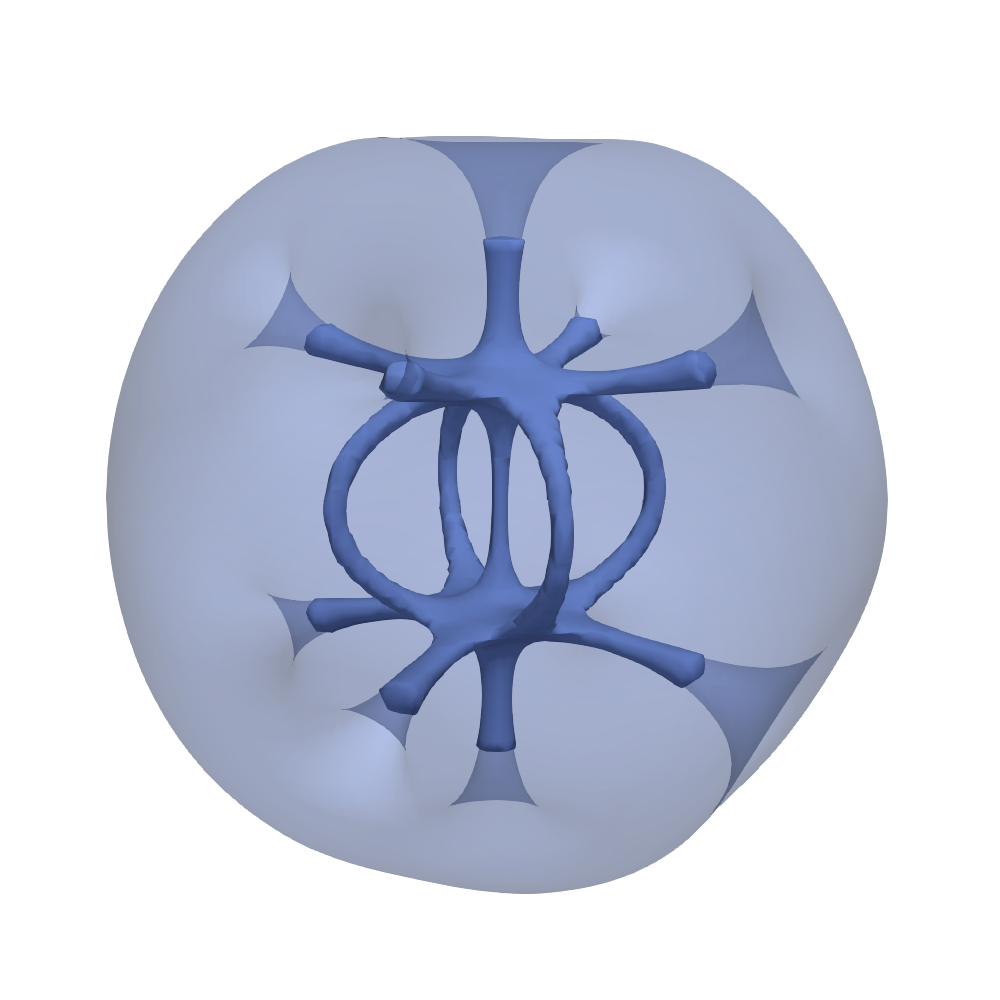}
\put(40,-13){(73)}
\end{overpic}
\begin{overpic}[width=\linfigwidth]{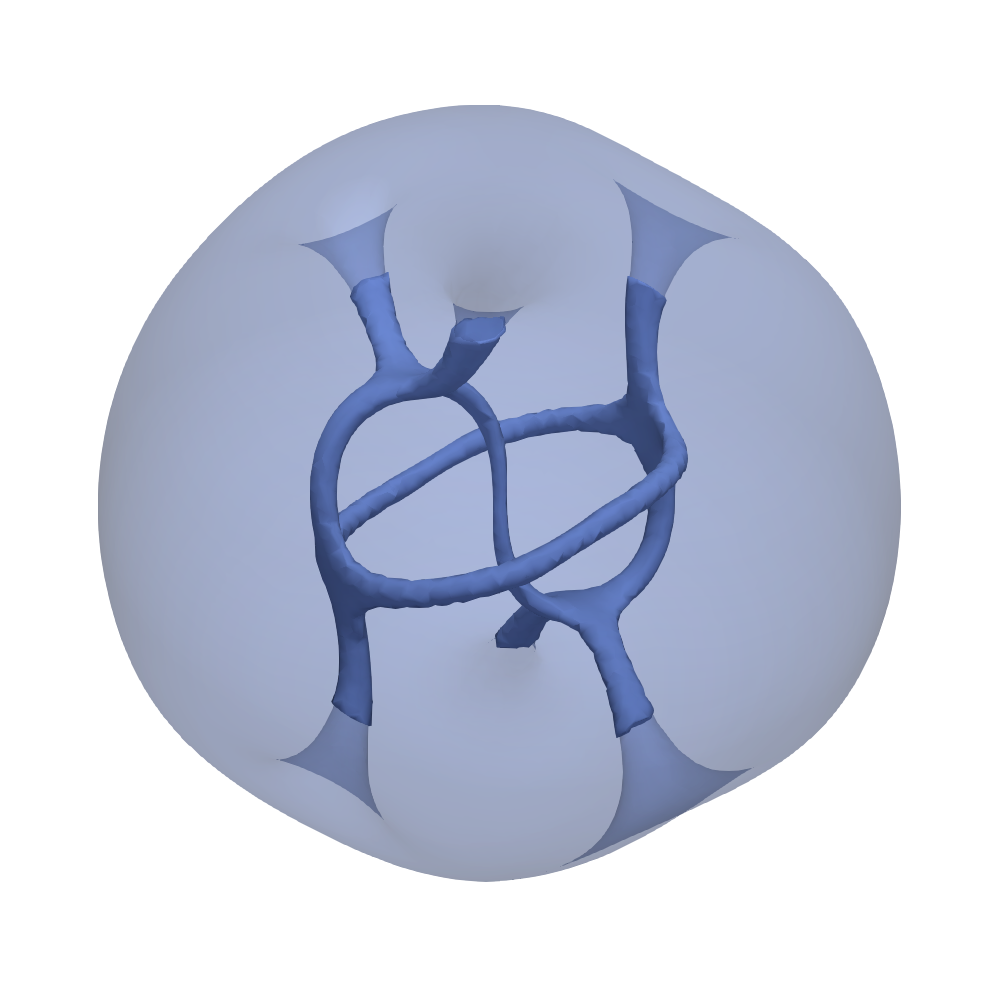}
\put(40,-13){(75)}
\end{overpic}\\
\begin{overpic}[width=\linfigwidth]{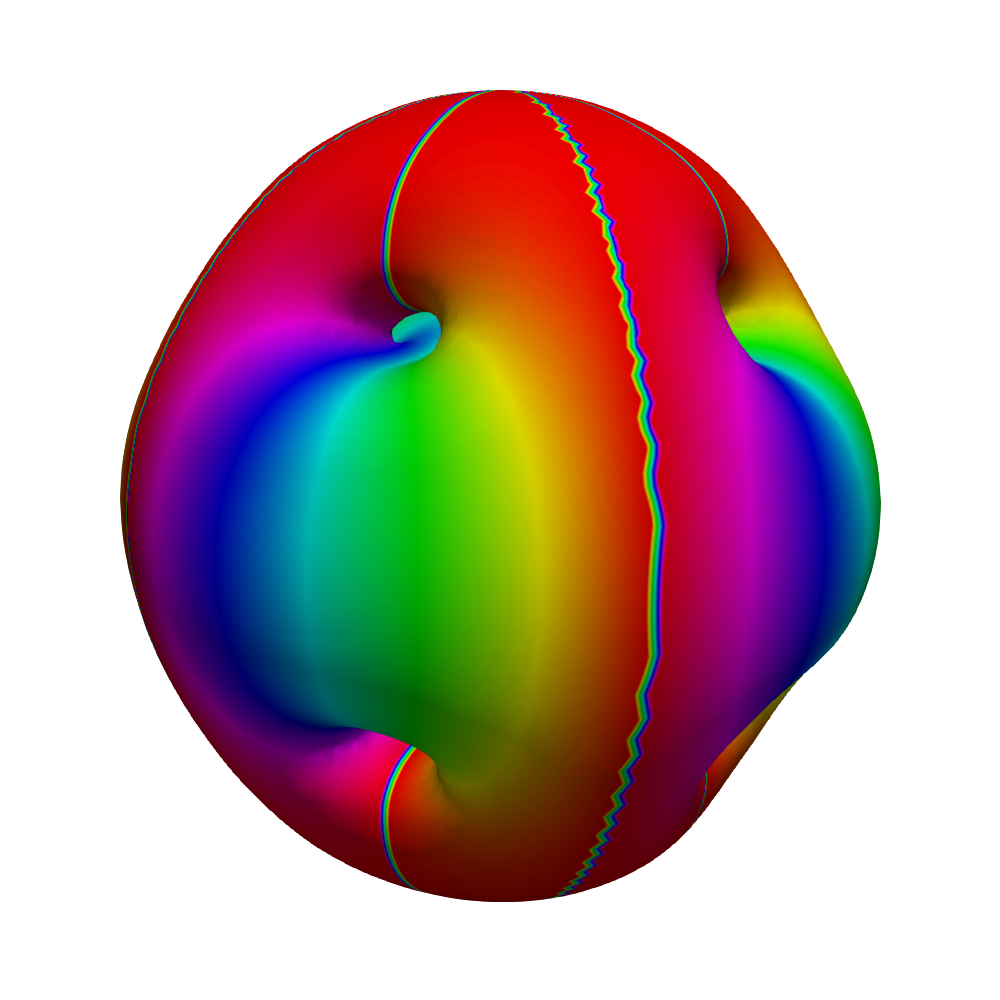}
\put(0,85){(d)}
\end{overpic}
\begin{overpic}[width=\linfigwidth]{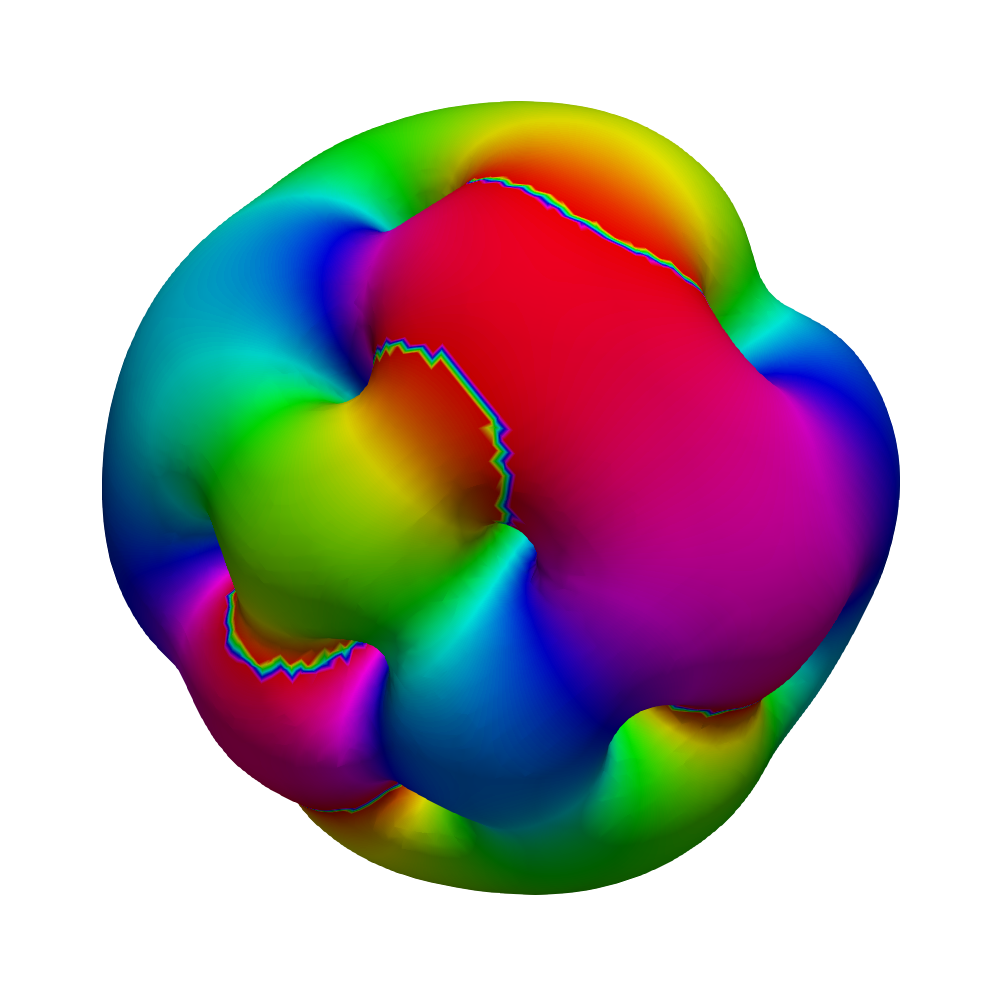}
\put(0,85){(e)}
\end{overpic}
\begin{overpic}[width=\linfigwidth]{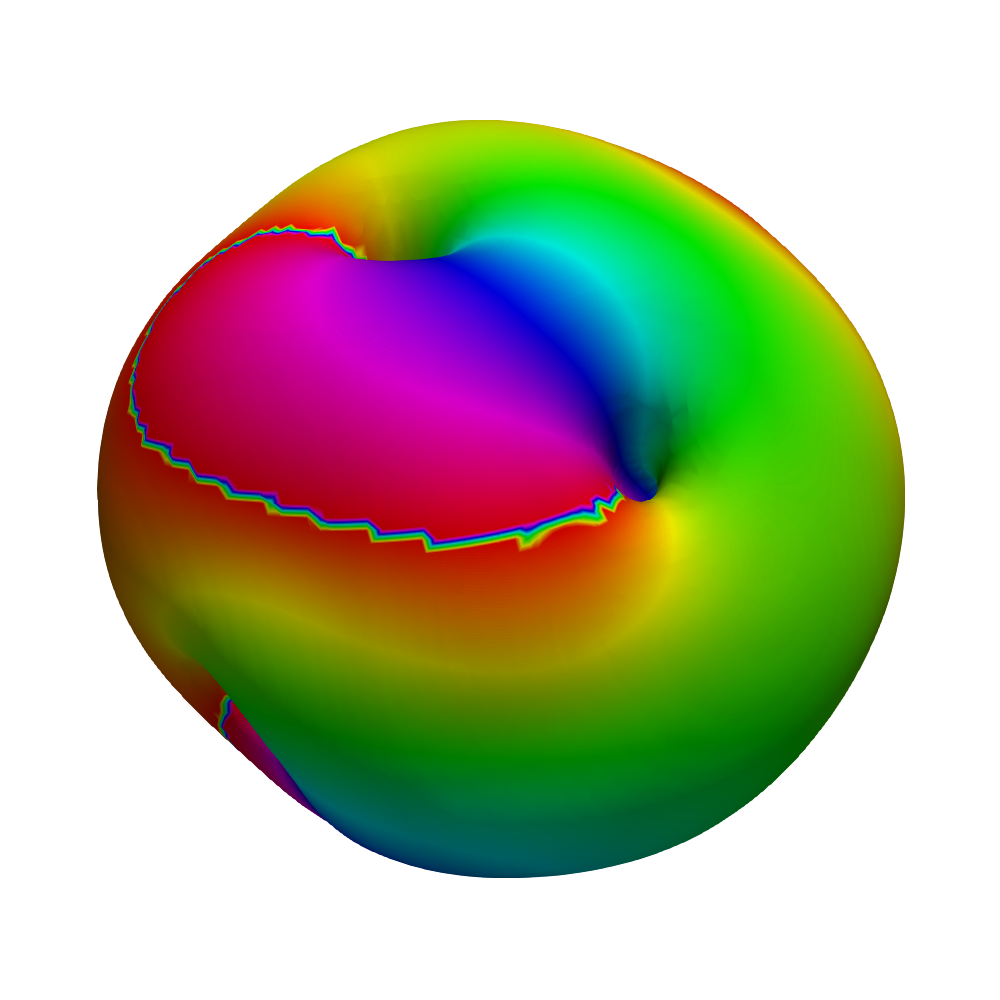}
\put(0,85){(f)}
\end{overpic}\\
\begin{overpic}[width=\linfigwidth]{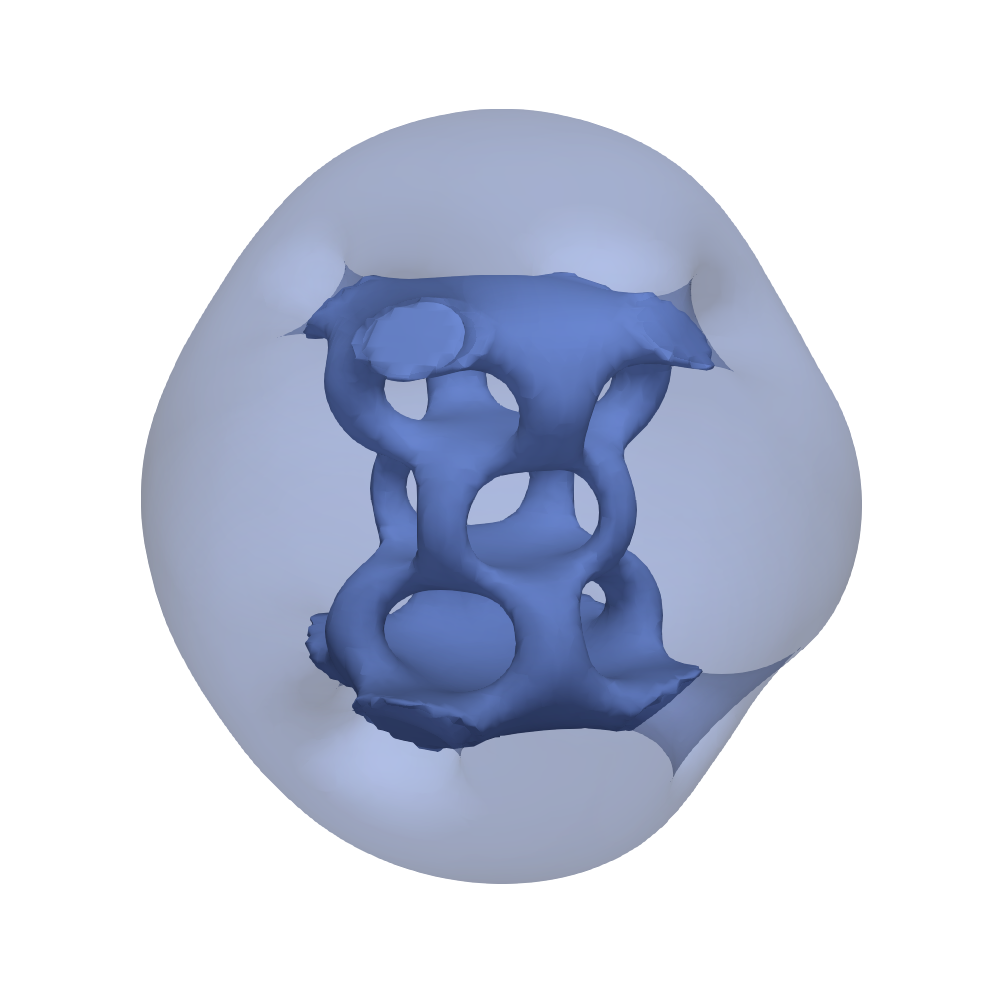}
\end{overpic}
\begin{overpic}[width=\linfigwidth]{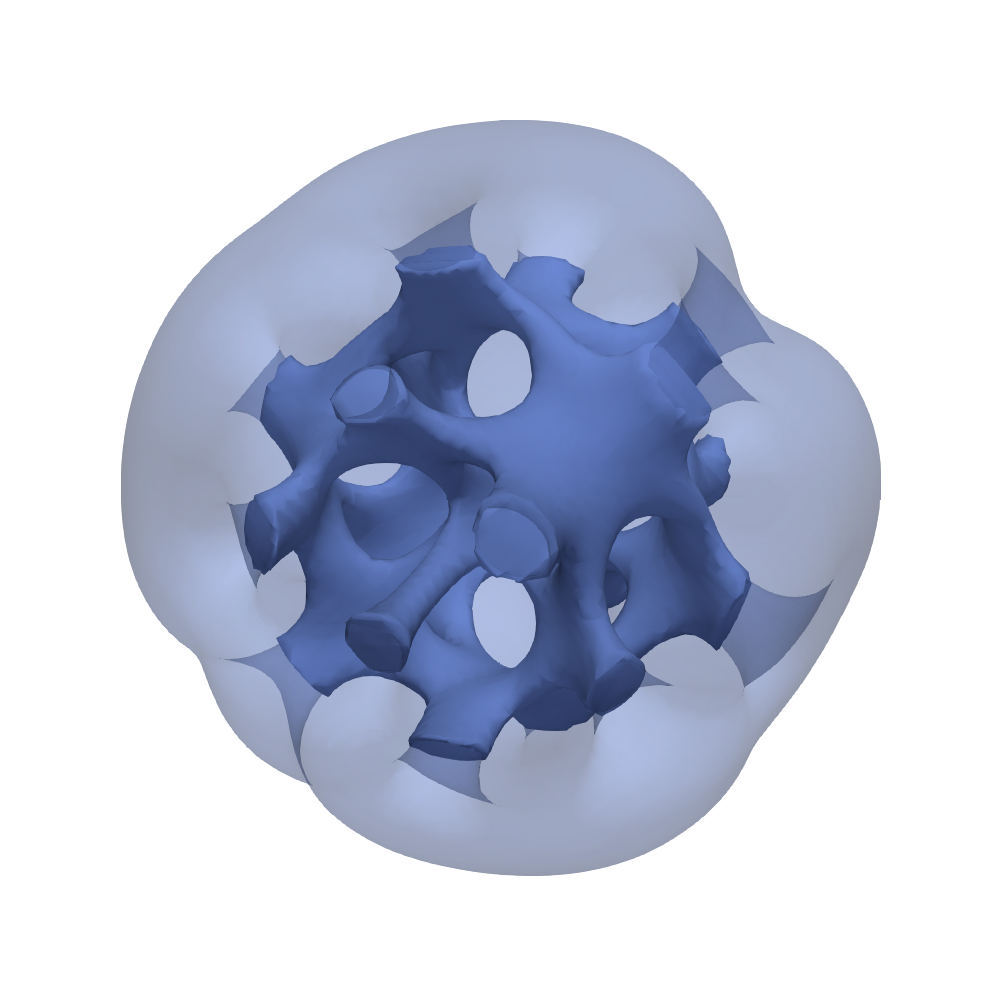}
\end{overpic}
\begin{overpic}[width=\linfigwidth]{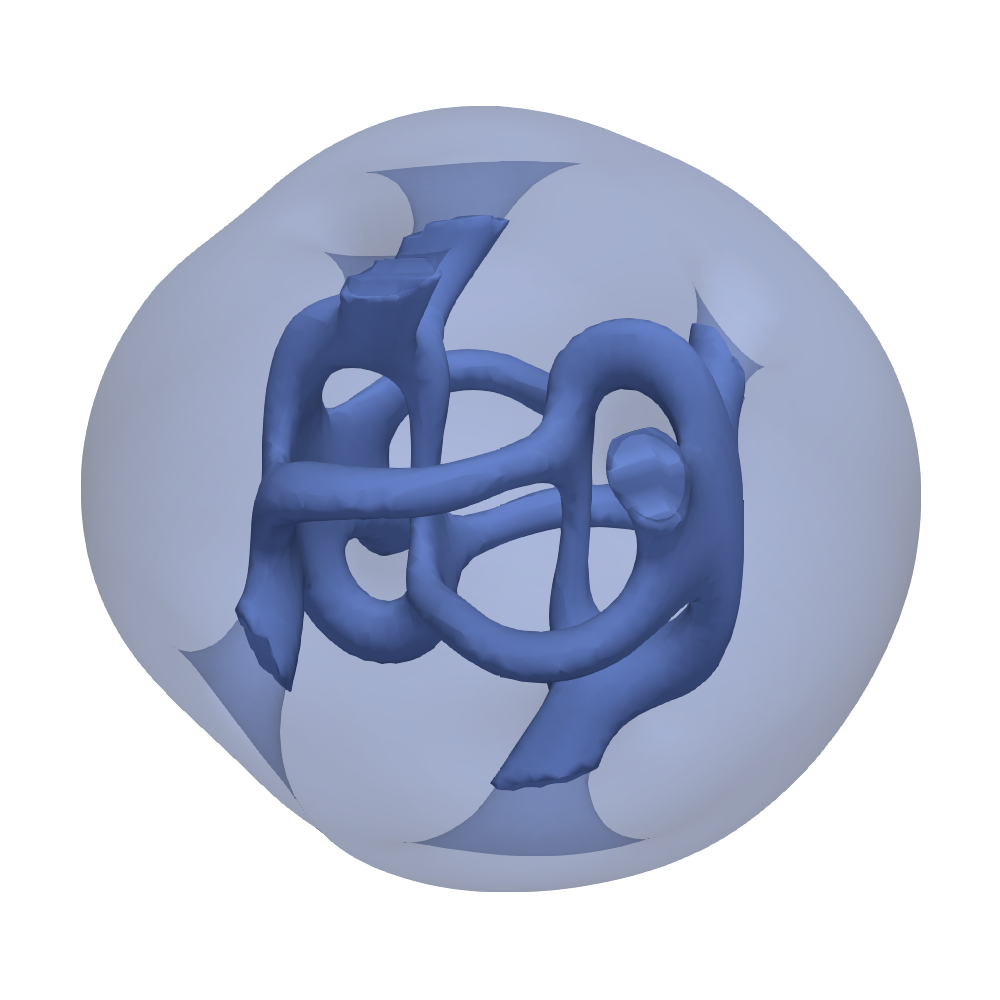}
\end{overpic}
\end{center}
\caption{Some of the solutions discovered by deflation that emanate 
from the 3rd (panels (a)-(c)) and 4th (panels (d)-(f)) excited states
at $\mu=9/2$ and $\mu=11/2$. The first and third rows show the phase 
of the solutions while the second and fourth rows represent the density 
isosurfaces at densities $0.30$ and $0.35$.}
\label{fig_complex_states}
\end{figure}

We close our presentation of the numerical results of deflation by offering 
a glimpse into the capabilities of the method for discovering higher excited 
states, i.e., ones that are initiated not from the 2nd (as up to now), but 
rather from the 3rd and 4th excited states at $\mu=9/2$ and $\mu=11/2$, respectively.
Some of the relevant nonlinear states discovered with deflation can be found 
in \cref{fig_complex_states}. The first examples, such as those of panels (a) 
and (b) can be identified straightforwardly: panel (a) represents a nonlinear
configuration bearing 9 vortex lines in a generalization of the star-shaped 
configuration of \cref{fig_new_states}(c) and~\cite{crasovan2004three}. 
The state of \cref{fig_complex_states}(b) appears to be a configuration bearing two perpendicular 
vortex rings (now joined--cf.~panel \cref{fig_new_states}(f)), along with 5 vortex 
lines: 4 of these are tangent, similar to the ones of \cref{fig_new_states}(e),
while one is piercing through the planes of the two rings. Going beyond these, 
however, the states become highly complex. \cref{fig_complex_states}(c) shows what 
appears to be a combination of an S-shaped and 2 U-shaped vortex lines (in the 
terminology of~\cite{danaila3}) along with a clearly discernible vortex ring.
Labyrinthine patterns of conjoined vortex rings and vortex lines appear; at the 
moment we do not have an immediate classification. \cref{fig_complex_states}(d) 
displays an apparent lattice of vortex rings, while panel \cref{fig_complex_states}(e)
is reminiscent of the vortex ring cages that appear in the dynamical instabilities 
of other states such as the spherical dark shell solitary wave of \cref{fig_linear_states}(f)~\cite{wenl1}. 
\cref{fig_complex_states}(f) displays a conglomeration of bent vortex
lines. Once again, all of these solutions have not been previously
identified as stationary states of the 3D NLS/GP model, to the best
of our knowledge.

\section{Conclusions \& Future Challenges}
Deflation reveals unknown and intriguing dynamical states of a fundamental 
model for 3D Bose-Einstein condensates. By building a priori knowledge of 
the linear eigenstates into the deflation procedure, we are able to identify 
a wide range of solutions. Many of the solutions found can be characterized 
using these underlying linear limits. However, deflation can also discover 
numerous unexpected topological nonlinear states such as the vortex ring with 
2 vortex lines, or the coupled bent vortex rings of \cref{fig_new_states}. 
Despite their complexity, such states may only be weakly unstable (thus potentially 
tractable experimentally) and feature long-time dynamics consisting of splittings 
and recombinations towards the original state. Recent experimental advances have 
enabled the formulation (painting) of arbitrary potentials~\cite{boshier}, the 
establishment of arbitrary density~\cite{optica} or imposition of controlled 
phase~\cite{yefsah0,spiel,markus} patterns, and even the realization of unstable 
(but sufficiently long-lived) complex topological states such as vortex knots~\cite{dsh2}. 
In light of all these developments and their impact on vortex ring and line 
dynamics~\cite{yefsah,lampo1,lampo2}, we expect that the states identified in this 
work to be within reach of current state-of-the-art experimental efforts.

As the nonlinearity of the model is increased, so is the complexity of the available
topological states; yet the numerical methods discussed here appear to remain efficient 
in this regime. They reveal not only solutions that are generalizations of previous ones,
but also vortex ring lattices, cages, bent-connected-multivortex ring and line patterns, 
and more. These warrant further study, topological classification and deeper physical 
understanding. We believe that this technique paves the way for a wide range of future 
exciting explorations in this and related fields.

\section*{Acknowledgements} This work is supported by the EPSRC Centre For 
Doctoral Training in Industrially Focused Mathematical Modelling (EP/L015803/1) 
in collaboration with Simula Research Laboratory (NB), EPSRC grants EP/R029423/1 and EP/V001493/1 (PEF), and by the U.S.~National Science Foundation under Grant no.~PHY-1602994 (PGK).

\appendix

\section{BdG spectral decomposition} \label{app_spec}
\begin{figure}[htbp]
\begin{center}
\hspace{0.1cm}
\begin{overpic}[width=\linfigwidth,trim={50 100 50 40}, clip]{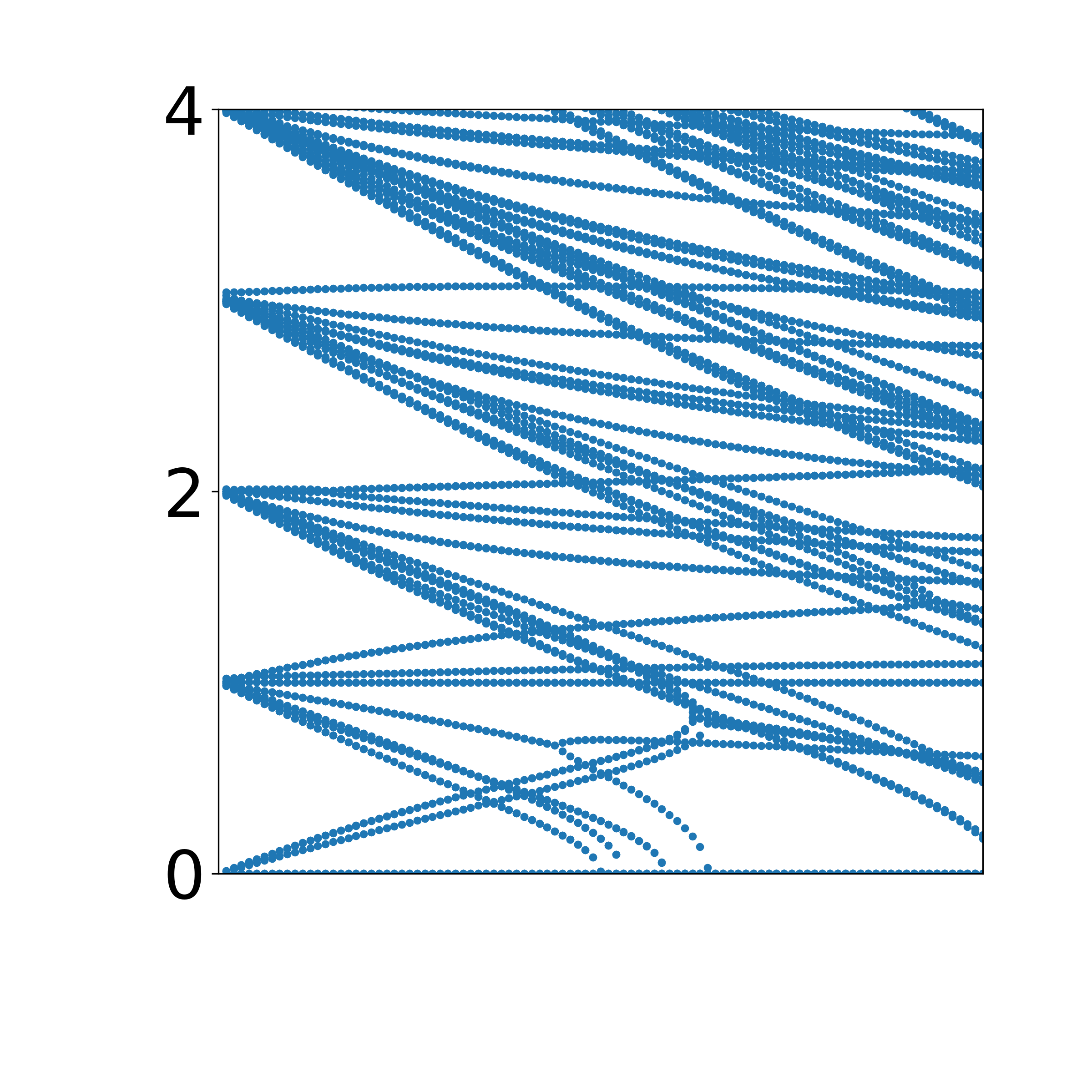}
\put(-14,32){\rotatebox{90}{$\mathcal{R}(\omega)$}}
\end{overpic}
\begin{overpic}[width=\linfigwidth,trim={50 100 50 40}, clip]{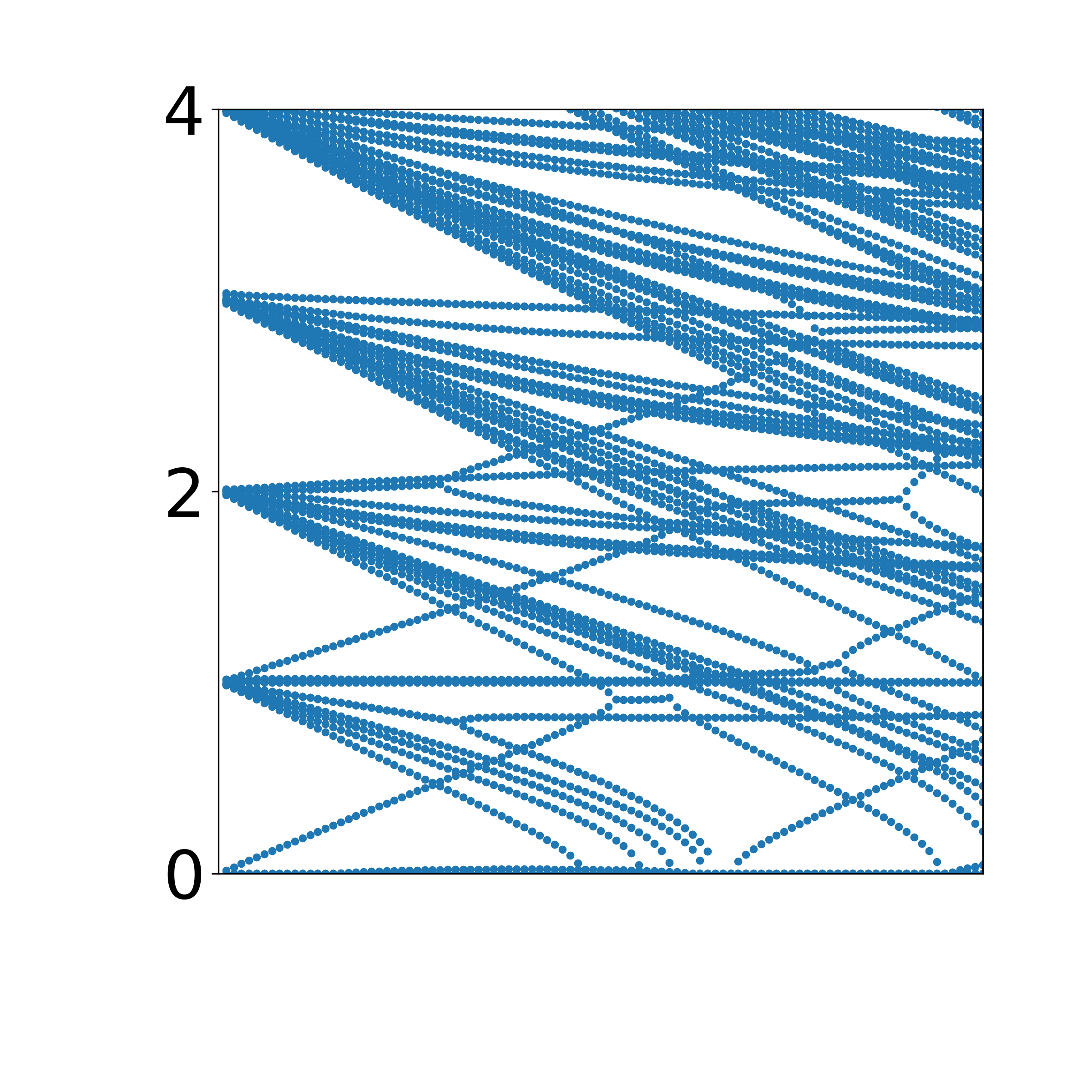}
\end{overpic}
\begin{overpic}[width=\linfigwidth,trim={50 100 50 40}, clip]{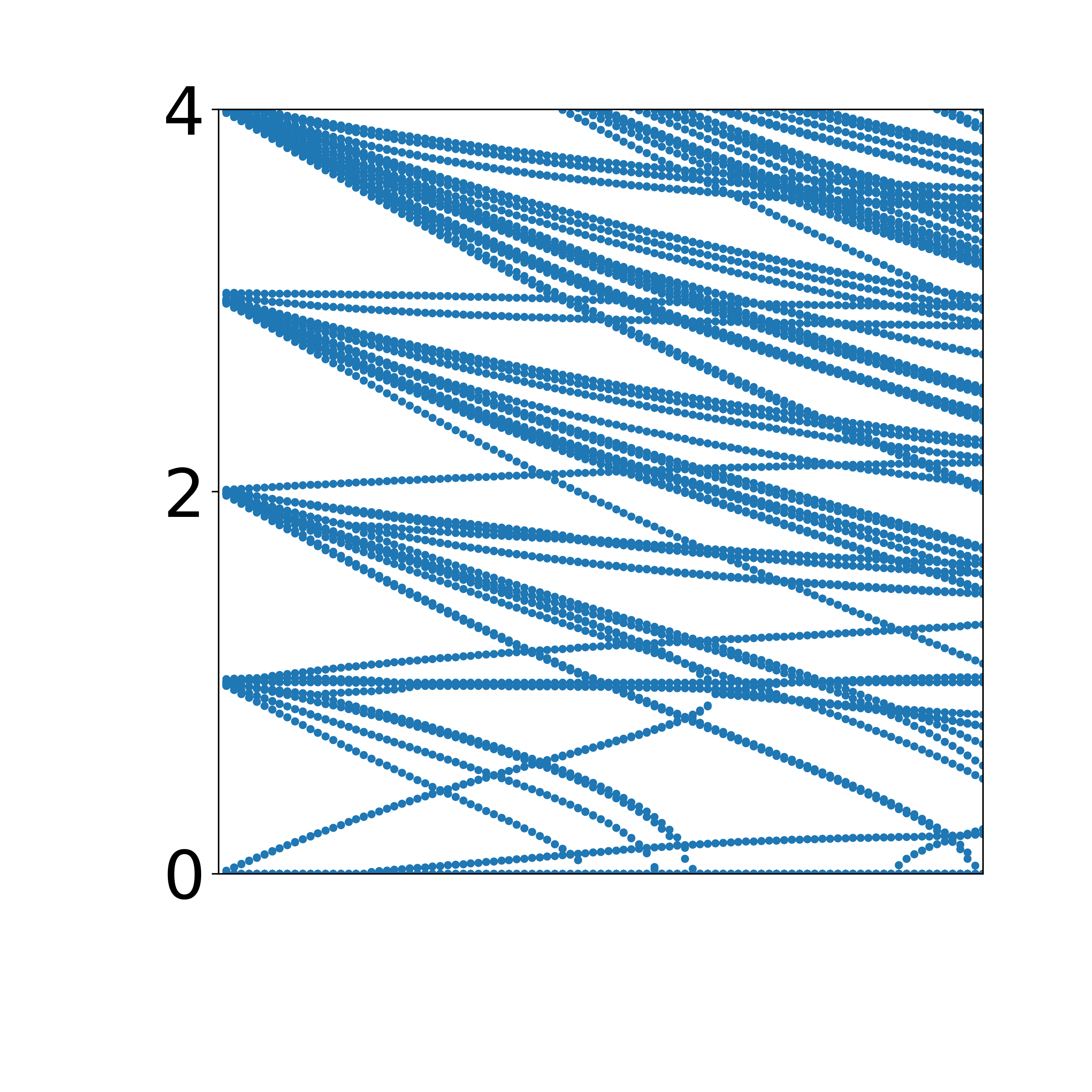}
\end{overpic}
\\
\vspace{0.1cm}
\hspace{0.1cm}
\begin{overpic}[width=\linfigwidth,trim={50 75 50 40}, clip]{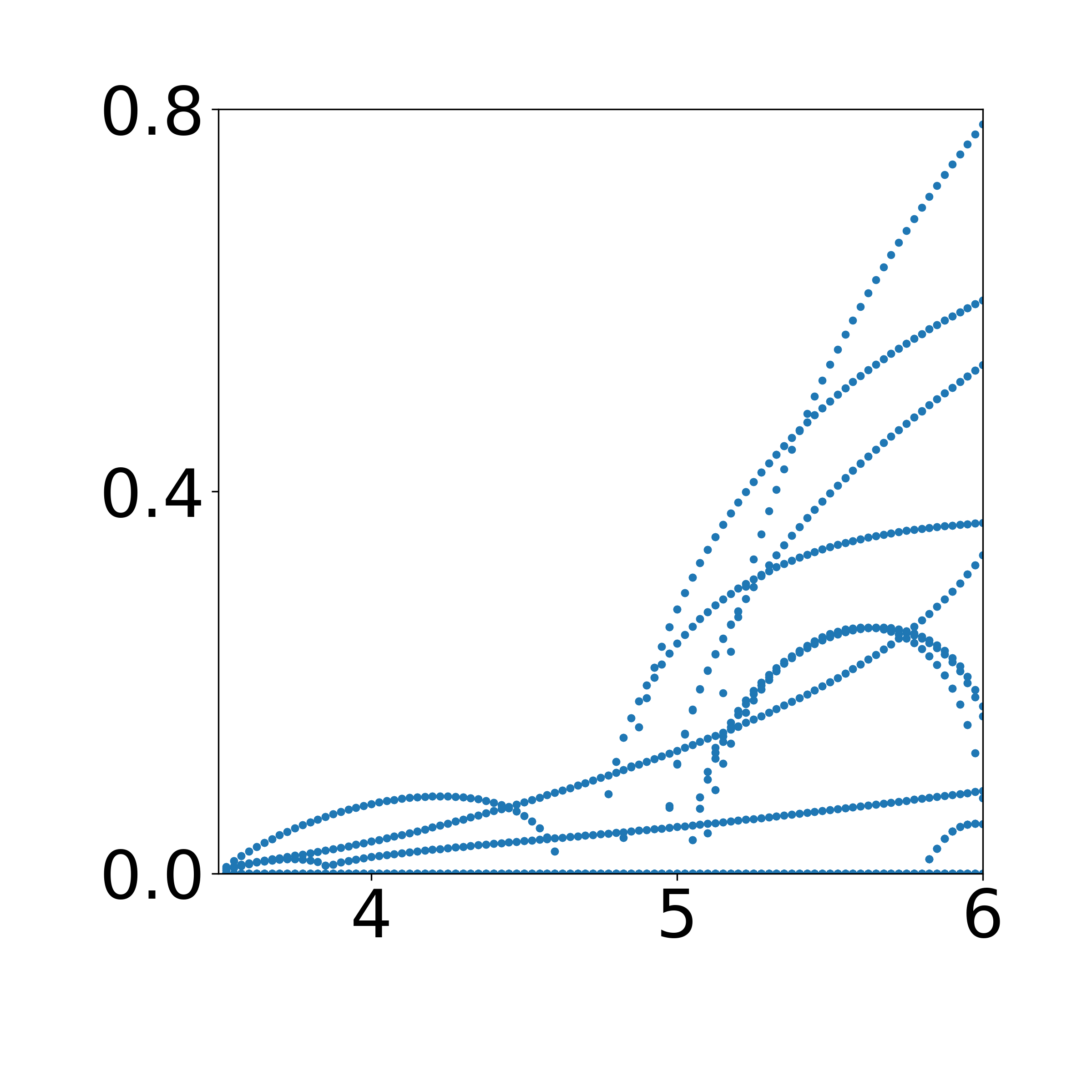}
\put(55,-6){$\mu$}
\put(15,82){(a)}
\put(-14,38){\rotatebox{90}{$\mathcal{I}(\omega)$}}
\end{overpic}
\begin{overpic}[width=\linfigwidth,trim={50 75 50 40}, clip]{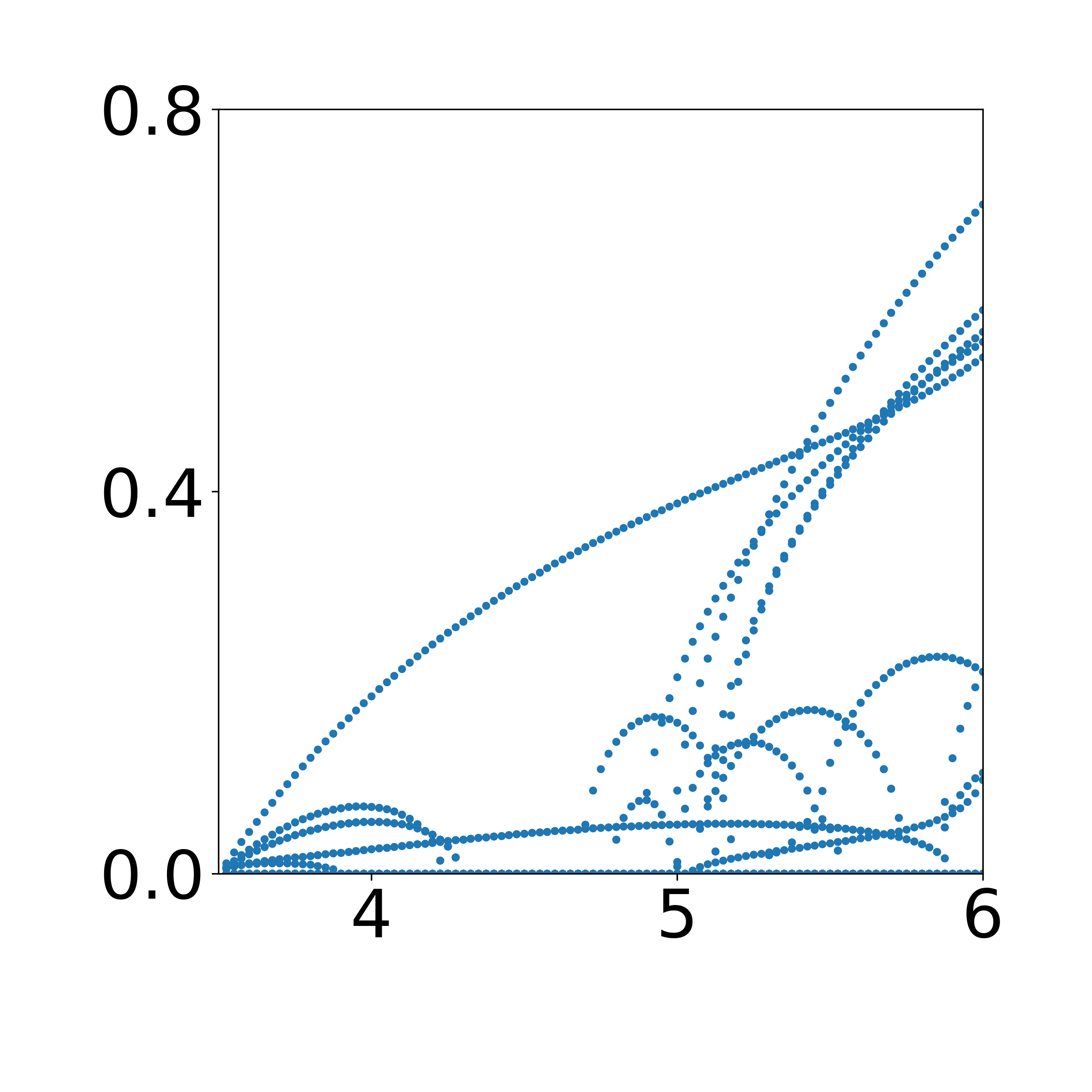}
\put(55,-6){$\mu$}
\put(15,82){(b)}
\end{overpic}
\begin{overpic}[width=\linfigwidth,trim={50 75 50 40}, clip]{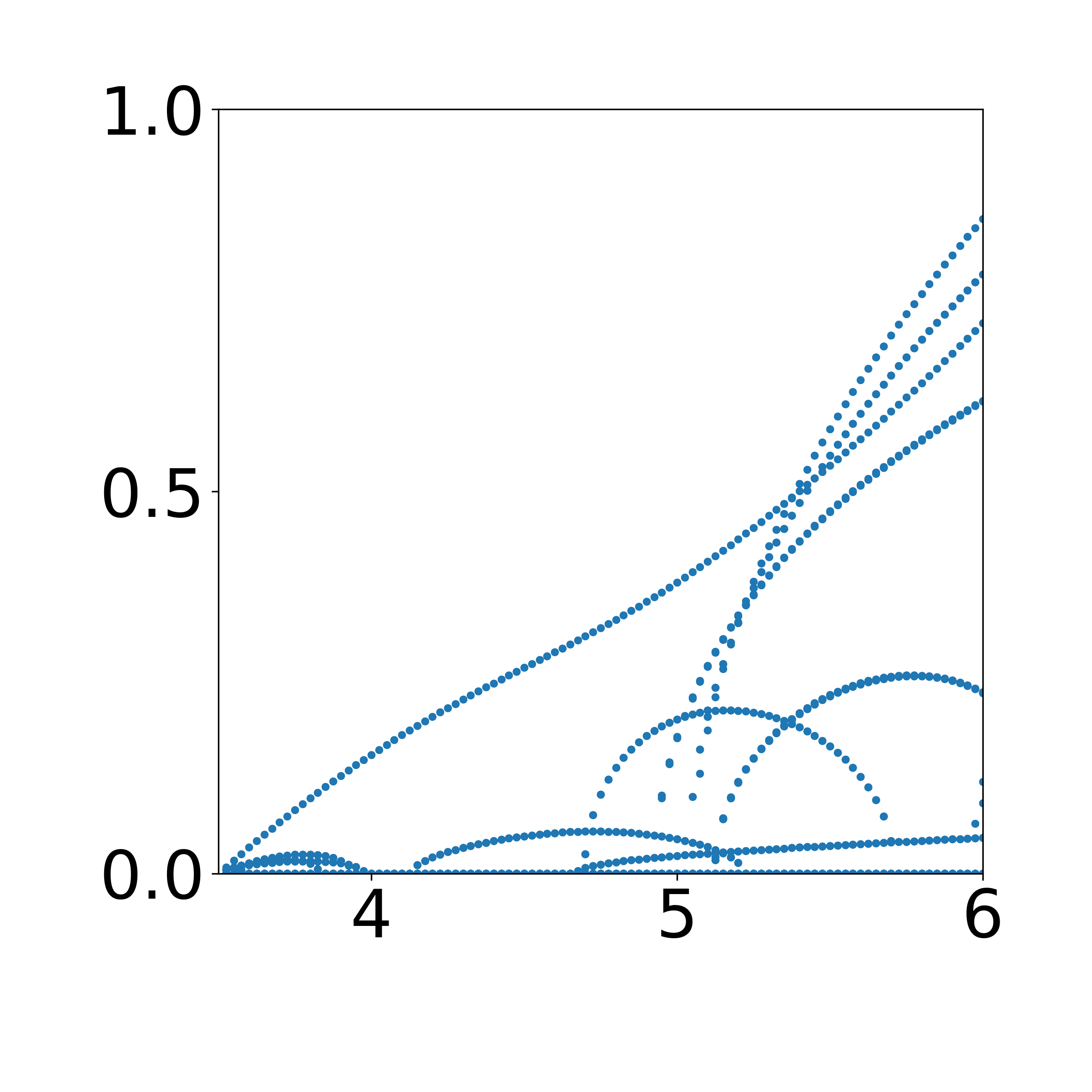}
\put(55,-6){$\mu$}
\put(15,82){(c)}
\end{overpic}
\\
\vspace{0.3cm}
\hspace{0.1cm}
\begin{overpic}[width=\linfigwidth,trim={50 100 50 40}, clip]{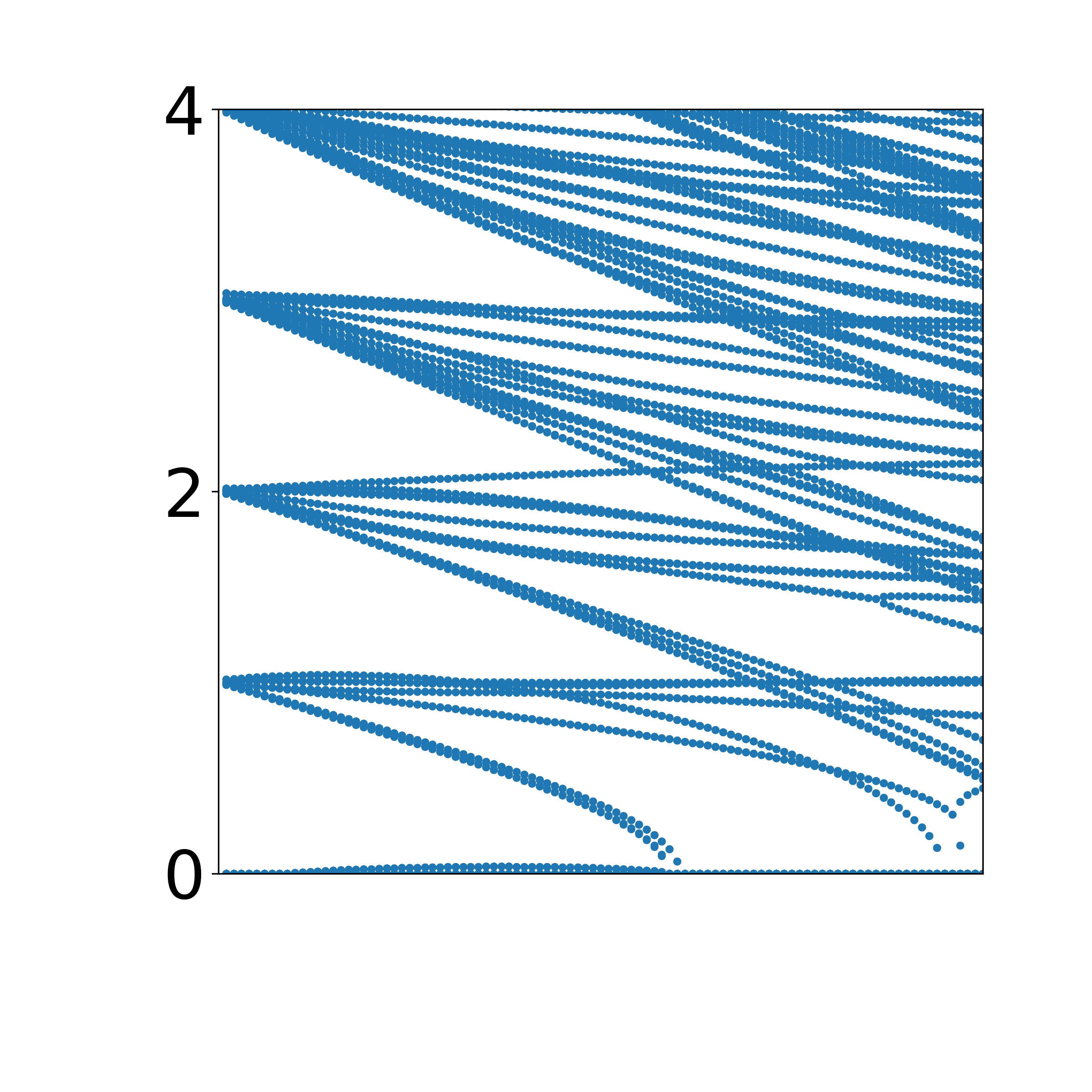}
\put(-14,32){\rotatebox{90}{$\mathcal{R}(\omega)$}}
\end{overpic}
\begin{overpic}[width=\linfigwidth,trim={50 100 50 40}, clip]{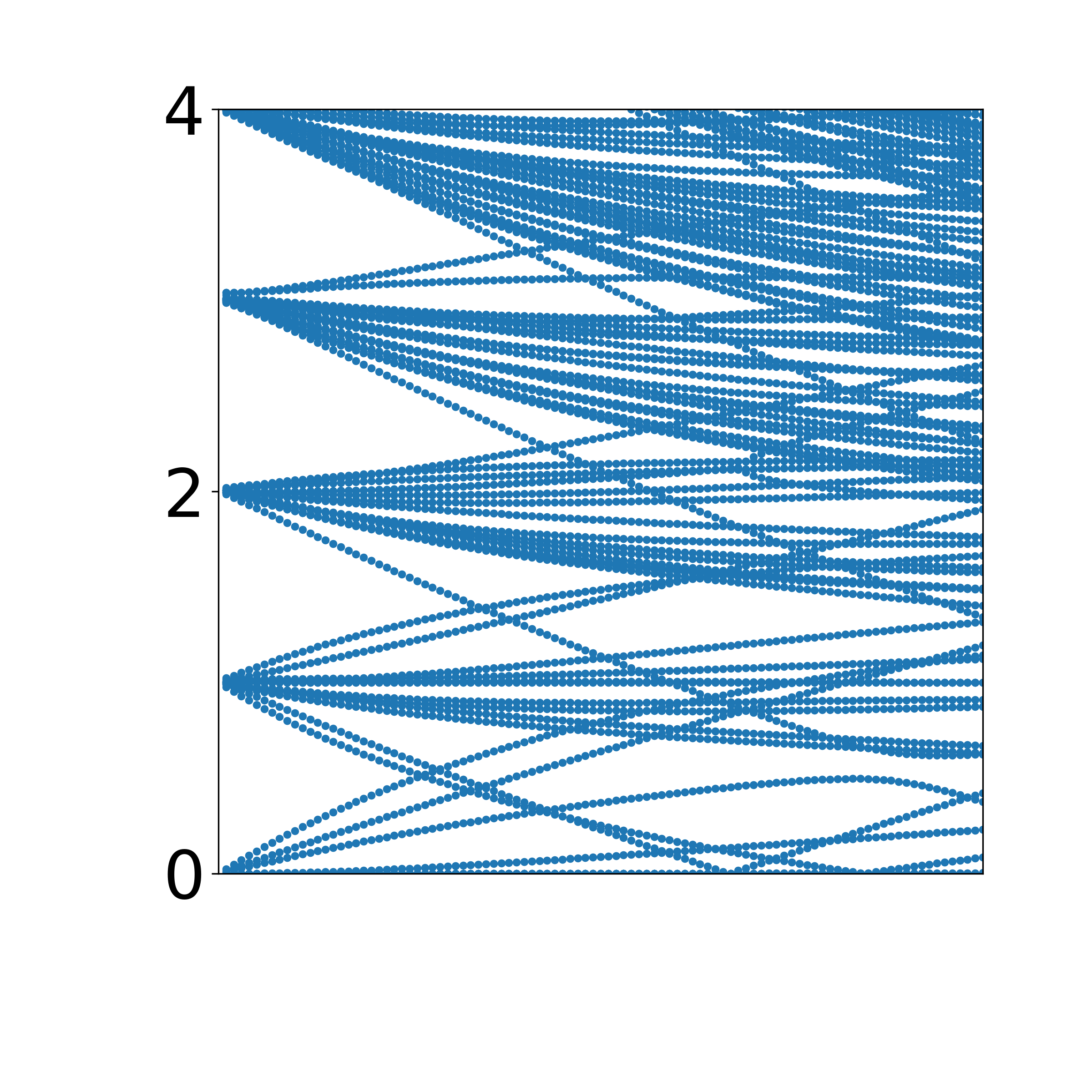}
\end{overpic}
\begin{overpic}[width=\linfigwidth,trim={50 100 50 40}, clip]{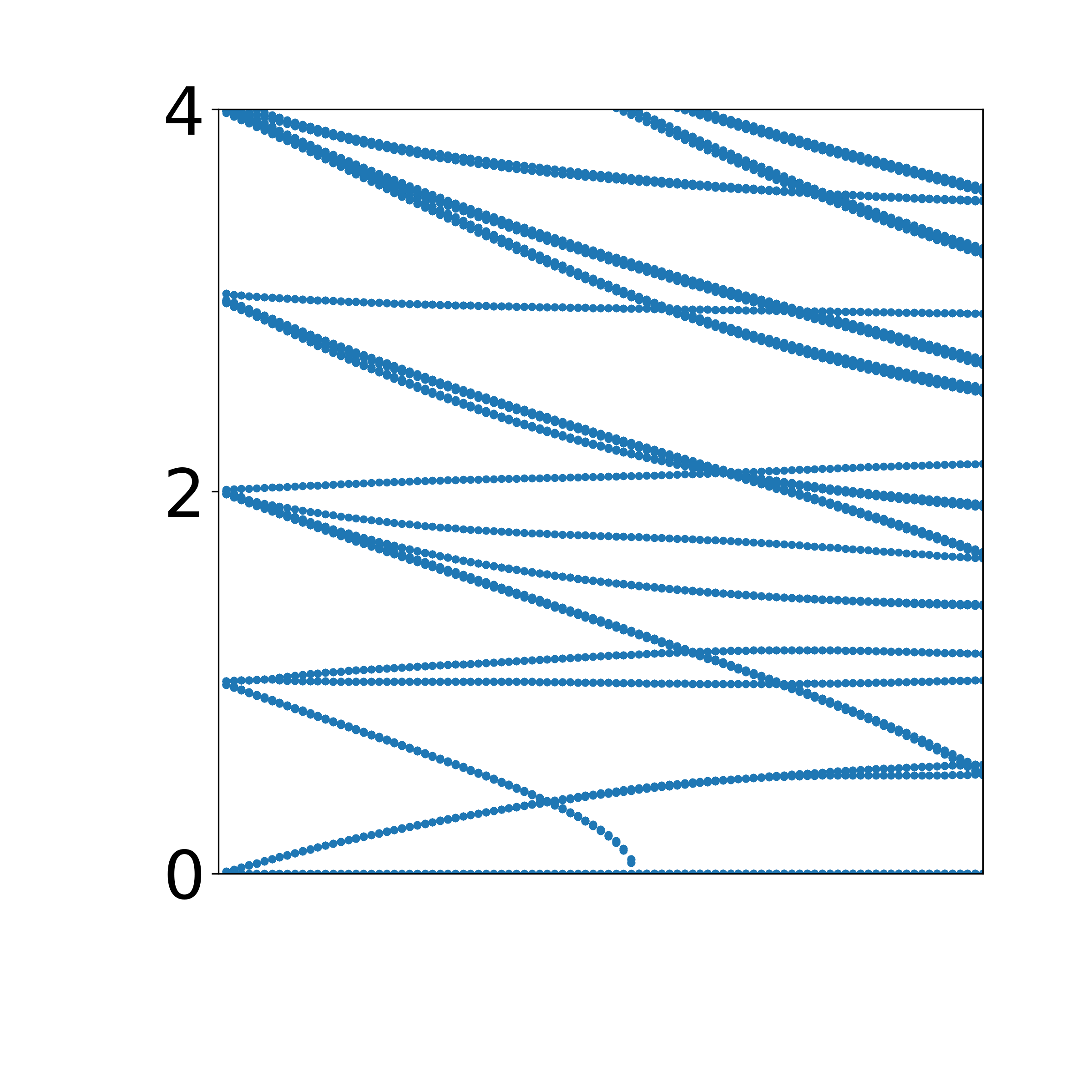}
\end{overpic}
\\
\vspace{0.1cm}
\hspace{0.1cm}
\begin{overpic}[width=\linfigwidth,trim={50 75 50 40}, clip]{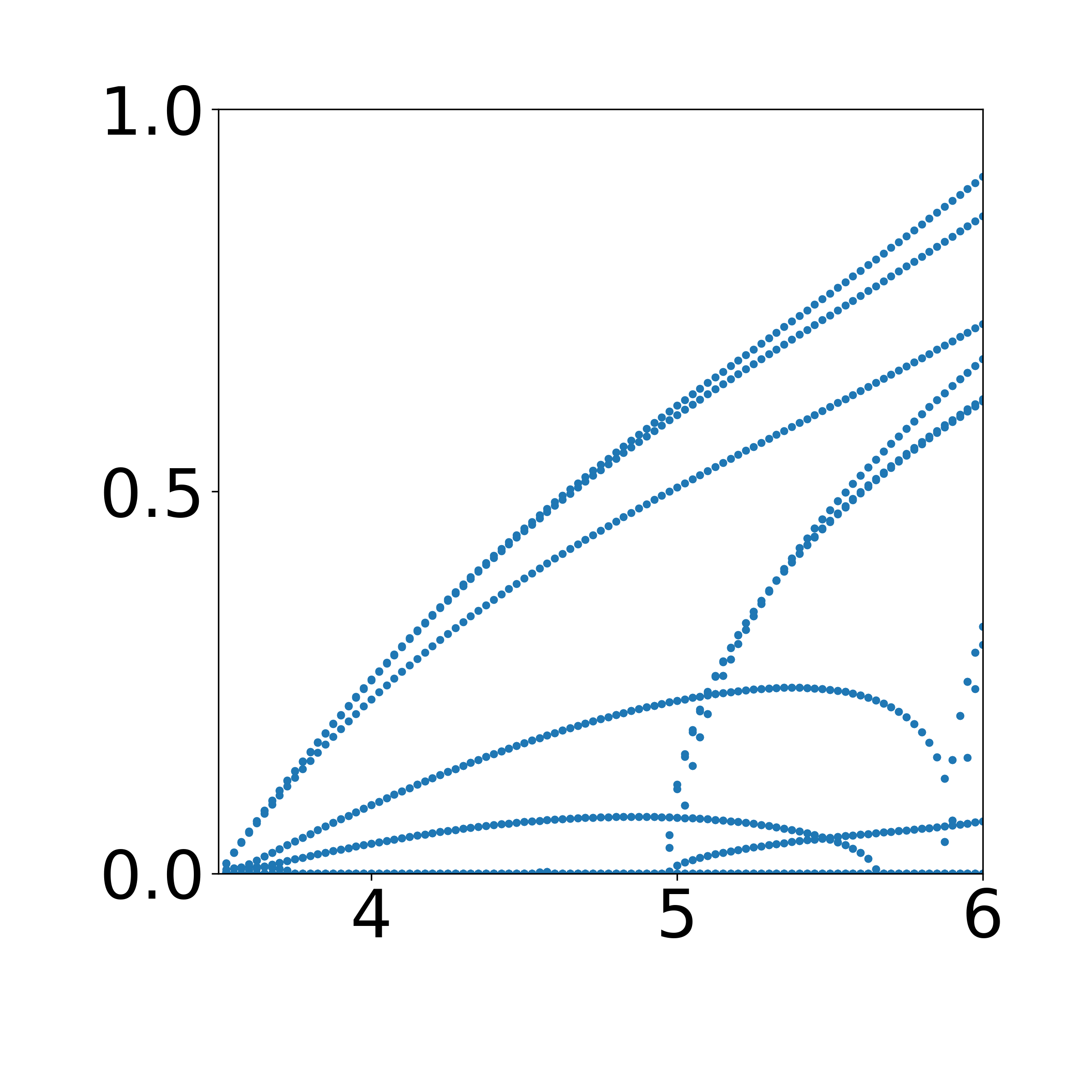}
\put(55,-6){$\mu$}
\put(15,82){(d)}
\put(-14,38){\rotatebox{90}{$\mathcal{I}(\omega)$}}
\end{overpic}
\begin{overpic}[width=\linfigwidth,trim={50 75 50 40}, clip]{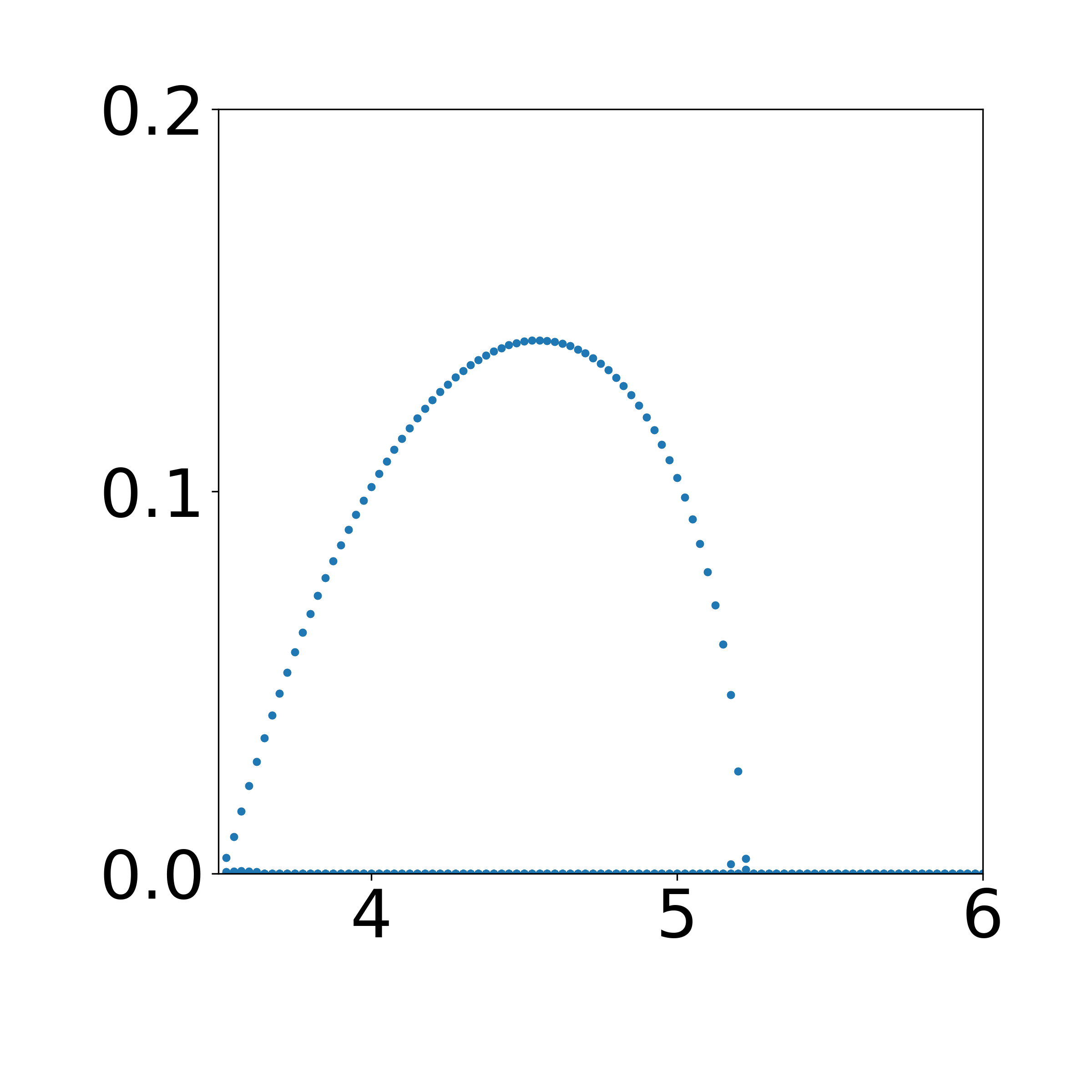}
\put(55,-6){$\mu$}
\put(15,82){(e)}
\end{overpic}
\begin{overpic}[width=\linfigwidth,trim={50 75 50 40}, clip]{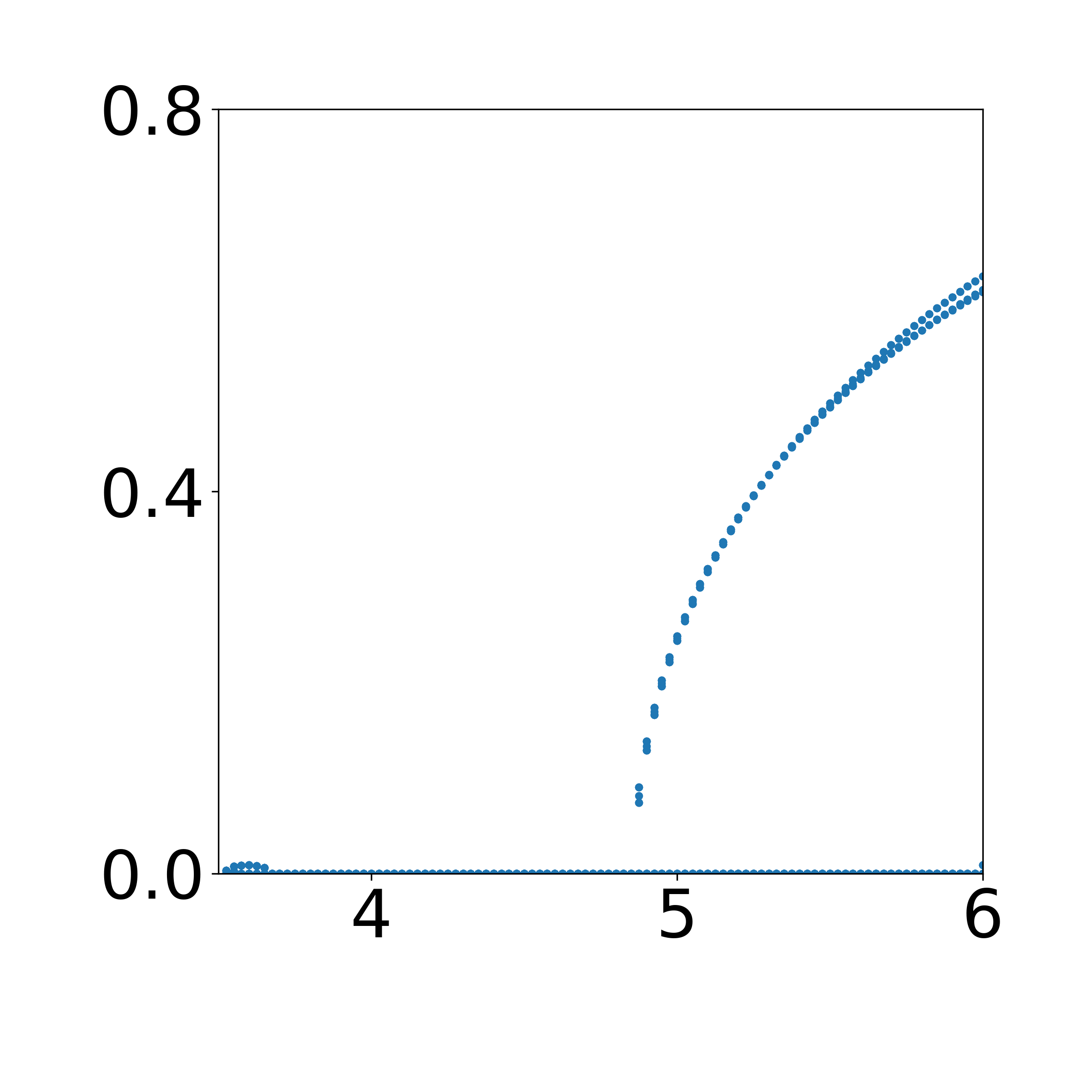}
\put(55,-6){$\mu$}
\put(15,82){(f)}
\end{overpic}
\end{center}
\caption{Spectra of the solutions presented in \cref{fig_linear_states} 
(e.g., the panel (a) corresponds to the branch 1(a), illustrated in the 
panel (a) of \cref{fig_linear_states}). The real and imaginary parts of 
the corresponding eigenfrequencies $\omega$ are respectively depicted in 
the top and bottom panels.}
\label{fig_stab_1}
\end{figure}

\begin{figure}[htbp]
\begin{center}
\hspace{0.1cm}
\begin{overpic}[width=\linfigwidth,trim={50 100 50 40}, clip]{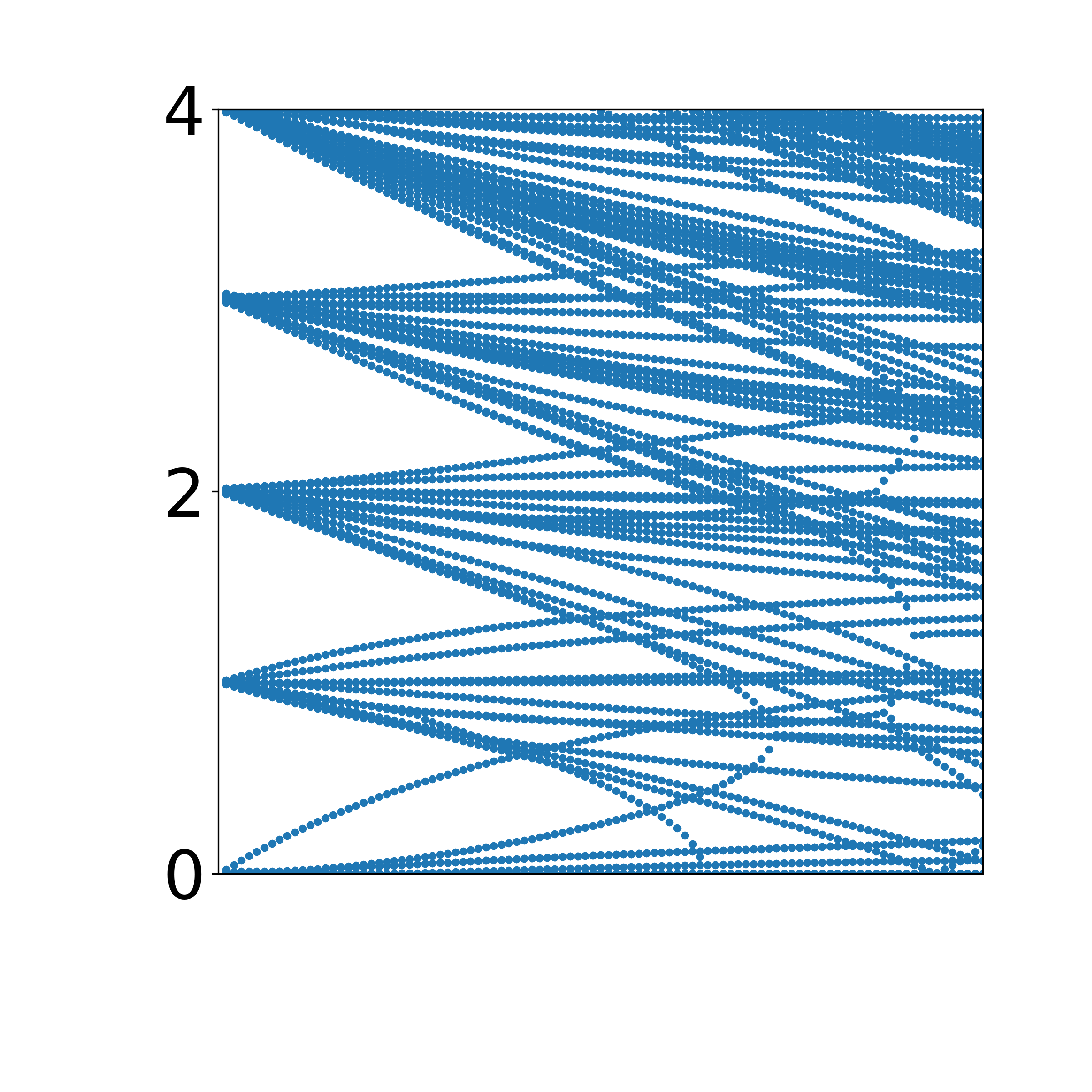}
\put(-14,32){\rotatebox{90}{$\mathcal{R}(\omega)$}}
\end{overpic}
\begin{overpic}[width=\linfigwidth,trim={50 100 50 40}, clip]{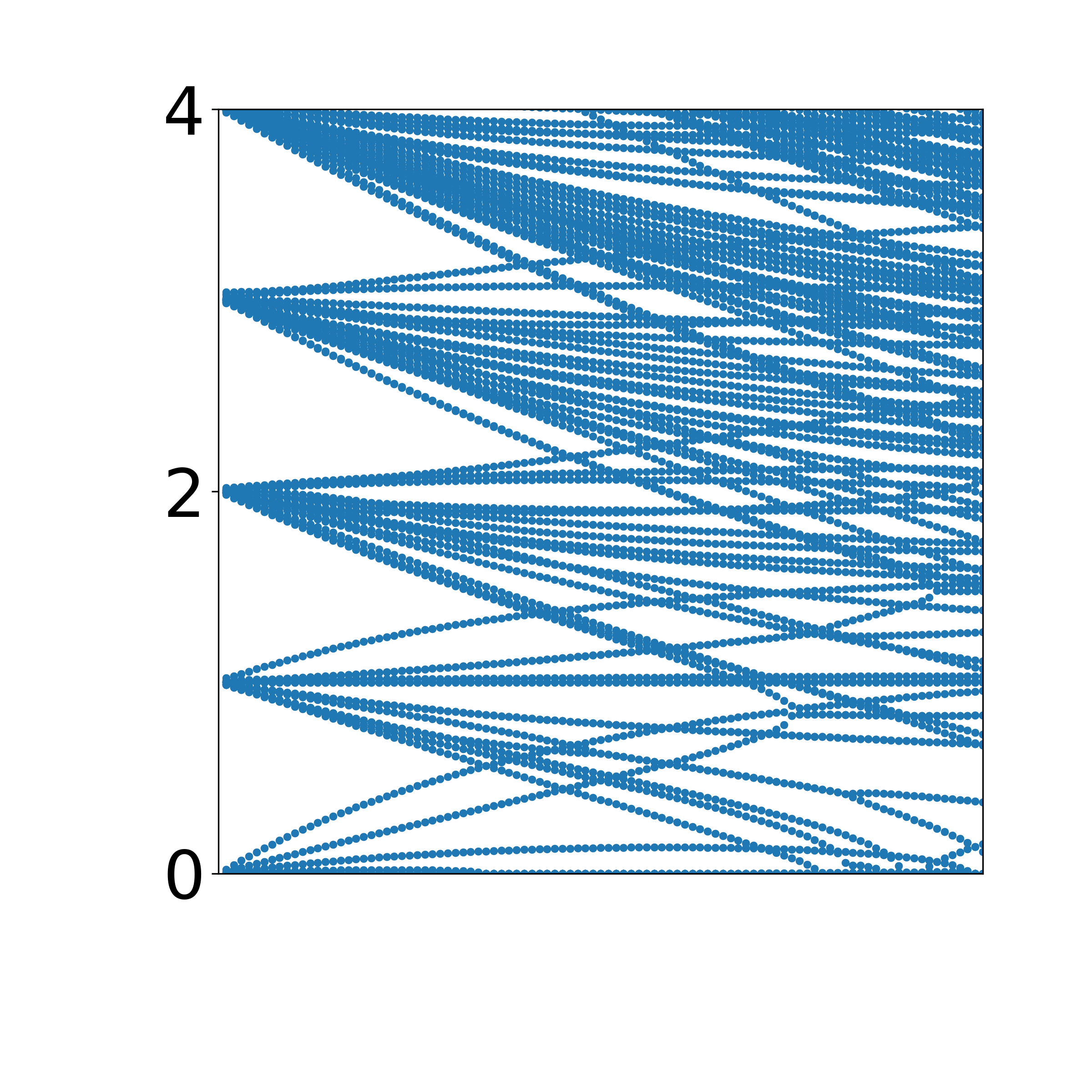}
\end{overpic}
\begin{overpic}[width=\linfigwidth,trim={50 100 50 40}, clip]{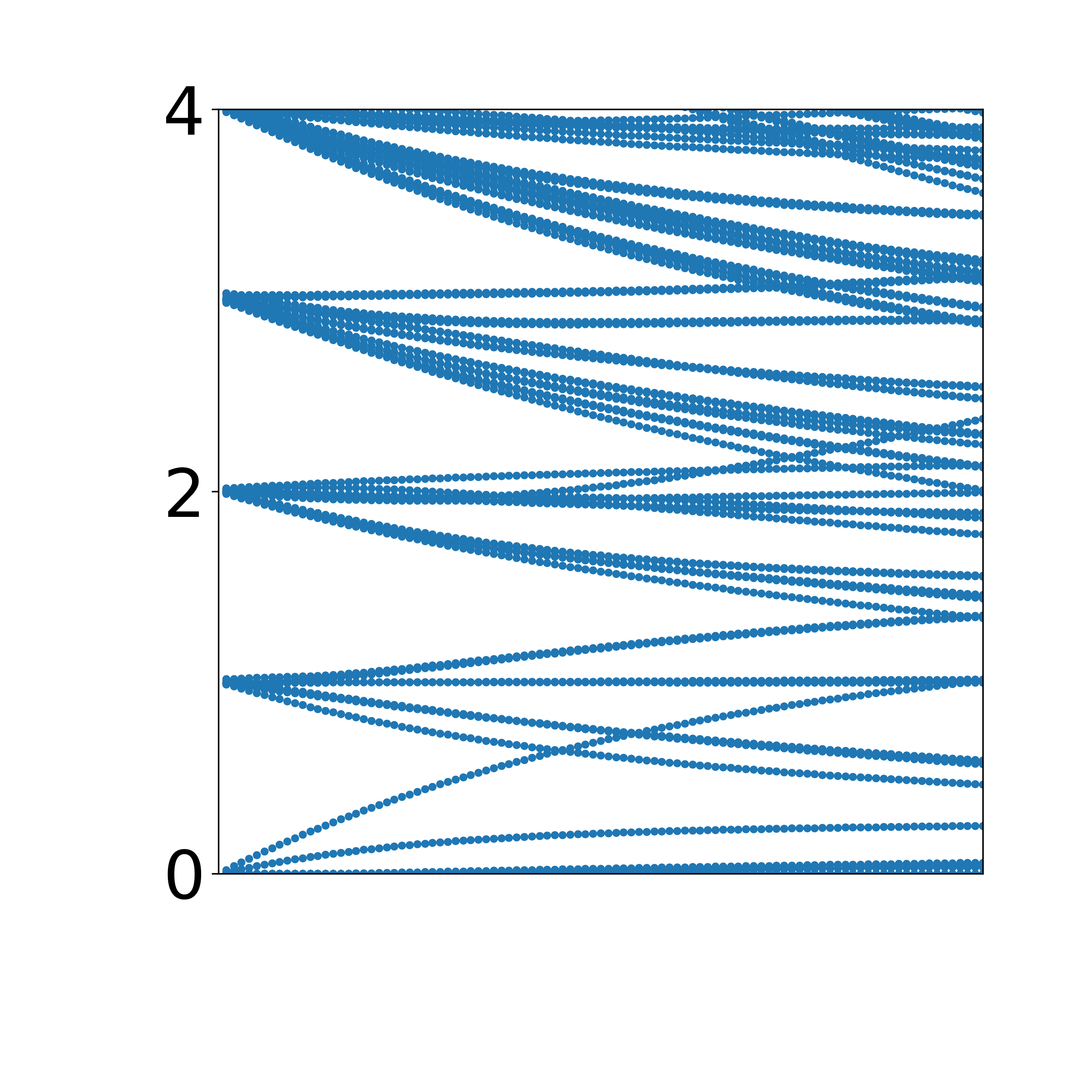}
\end{overpic}
\\
\vspace{0.1cm}
\hspace{0.1cm}
\begin{overpic}[width=\linfigwidth,trim={50 75 50 40}, clip]{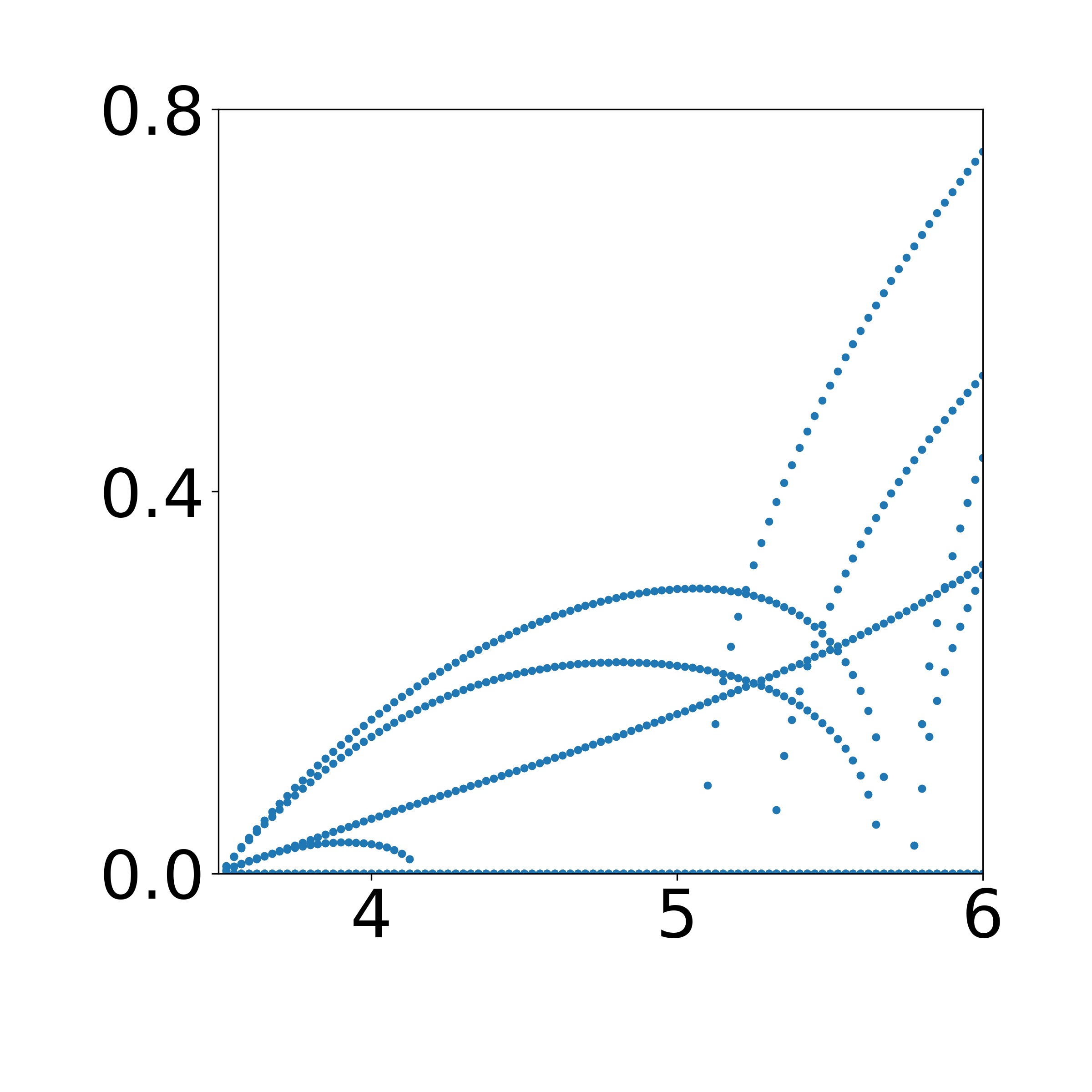}
\put(55,-6){$\mu$}
\put(15,82){(a)}
\put(-14,38){\rotatebox{90}{$\mathcal{I}(\omega)$}}
\end{overpic}
\begin{overpic}[width=\linfigwidth,trim={50 75 50 40}, clip]{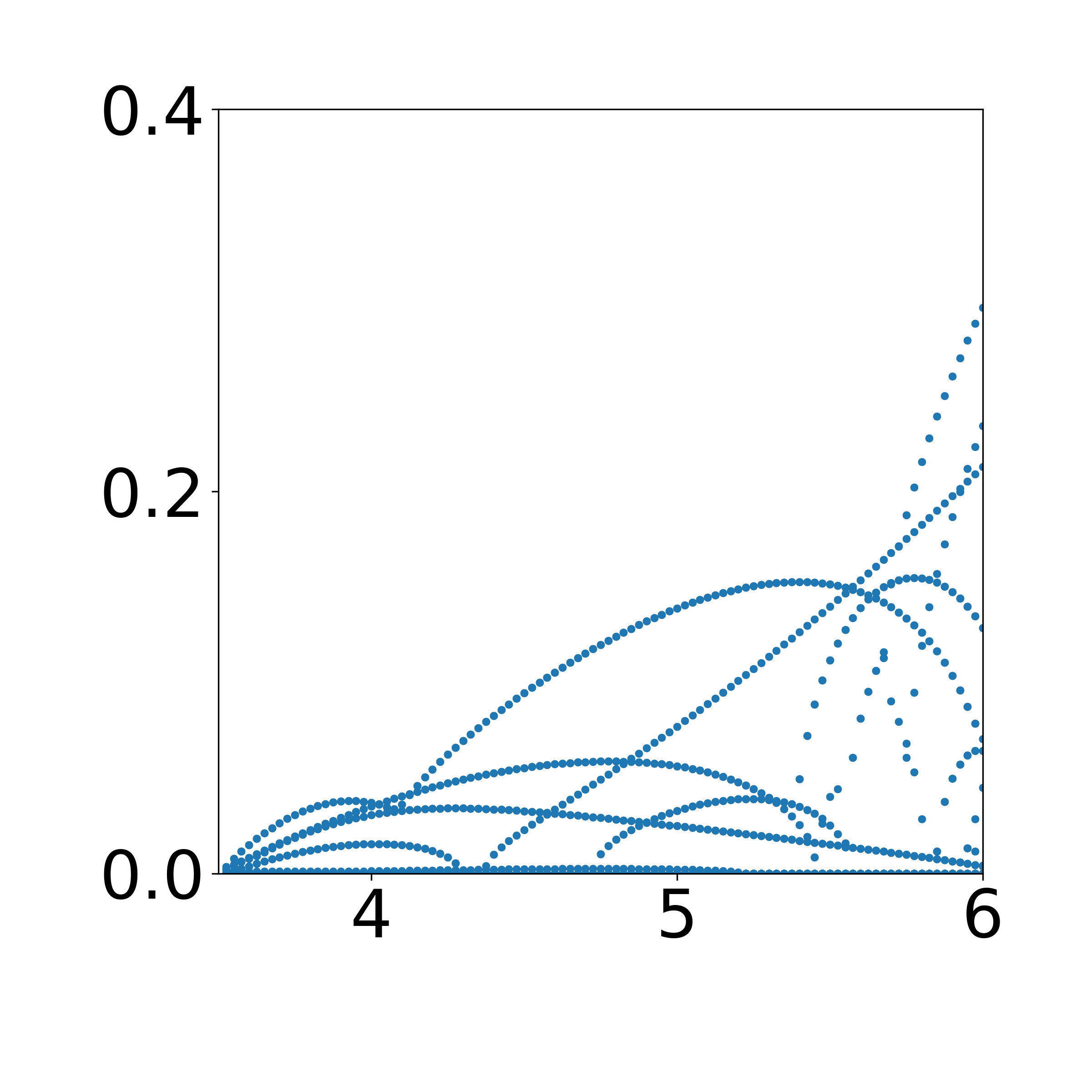}
\put(55,-6){$\mu$}
\put(15,82){(b)}
\end{overpic}
\begin{overpic}[width=\linfigwidth,trim={50 75 50 40}, clip]{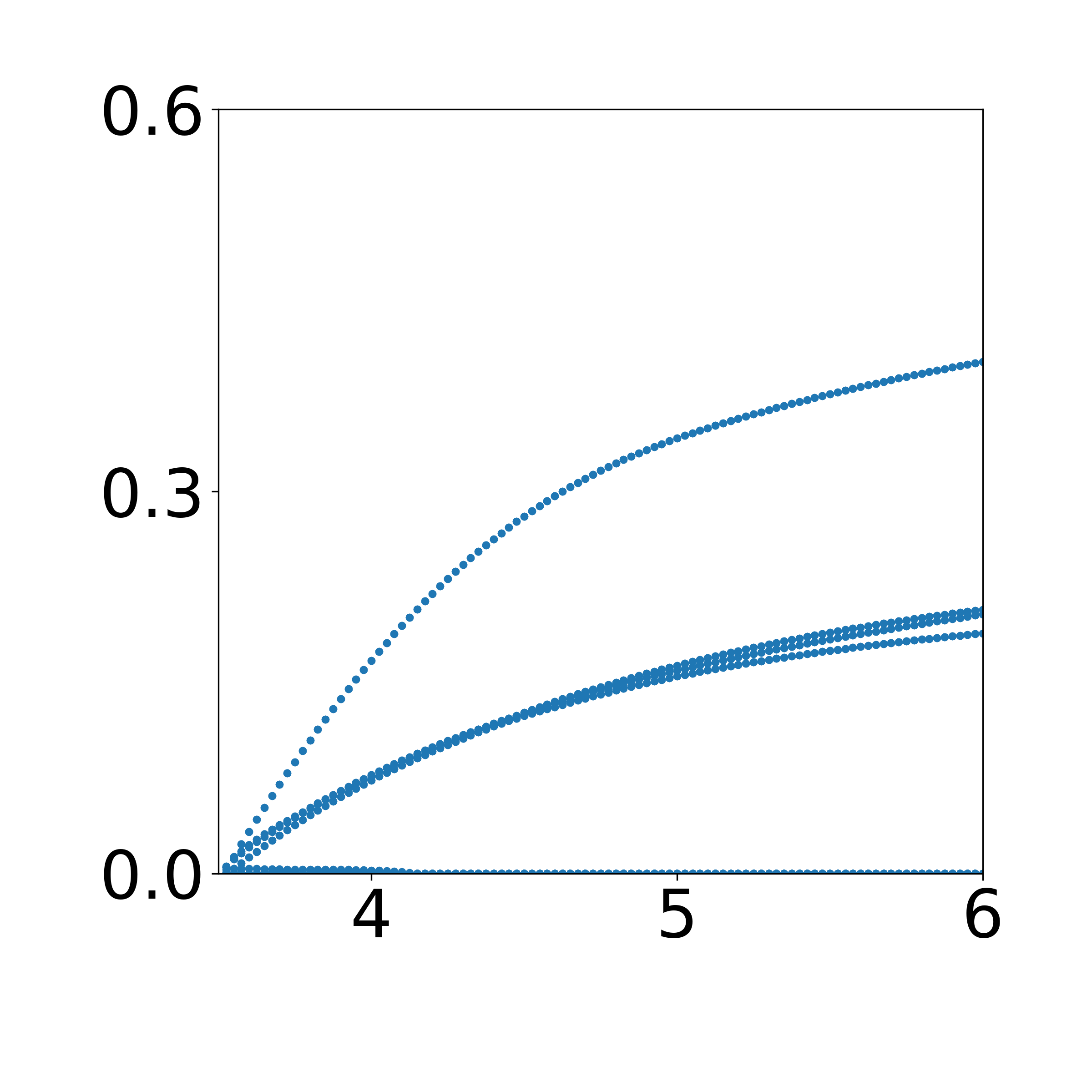}
\put(55,-6){$\mu$}
\put(15,82){(c)}
\end{overpic}
\\
\vspace{0.3cm}
\hspace{0.1cm}
\begin{overpic}[width=\linfigwidth,trim={50 100 50 40}, clip]{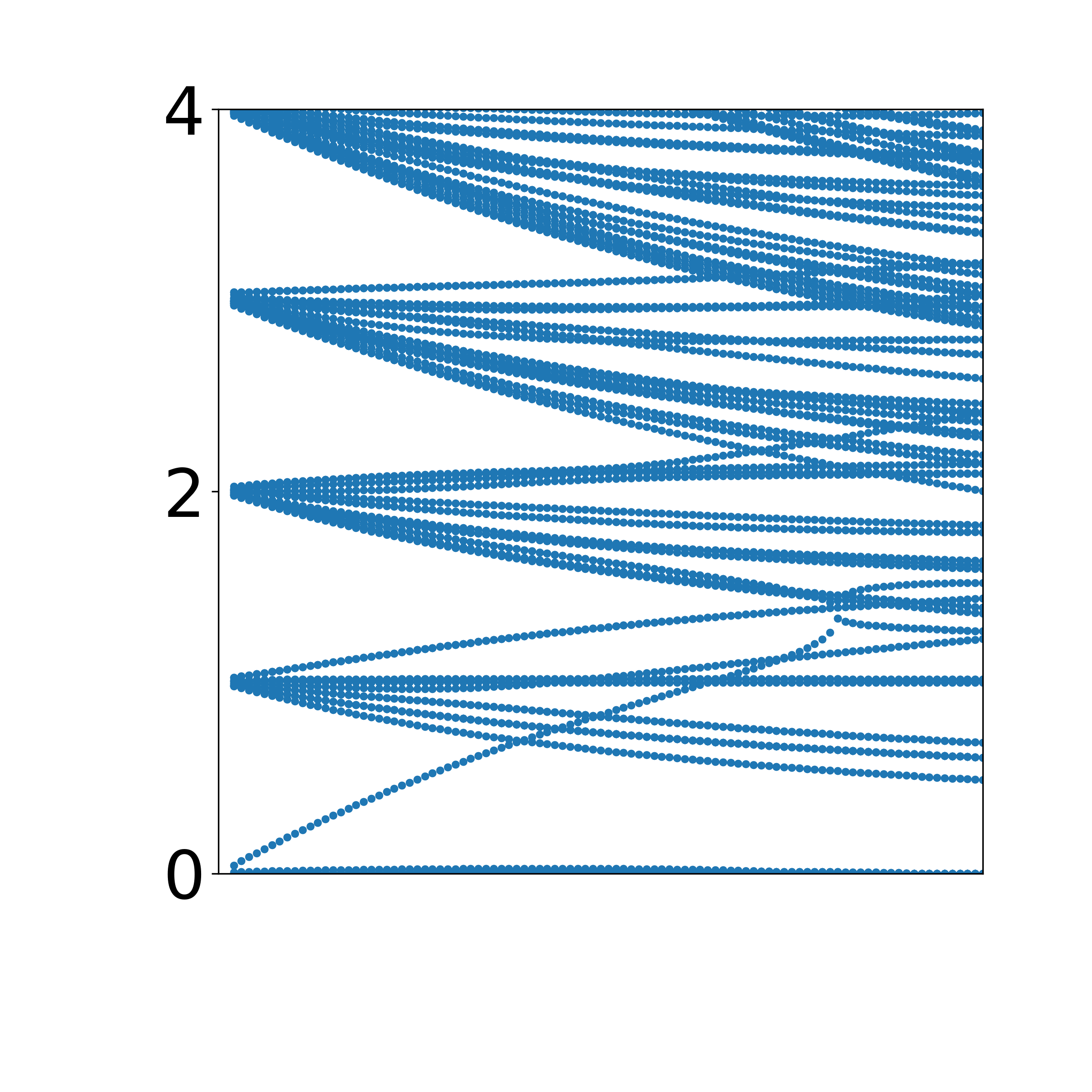}
\put(-14,32){\rotatebox{90}{$\mathcal{R}(\omega)$}}
\end{overpic}
\begin{overpic}[width=\linfigwidth,trim={50 100 50 40}, clip]{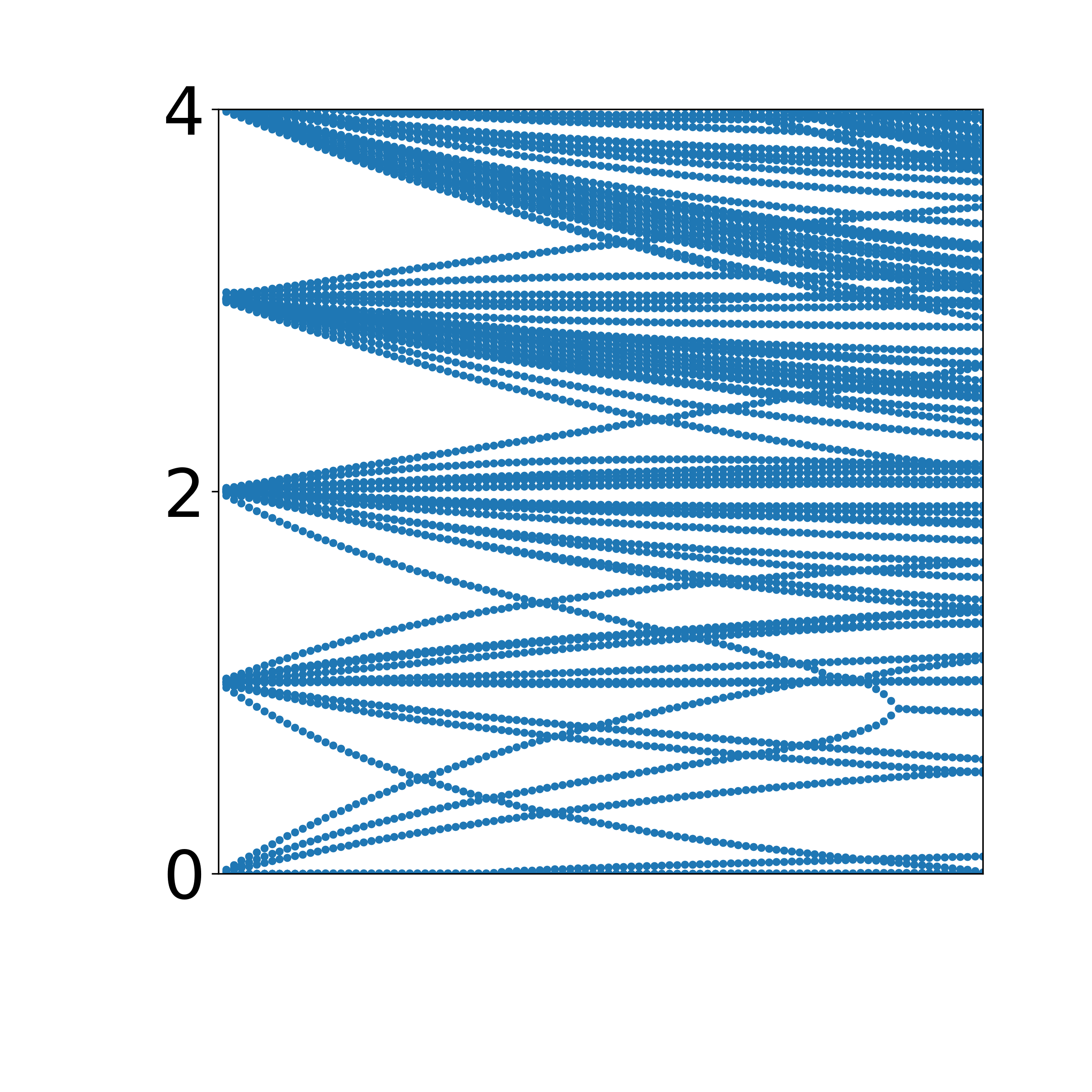}
\end{overpic}
\\
\vspace{0.1cm}
\hspace{0.1cm}
\begin{overpic}[width=\linfigwidth,trim={50 75 50 40}, clip]{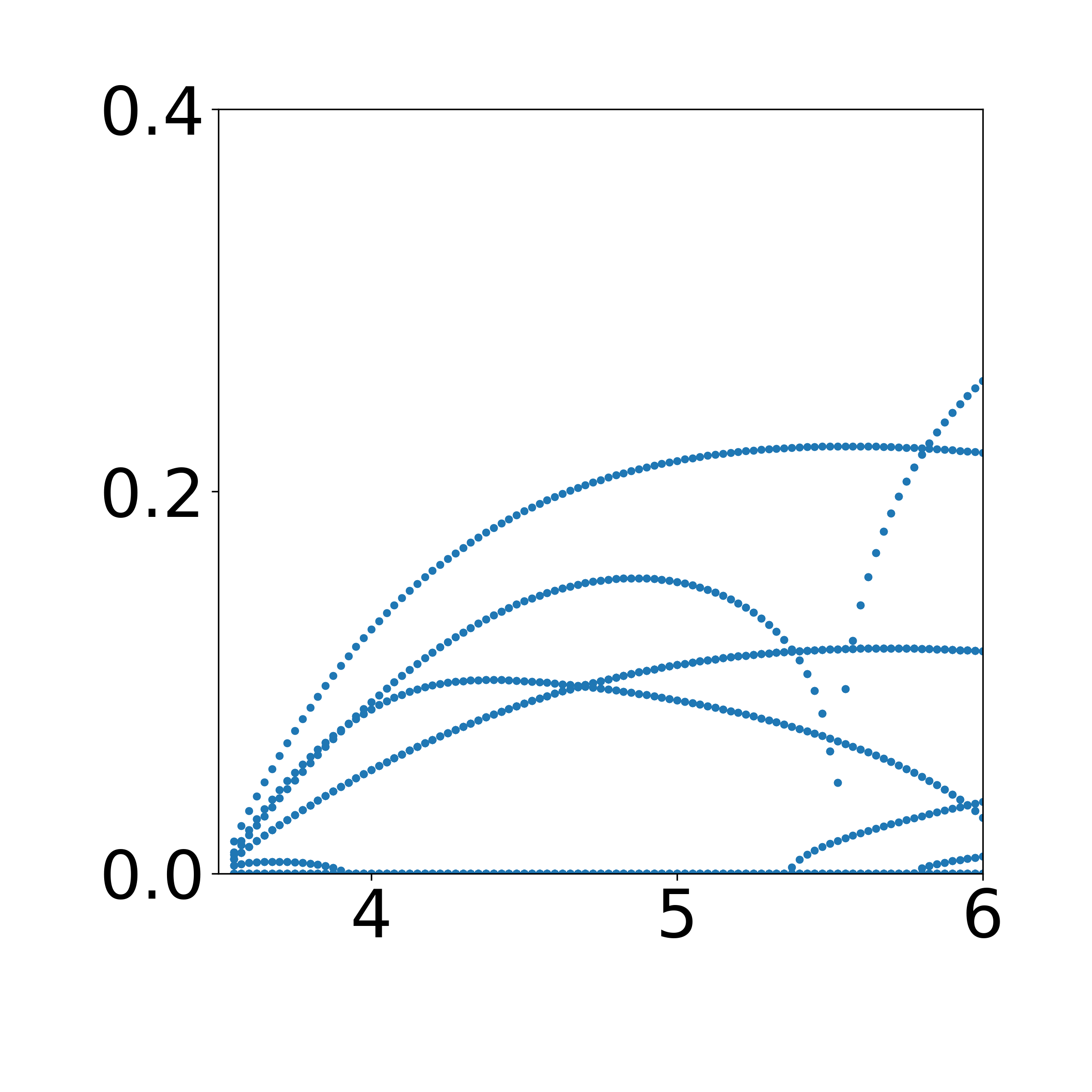}
\put(55,-6){$\mu$}
\put(15,82){(d)}
\put(-14,38){\rotatebox{90}{$\mathcal{I}(\omega)$}}
\end{overpic}
\begin{overpic}[width=\linfigwidth,trim={50 75 50 40}, clip]{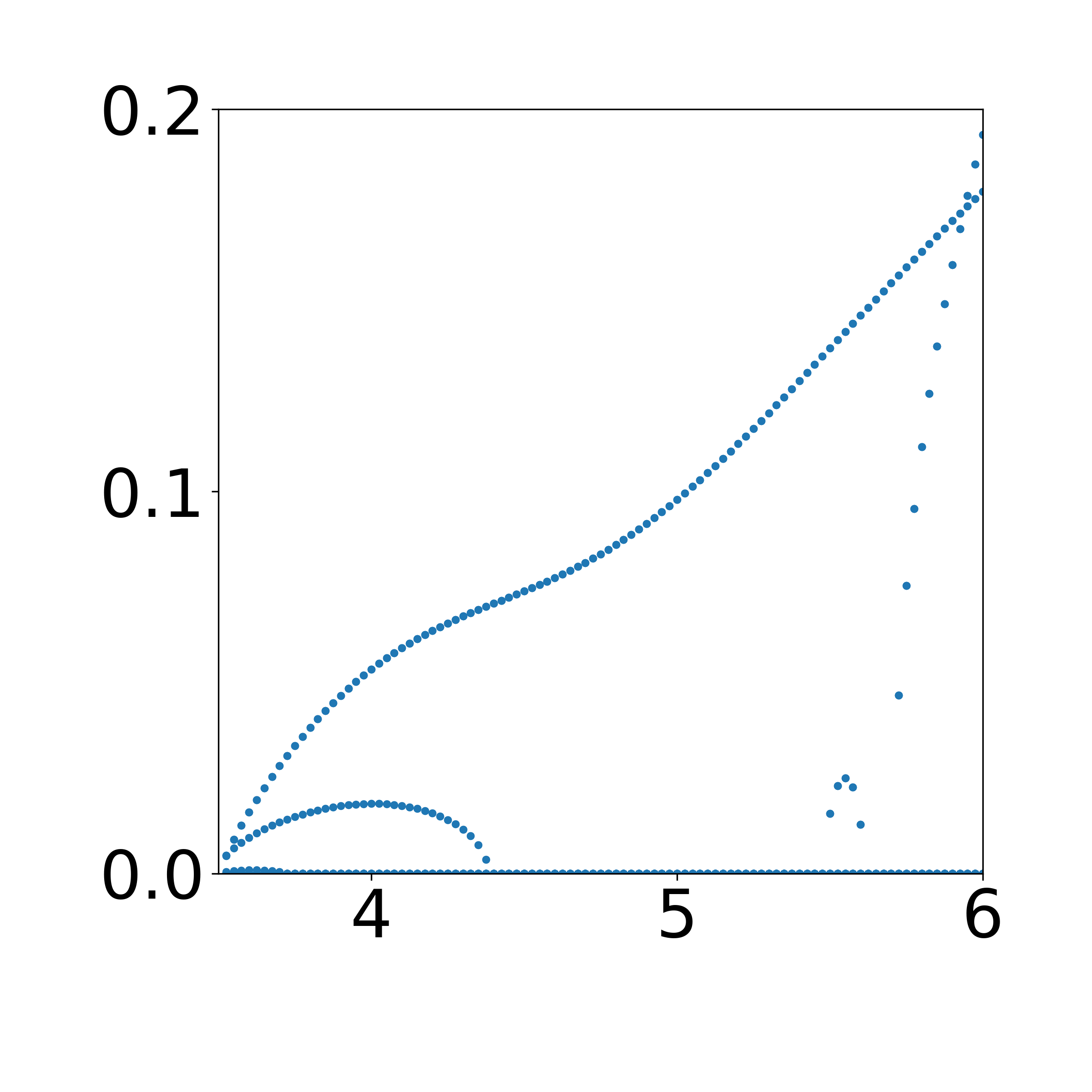}
\put(55,-6){$\mu$}
\put(15,82){(e)}
\end{overpic}
\end{center}
\caption{Spectra of the solutions presented in \cref{fig_new_states}(a)-(e).}
\label{fig_stab_2}
\end{figure}
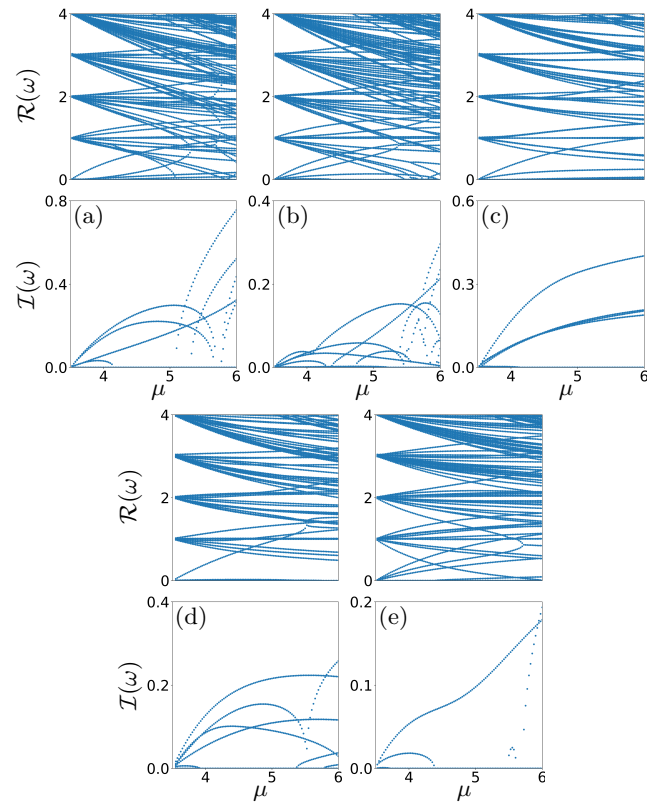

We decompose the steady-state solution $\phi^{0}$, eigenvector 
$(a,b)^\top$, and eigenvalue $\rho$ into real and imaginary 
components as $\phi^{0}=\phi^{0}_{r}+i\phi^{0}_{c}$, $a=a_r+i a_c$, 
$b = b_r+i b_c$, and $\rho = \rho_r+i\rho_c$, respectively, 
and rewrite \eqref{eq_eig} as
\begin{equation} \label{eq_real_eigenproblem}
\begin{aligned}
&\begin{pmatrix}
A_{11} & 0 & B_1 & -B_2\\
0 & A_{11} & B_2 & B_1\\
-B_1 & -B_2 & -A_{11} & 0\\
B_2 & -B_1 & 0 & -A_{11}
\end{pmatrix}
\begin{pmatrix}
a_r\\
a_c\\
b_r\\
b_c
\end{pmatrix} = \\
&\begin{pmatrix}
\rho_r & -\rho_c & 0 & 0\\
\rho_c & \rho_r & 0 & 0\\
0 & 0 & \rho_r & -\rho_c\\
0 & 0 & \rho_c & \rho_r\\
\end{pmatrix}
\begin{pmatrix}
a_r\\
a_c\\
b_r\\
b_c
\end{pmatrix},
\end{aligned}
\end{equation}
where $B_1= (\phi^0_r)^2-(\phi^0_c)^2$ and $B_2=2\phi^0_r \phi^0_c$. 
The eigenvalues of the matrix on the right-hand side of \eqref{eq_real_eigenproblem} 
are $\rho_r\pm i\rho_c$ (with multiplicity two). Therefore, solving a 
real eigenvalue problem with the left-hand matrix of \eqref{eq_real_eigenproblem} 
yields the same eigenvalues and eigenvectors as the complex eigenvalue 
problem \eqref{eq_eig}. We use a Krylov--Schur algorithm with a shift-and-invert 
spectral transformation~\cite{stewart2002}, implemented in the SLEPc 
library~\cite{hernandez2005slepc}, to solve the following eigenvalue problem:
\begin{align} \label{eq_real_eigenproblem_2}
\begin{pmatrix}
A_{11} & 0 & B_1 & -B_2\\
0 & A_{11} & B_2 & B_1\\
-B_1 & -B_2 & -A_{11} & 0\\
B_2 & -B_1 & 0 & -A_{11}
\end{pmatrix}
\begin{pmatrix}
a_r\\
a_c\\
b_r\\
b_c
\end{pmatrix} =
\rho
\begin{pmatrix}
a_r\\
a_c\\
b_r\\
b_c
\end{pmatrix},
\end{align}
where the matrices are real and $\rho=\rho_r\pm i\rho_c$ is complex. 
This problem is discretized with the same piecewise cubic finite element 
method used to find multiple solutions with deflation. The spectra of 
the states of Figs.~\ref{fig_linear_states} and \ref{fig_new_states}
are shown in Figs.~\ref{fig_stab_1} and~\ref{fig_stab_2}, respectively.

\bibliography{biblio}

\begin{thebibliography}{41}
\expandafter\ifx\csname natexlab\endcsname\relax\def\natexlab#1{#1}\fi
\expandafter\ifx\csname bibnamefont\endcsname\relax
  \def\bibnamefont#1{#1}\fi
\expandafter\ifx\csname bibfnamefont\endcsname\relax
  \def\bibfnamefont#1{#1}\fi
\expandafter\ifx\csname citenamefont\endcsname\relax
  \def\citenamefont#1{#1}\fi
\expandafter\ifx\csname url\endcsname\relax
  \def\url#1{\texttt{#1}}\fi
\expandafter\ifx\csname urlprefix\endcsname\relax\def\urlprefix{URL }\fi
\providecommand{\bibinfo}[2]{#2}
\providecommand{\eprint}[2][]{\url{#2}}

\bibitem[{\citenamefont{Ablowitz et~al.}(2004)\citenamefont{Ablowitz, Prinari,
  and Trubatch}}]{ablowitz}
\bibinfo{author}{\bibfnamefont{M.}~\bibnamefont{Ablowitz}},
  \bibinfo{author}{\bibfnamefont{B.}~\bibnamefont{Prinari}}, \bibnamefont{and}
  \bibinfo{author}{\bibfnamefont{A.}~\bibnamefont{Trubatch}},
  \emph{\bibinfo{title}{Discrete and Continuous Nonlinear Schr\"odinger
  Systems}} (\bibinfo{publisher}{Cambridge University Press, Cambridge},
  \bibinfo{year}{2004}).

\bibitem[{\citenamefont{Ablowitz}(2011)}]{ablowitz1}
\bibinfo{author}{\bibfnamefont{M.}~\bibnamefont{Ablowitz}},
  \emph{\bibinfo{title}{Nonlinear Dispersive Waves}}
  (\bibinfo{publisher}{Cambridge University Press, Cambridge},
  \bibinfo{year}{2011}).

\bibitem[{\citenamefont{Sulem and Sulem}(1999)}]{sulem}
\bibinfo{author}{\bibfnamefont{C.}~\bibnamefont{Sulem}} \bibnamefont{and}
  \bibinfo{author}{\bibfnamefont{P.}~\bibnamefont{Sulem}},
  \emph{\bibinfo{title}{The nonlinear {Schr\"odinger} equation: self-focusing
  and wave collapse}} (\bibinfo{publisher}{Springer}, \bibinfo{address}{New
  York}, \bibinfo{year}{1999}).

\bibitem[{\citenamefont{Pethick and Smith}(2002)}]{pethick}
\bibinfo{author}{\bibfnamefont{C.~J.} \bibnamefont{Pethick}} \bibnamefont{and}
  \bibinfo{author}{\bibfnamefont{H.}~\bibnamefont{Smith}},
  \emph{\bibinfo{title}{Bose--{E}instein {C}ondensation in {D}ilute {G}ases}}
  (\bibinfo{publisher}{Cambridge University Press},
  \bibinfo{address}{Cambridge, United Kingdom}, \bibinfo{year}{2002}).

\bibitem[{\citenamefont{Stringari and Pitaevskii}(2003)}]{stringari}
\bibinfo{author}{\bibfnamefont{S.}~\bibnamefont{Stringari}} \bibnamefont{and}
  \bibinfo{author}{\bibfnamefont{L.}~\bibnamefont{Pitaevskii}},
  \emph{\bibinfo{title}{Bose--Einstein Condensation}}
  (\bibinfo{publisher}{Oxford University Press}, \bibinfo{address}{Oxford,
  United Kingdom}, \bibinfo{year}{2003}).

\bibitem[{\citenamefont{Kevrekidis et~al.}(2015)\citenamefont{Kevrekidis,
  Frantzeskakis, and Carretero-González}}]{siambook}
\bibinfo{author}{\bibfnamefont{P.}~\bibnamefont{Kevrekidis}},
  \bibinfo{author}{\bibfnamefont{D.}~\bibnamefont{Frantzeskakis}},
  \bibnamefont{and}
  \bibinfo{author}{\bibfnamefont{R.}~\bibnamefont{Carretero-González}},
  \emph{\bibinfo{title}{The Defocusing Nonlinear Schrodinger Equation}}
  (\bibinfo{publisher}{Society for Industrial and Applied Mathematics},
  \bibinfo{address}{Philadelphia, PA}, \bibinfo{year}{2015}).

\bibitem[{\citenamefont{Hasegawa and Kodama}(1995)}]{hasegawa}
\bibinfo{author}{\bibfnamefont{A.}~\bibnamefont{Hasegawa}} \bibnamefont{and}
  \bibinfo{author}{\bibfnamefont{Y.}~\bibnamefont{Kodama}},
  \emph{\bibinfo{title}{Solitons in {O}ptical {C}ommunications}}
  (\bibinfo{publisher}{Clarendon Press}, \bibinfo{address}{Oxford},
  \bibinfo{year}{1995}).

\bibitem[{\citenamefont{Kono and \v{S}kori{\'c}}(2010)}]{zakh1}
\bibinfo{author}{\bibfnamefont{M.}~\bibnamefont{Kono}} \bibnamefont{and}
  \bibinfo{author}{\bibfnamefont{M.}~\bibnamefont{\v{S}kori{\'c}}},
  \emph{\bibinfo{title}{Nonlinear Physics of Plasmas}}
  (\bibinfo{publisher}{Springer Verlag}, \bibinfo{address}{Heidelberg},
  \bibinfo{year}{2010}).

\bibitem[{\citenamefont{Kharif et~al.}(2009)\citenamefont{Kharif, Pelinovsky,
  and Slunyaev}}]{slunaev}
\bibinfo{author}{\bibfnamefont{C.}~\bibnamefont{Kharif}},
  \bibinfo{author}{\bibfnamefont{E.}~\bibnamefont{Pelinovsky}},
  \bibnamefont{and} \bibinfo{author}{\bibfnamefont{A.}~\bibnamefont{Slunyaev}},
  \emph{\bibinfo{title}{Rogue Waves in the Ocean}}
  (\bibinfo{publisher}{Springer Verlag}, \bibinfo{address}{Berlin},
  \bibinfo{year}{2009}).

\bibitem[{\citenamefont{Frantzeskakis}(2010)}]{djf}
\bibinfo{author}{\bibfnamefont{D.~J.} \bibnamefont{Frantzeskakis}},
  \bibinfo{journal}{J. Phys. A-Math. Theor.} \textbf{\bibinfo{volume}{43}},
  \bibinfo{pages}{213001} (\bibinfo{year}{2010}), ISSN
  \bibinfo{issn}{1751-8121},
  \urlprefix\url{http://iopscience.iop.org/1751-8121/43/21/213001}.

\bibitem[{\citenamefont{Abdullaev et~al.}(2005)\citenamefont{Abdullaev, Gammal,
  Kamchatnov, and Tomio}}]{tomio}
\bibinfo{author}{\bibfnamefont{F.}~\bibnamefont{Abdullaev}},
  \bibinfo{author}{\bibfnamefont{A.}~\bibnamefont{Gammal}},
  \bibinfo{author}{\bibfnamefont{A.}~\bibnamefont{Kamchatnov}},
  \bibnamefont{and} \bibinfo{author}{\bibfnamefont{L.}~\bibnamefont{Tomio}},
  \bibinfo{journal}{Int. J. Mod. Phys. B} \textbf{\bibinfo{volume}{19}},
  \bibinfo{pages}{3415} (\bibinfo{year}{2005}),
  \urlprefix\url{https://doi.org/10.1142/S0217979205032279}.

\bibitem[{\citenamefont{Fetter and Svidzinsky}(2001)}]{fetter1}
\bibinfo{author}{\bibfnamefont{A.~L.} \bibnamefont{Fetter}} \bibnamefont{and}
  \bibinfo{author}{\bibfnamefont{A.~A.} \bibnamefont{Svidzinsky}},
  \bibinfo{journal}{J. Phys.-Condens. Mat.} \textbf{\bibinfo{volume}{13}},
  \bibinfo{pages}{R135} (\bibinfo{year}{2001}),
  \urlprefix\url{https://iopscience.iop.org/article/10.1088/0953-8984/13/12/201}.

\bibitem[{\citenamefont{Fetter}(2009)}]{fetter2}
\bibinfo{author}{\bibfnamefont{A.~L.} \bibnamefont{Fetter}},
  \bibinfo{journal}{Rev. Mod. Phys.} \textbf{\bibinfo{volume}{81}},
  \bibinfo{pages}{647} (\bibinfo{year}{2009}),
  \urlprefix\url{https://link.aps.org/doi/10.1103/RevModPhys.81.647}.

\bibitem[{\citenamefont{Komineas}(2007)}]{komineas}
\bibinfo{author}{\bibfnamefont{S.}~\bibnamefont{Komineas}},
  \bibinfo{journal}{Eur. Phys. J.-Spec. Top.} \textbf{\bibinfo{volume}{147}},
  \bibinfo{pages}{133} (\bibinfo{year}{2007}), ISSN \bibinfo{issn}{1951-6401},
  \urlprefix\url{https://doi.org/10.1140/epjst/e2007-00206-8}.

\bibitem[{\citenamefont{Mateo and Brand}(2014)}]{mateo2014chladni}
\bibinfo{author}{\bibfnamefont{A.~M.} \bibnamefont{Mateo}} \bibnamefont{and}
  \bibinfo{author}{\bibfnamefont{J.}~\bibnamefont{Brand}},
  \bibinfo{journal}{Phys. Rev. Lett.} \textbf{\bibinfo{volume}{113}},
  \bibinfo{pages}{255302} (\bibinfo{year}{2014}),
  \urlprefix\url{https://link.aps.org/doi/10.1103/PhysRevLett.113.255302}.

\bibitem[{\citenamefont{Ruostekoski and Anglin}(2001)}]{ruost1}
\bibinfo{author}{\bibfnamefont{J.}~\bibnamefont{Ruostekoski}} \bibnamefont{and}
  \bibinfo{author}{\bibfnamefont{J.~R.} \bibnamefont{Anglin}},
  \bibinfo{journal}{Phys. Rev. Lett.} \textbf{\bibinfo{volume}{86}},
  \bibinfo{pages}{3934} (\bibinfo{year}{2001}),
  \urlprefix\url{https://link.aps.org/doi/10.1103/PhysRevLett.86.3934}.

\bibitem[{\citenamefont{Ruostekoski and Anglin}(2003)}]{ruost2}
\bibinfo{author}{\bibfnamefont{J.}~\bibnamefont{Ruostekoski}} \bibnamefont{and}
  \bibinfo{author}{\bibfnamefont{J.~R.} \bibnamefont{Anglin}},
  \bibinfo{journal}{Phys. Rev. Lett.} \textbf{\bibinfo{volume}{91}},
  \bibinfo{pages}{190402} (\bibinfo{year}{2003}),
  \urlprefix\url{https://link.aps.org/doi/10.1103/PhysRevLett.91.190402}.

\bibitem[{\citenamefont{Kleckner and Irvine}(2013)}]{irvine}
\bibinfo{author}{\bibfnamefont{D.}~\bibnamefont{Kleckner}} \bibnamefont{and}
  \bibinfo{author}{\bibfnamefont{W.}~\bibnamefont{Irvine}},
  \bibinfo{journal}{Nat. Phys.} \textbf{\bibinfo{volume}{9}},
  \bibinfo{pages}{253} (\bibinfo{year}{2013}),
  \urlprefix\url{https://doi.org/10.1038/nphys2560}.

\bibitem[{\citenamefont{Farrell et~al.}(2015)\citenamefont{Farrell, Birkisson,
  and Funke}}]{farrell2015deflation}
\bibinfo{author}{\bibfnamefont{P.~E.} \bibnamefont{Farrell}},
  \bibinfo{author}{\bibfnamefont{A.}~\bibnamefont{Birkisson}},
  \bibnamefont{and} \bibinfo{author}{\bibfnamefont{S.~W.} \bibnamefont{Funke}},
  \bibinfo{journal}{SIAM J. Sci. Comput.} \textbf{\bibinfo{volume}{37}},
  \bibinfo{pages}{A2026} (\bibinfo{year}{2015}),
  \urlprefix\url{https://doi.org/10.1137/140984798}.

\bibitem[{\citenamefont{Charalampidis et~al.}(2018)\citenamefont{Charalampidis,
  Kevrekidis, and Farrell}}]{charalampidis2018computing}
\bibinfo{author}{\bibfnamefont{E.~G.} \bibnamefont{Charalampidis}},
  \bibinfo{author}{\bibfnamefont{P.~G.} \bibnamefont{Kevrekidis}},
  \bibnamefont{and} \bibinfo{author}{\bibfnamefont{P.~E.}
  \bibnamefont{Farrell}}, \bibinfo{journal}{Commun. Nonlinear Sci. Numer.
  Simulat.} \textbf{\bibinfo{volume}{54}}, \bibinfo{pages}{482}
  (\bibinfo{year}{2018}),
  \urlprefix\url{https://doi.org/10.1016/j.cnsns.2017.05.024}.

\bibitem[{\citenamefont{Charalampidis et~al.}(2020)\citenamefont{Charalampidis,
  Boull{\'e}, Farrell, and Kevrekidis}}]{charalampidis2019bifurcation}
\bibinfo{author}{\bibfnamefont{E.}~\bibnamefont{Charalampidis}},
  \bibinfo{author}{\bibfnamefont{N.}~\bibnamefont{Boull{\'e}}},
  \bibinfo{author}{\bibfnamefont{P.}~\bibnamefont{Farrell}}, \bibnamefont{and}
  \bibinfo{author}{\bibfnamefont{P.}~\bibnamefont{Kevrekidis}},
  \bibinfo{journal}{Commun. Nonlinear Sci. Numer. Simulat.}
  \textbf{\bibinfo{volume}{87}}, \bibinfo{pages}{105255}
  (\bibinfo{year}{2020}),
  \urlprefix\url{https://doi.org/10.1016/j.cnsns.2020.105255}.

\bibitem[{\citenamefont{Rathgeber et~al.}(2016)\citenamefont{Rathgeber, Ham,
  Mitchell, Lange, Luporini, Mcrae, Bercea, Markall, and
  Kelly}}]{rathgeber2016}
\bibinfo{author}{\bibfnamefont{F.}~\bibnamefont{Rathgeber}},
  \bibinfo{author}{\bibfnamefont{D.~A.} \bibnamefont{Ham}},
  \bibinfo{author}{\bibfnamefont{L.}~\bibnamefont{Mitchell}},
  \bibinfo{author}{\bibfnamefont{M.}~\bibnamefont{Lange}},
  \bibinfo{author}{\bibfnamefont{F.}~\bibnamefont{Luporini}},
  \bibinfo{author}{\bibfnamefont{A.~T.~T.} \bibnamefont{Mcrae}},
  \bibinfo{author}{\bibfnamefont{G.-T.} \bibnamefont{Bercea}},
  \bibinfo{author}{\bibfnamefont{G.~R.} \bibnamefont{Markall}},
  \bibnamefont{and} \bibinfo{author}{\bibfnamefont{P.~H.~J.}
  \bibnamefont{Kelly}}, \bibinfo{journal}{ACM Trans. Math. Softw.}
  \textbf{\bibinfo{volume}{43}}, \bibinfo{pages}{1} (\bibinfo{year}{2016}),
  \urlprefix\url{https://doi.org/10.1145/2998441}.

\bibitem[{\citenamefont{Stewart}(2002)}]{stewart2002}
\bibinfo{author}{\bibfnamefont{G.~W.} \bibnamefont{Stewart}},
  \bibinfo{journal}{SIAM J. Matrix Anal. A.} \textbf{\bibinfo{volume}{23}},
  \bibinfo{pages}{601} (\bibinfo{year}{2002}),
  \urlprefix\url{https://doi.org/10.1137/S0895479800371529}.

\bibitem[{\citenamefont{Hernandez et~al.}(2005)\citenamefont{Hernandez, Roman,
  and Vidal}}]{hernandez2005slepc}
\bibinfo{author}{\bibfnamefont{V.}~\bibnamefont{Hernandez}},
  \bibinfo{author}{\bibfnamefont{J.~E.} \bibnamefont{Roman}}, \bibnamefont{and}
  \bibinfo{author}{\bibfnamefont{V.}~\bibnamefont{Vidal}},
  \bibinfo{journal}{ACM Trans. Math. Softw.} \textbf{\bibinfo{volume}{31}},
  \bibinfo{pages}{351} (\bibinfo{year}{2005}),
  \urlprefix\url{https://doi.org/10.1145/1089014.1089019}.

\bibitem[{\citenamefont{Delfour et~al.}(1981)\citenamefont{Delfour, Fortin, and
  Payre}}]{delfour1981finite}
\bibinfo{author}{\bibfnamefont{M.}~\bibnamefont{Delfour}},
  \bibinfo{author}{\bibfnamefont{M.}~\bibnamefont{Fortin}}, \bibnamefont{and}
  \bibinfo{author}{\bibfnamefont{G.}~\bibnamefont{Payre}}, \bibinfo{journal}{J.
  Comput. Phys.} \textbf{\bibinfo{volume}{44}}, \bibinfo{pages}{277}
  (\bibinfo{year}{1981}),
  \urlprefix\url{https://doi.org/10.1016/0021-9991(81)90052-8}.

\bibitem[{\citenamefont{Wang et~al.}(2019)\citenamefont{Wang, Kevrekidis, and
  Babaev}}]{wenl}
\bibinfo{author}{\bibfnamefont{W.}~\bibnamefont{Wang}},
  \bibinfo{author}{\bibfnamefont{P.~G.} \bibnamefont{Kevrekidis}},
  \bibnamefont{and} \bibinfo{author}{\bibfnamefont{E.}~\bibnamefont{Babaev}},
  \bibinfo{journal}{Phys. Rev. A} \textbf{\bibinfo{volume}{100}},
  \bibinfo{pages}{053621} (\bibinfo{year}{2019}),
  \urlprefix\url{https://link.aps.org/doi/10.1103/PhysRevA.100.053621}.

\bibitem[{\citenamefont{Wang et~al.}(2016)\citenamefont{Wang, Kevrekidis,
  Carretero-Gonz\'alez, and Frantzeskakis}}]{wenl1}
\bibinfo{author}{\bibfnamefont{W.}~\bibnamefont{Wang}},
  \bibinfo{author}{\bibfnamefont{P.~G.} \bibnamefont{Kevrekidis}},
  \bibinfo{author}{\bibfnamefont{R.}~\bibnamefont{Carretero-Gonz\'alez}},
  \bibnamefont{and} \bibinfo{author}{\bibfnamefont{D.~J.}
  \bibnamefont{Frantzeskakis}}, \bibinfo{journal}{Phys. Rev. A}
  \textbf{\bibinfo{volume}{93}}, \bibinfo{pages}{023630}
  (\bibinfo{year}{2016}),
  \urlprefix\url{https://link.aps.org/doi/10.1103/PhysRevA.93.023630}.

\bibitem[{\citenamefont{Mateo and Brand}(2015)}]{mateo2015stability}
\bibinfo{author}{\bibfnamefont{A.~M.} \bibnamefont{Mateo}} \bibnamefont{and}
  \bibinfo{author}{\bibfnamefont{J.}~\bibnamefont{Brand}},
  \bibinfo{journal}{New J. Phys.} \textbf{\bibinfo{volume}{17}},
  \bibinfo{pages}{125013} (\bibinfo{year}{2015}),
  \urlprefix\url{https://iopscience.iop.org/article/10.1088/1367-2630/17/12/125013}.

\bibitem[{\citenamefont{Seydel}(2010)}]{seydel2010}
\bibinfo{author}{\bibfnamefont{R.}~\bibnamefont{Seydel}},
  \emph{\bibinfo{title}{Practical Bifurcation and Stability Analysis}},
  vol.~\bibinfo{volume}{5} of \emph{\bibinfo{series}{Interdisciplinary Applied
  Mathematics}} (\bibinfo{publisher}{Springer}, \bibinfo{year}{2010}),
  \bibinfo{edition}{3rd} ed.

\bibitem[{\citenamefont{Crasovan et~al.}(2004)\citenamefont{Crasovan,
  P\'erez-Garc\'{\i}a, Danaila, Mihalache, and Torner}}]{crasovan2004three}
\bibinfo{author}{\bibfnamefont{L.-C.} \bibnamefont{Crasovan}},
  \bibinfo{author}{\bibfnamefont{V.~M.} \bibnamefont{P\'erez-Garc\'{\i}a}},
  \bibinfo{author}{\bibfnamefont{I.}~\bibnamefont{Danaila}},
  \bibinfo{author}{\bibfnamefont{D.}~\bibnamefont{Mihalache}},
  \bibnamefont{and} \bibinfo{author}{\bibfnamefont{L.}~\bibnamefont{Torner}},
  \bibinfo{journal}{Phys. Rev. A} \textbf{\bibinfo{volume}{70}},
  \bibinfo{pages}{033605} (\bibinfo{year}{2004}),
  \urlprefix\url{https://link.aps.org/doi/10.1103/PhysRevA.70.033605}.

\bibitem[{\citenamefont{M\"ott\"onen et~al.}(2005)\citenamefont{M\"ott\"onen,
  Virtanen, Isoshima, and Salomaa}}]{mikko}
\bibinfo{author}{\bibfnamefont{M.}~\bibnamefont{M\"ott\"onen}},
  \bibinfo{author}{\bibfnamefont{S.~M.~M.} \bibnamefont{Virtanen}},
  \bibinfo{author}{\bibfnamefont{T.}~\bibnamefont{Isoshima}}, \bibnamefont{and}
  \bibinfo{author}{\bibfnamefont{M.~M.} \bibnamefont{Salomaa}},
  \bibinfo{journal}{Phys. Rev. A} \textbf{\bibinfo{volume}{71}},
  \bibinfo{pages}{033626} (\bibinfo{year}{2005}),
  \urlprefix\url{https://link.aps.org/doi/10.1103/PhysRevA.71.033626}.

\bibitem[{\citenamefont{Aftalion and Danaila}(2003)}]{danaila3}
\bibinfo{author}{\bibfnamefont{A.}~\bibnamefont{Aftalion}} \bibnamefont{and}
  \bibinfo{author}{\bibfnamefont{I.}~\bibnamefont{Danaila}},
  \bibinfo{journal}{Phys. Rev. A} \textbf{\bibinfo{volume}{68}},
  \bibinfo{pages}{023603} (\bibinfo{year}{2003}),
  \urlprefix\url{https://link.aps.org/doi/10.1103/PhysRevA.68.023603}.

\bibitem[{\citenamefont{Henderson et~al.}(2009)\citenamefont{Henderson, Ryu,
  MacCormick, and Boshier}}]{boshier}
\bibinfo{author}{\bibfnamefont{K.}~\bibnamefont{Henderson}},
  \bibinfo{author}{\bibfnamefont{C.}~\bibnamefont{Ryu}},
  \bibinfo{author}{\bibfnamefont{C.}~\bibnamefont{MacCormick}},
  \bibnamefont{and} \bibinfo{author}{\bibfnamefont{M.~G.}
  \bibnamefont{Boshier}}, \bibinfo{journal}{New J. Phys.}
  \textbf{\bibinfo{volume}{11}}, \bibinfo{pages}{043030}
  (\bibinfo{year}{2009}),
  \urlprefix\url{https://doi.org/10.1088%2F1367-2630%2F11%2F4%2F043030}.

\bibitem[{\citenamefont{Gauthier et~al.}(2016)\citenamefont{Gauthier, Lenton,
  Parry, Baker, Davis, Rubinsztein-Dunlop, and Neely}}]{optica}
\bibinfo{author}{\bibfnamefont{G.}~\bibnamefont{Gauthier}},
  \bibinfo{author}{\bibfnamefont{I.}~\bibnamefont{Lenton}},
  \bibinfo{author}{\bibfnamefont{N.~M.} \bibnamefont{Parry}},
  \bibinfo{author}{\bibfnamefont{M.}~\bibnamefont{Baker}},
  \bibinfo{author}{\bibfnamefont{M.~J.} \bibnamefont{Davis}},
  \bibinfo{author}{\bibfnamefont{H.}~\bibnamefont{Rubinsztein-Dunlop}},
  \bibnamefont{and} \bibinfo{author}{\bibfnamefont{T.~W.} \bibnamefont{Neely}},
  \bibinfo{journal}{Optica} \textbf{\bibinfo{volume}{3}}, \bibinfo{pages}{1136}
  (\bibinfo{year}{2016}),
  \urlprefix\url{http://www.osapublishing.org/optica/abstract.cfm?URI=optica-3-10-1136}.

\bibitem[{\citenamefont{Ku et~al.}(2014)\citenamefont{Ku, Ji, Mukherjee,
  Guardado-Sanchez, Cheuk, Yefsah, and Zwierlein}}]{yefsah0}
\bibinfo{author}{\bibfnamefont{M.~J.~H.} \bibnamefont{Ku}},
  \bibinfo{author}{\bibfnamefont{W.}~\bibnamefont{Ji}},
  \bibinfo{author}{\bibfnamefont{B.}~\bibnamefont{Mukherjee}},
  \bibinfo{author}{\bibfnamefont{E.}~\bibnamefont{Guardado-Sanchez}},
  \bibinfo{author}{\bibfnamefont{L.~W.} \bibnamefont{Cheuk}},
  \bibinfo{author}{\bibfnamefont{T.}~\bibnamefont{Yefsah}}, \bibnamefont{and}
  \bibinfo{author}{\bibfnamefont{M.~W.} \bibnamefont{Zwierlein}},
  \bibinfo{journal}{Phys. Rev. Lett.} \textbf{\bibinfo{volume}{113}},
  \bibinfo{pages}{065301} (\bibinfo{year}{2014}),
  \urlprefix\url{https://link.aps.org/doi/10.1103/PhysRevLett.113.065301}.

\bibitem[{\citenamefont{Fritsch et~al.}(2020)\citenamefont{Fritsch, Lu, Reid,
  Pi\~neiro, and Spielman}}]{spiel}
\bibinfo{author}{\bibfnamefont{A.~R.} \bibnamefont{Fritsch}},
  \bibinfo{author}{\bibfnamefont{M.}~\bibnamefont{Lu}},
  \bibinfo{author}{\bibfnamefont{G.~H.} \bibnamefont{Reid}},
  \bibinfo{author}{\bibfnamefont{A.~M.} \bibnamefont{Pi\~neiro}},
  \bibnamefont{and} \bibinfo{author}{\bibfnamefont{I.~B.}
  \bibnamefont{Spielman}}, \bibinfo{journal}{Phys. Rev. A}
  \textbf{\bibinfo{volume}{101}}, \bibinfo{pages}{053629}
  (\bibinfo{year}{2020}),
  \urlprefix\url{https://link.aps.org/doi/10.1103/PhysRevA.101.053629}.

\bibitem[{\citenamefont{Lannig et~al.}(2020)\citenamefont{Lannig, Schmied,
  Pr{\"u}fer, Kunkel, Strohmaier, Strobel, Gasenzer, Kevrekidis, and
  Oberthaler}}]{markus}
\bibinfo{author}{\bibfnamefont{S.}~\bibnamefont{Lannig}},
  \bibinfo{author}{\bibfnamefont{C.-M.} \bibnamefont{Schmied}},
  \bibinfo{author}{\bibfnamefont{M.}~\bibnamefont{Pr{\"u}fer}},
  \bibinfo{author}{\bibfnamefont{P.}~\bibnamefont{Kunkel}},
  \bibinfo{author}{\bibfnamefont{R.}~\bibnamefont{Strohmaier}},
  \bibinfo{author}{\bibfnamefont{H.}~\bibnamefont{Strobel}},
  \bibinfo{author}{\bibfnamefont{T.}~\bibnamefont{Gasenzer}},
  \bibinfo{author}{\bibfnamefont{P.}~\bibnamefont{Kevrekidis}},
  \bibnamefont{and}
  \bibinfo{author}{\bibfnamefont{M.}~\bibnamefont{Oberthaler}},
  \bibinfo{journal}{arXiv preprint arXiv:2005.13278}  (\bibinfo{year}{2020}),
  \urlprefix\url{https://arxiv.org/abs/2005.13278}.

\bibitem[{\citenamefont{Hall et~al.}(2016)\citenamefont{Hall, Ray, Tiurev,
  Ruokokoski, Gheorghe, and M{\"o}tt{\"o}nen}}]{dsh2}
\bibinfo{author}{\bibfnamefont{D.~S.} \bibnamefont{Hall}},
  \bibinfo{author}{\bibfnamefont{M.~W.} \bibnamefont{Ray}},
  \bibinfo{author}{\bibfnamefont{K.}~\bibnamefont{Tiurev}},
  \bibinfo{author}{\bibfnamefont{E.}~\bibnamefont{Ruokokoski}},
  \bibinfo{author}{\bibfnamefont{A.~H.} \bibnamefont{Gheorghe}},
  \bibnamefont{and}
  \bibinfo{author}{\bibfnamefont{M.}~\bibnamefont{M{\"o}tt{\"o}nen}},
  \bibinfo{journal}{Nat. Phys.} \textbf{\bibinfo{volume}{12}},
  \bibinfo{pages}{478} (\bibinfo{year}{2016}),
  \urlprefix\url{https://doi.org/10.1038/nphys3624}.

\bibitem[{\citenamefont{Ku et~al.}(2016)\citenamefont{Ku, Mukherjee, Yefsah,
  and Zwierlein}}]{yefsah}
\bibinfo{author}{\bibfnamefont{M.~J.~H.} \bibnamefont{Ku}},
  \bibinfo{author}{\bibfnamefont{B.}~\bibnamefont{Mukherjee}},
  \bibinfo{author}{\bibfnamefont{T.}~\bibnamefont{Yefsah}}, \bibnamefont{and}
  \bibinfo{author}{\bibfnamefont{M.~W.} \bibnamefont{Zwierlein}},
  \bibinfo{journal}{Phys. Rev. Lett.} \textbf{\bibinfo{volume}{116}},
  \bibinfo{pages}{045304} (\bibinfo{year}{2016}),
  \urlprefix\url{https://link.aps.org/doi/10.1103/PhysRevLett.116.045304}.

\bibitem[{\citenamefont{Serafini et~al.}(2015)\citenamefont{Serafini, Barbiero,
  Debortoli, Donadello, Larcher, Dalfovo, Lamporesi, and Ferrari}}]{lampo1}
\bibinfo{author}{\bibfnamefont{S.}~\bibnamefont{Serafini}},
  \bibinfo{author}{\bibfnamefont{M.}~\bibnamefont{Barbiero}},
  \bibinfo{author}{\bibfnamefont{M.}~\bibnamefont{Debortoli}},
  \bibinfo{author}{\bibfnamefont{S.}~\bibnamefont{Donadello}},
  \bibinfo{author}{\bibfnamefont{F.}~\bibnamefont{Larcher}},
  \bibinfo{author}{\bibfnamefont{F.}~\bibnamefont{Dalfovo}},
  \bibinfo{author}{\bibfnamefont{G.}~\bibnamefont{Lamporesi}},
  \bibnamefont{and} \bibinfo{author}{\bibfnamefont{G.}~\bibnamefont{Ferrari}},
  \bibinfo{journal}{Phys. Rev. Lett.} \textbf{\bibinfo{volume}{115}},
  \bibinfo{pages}{170402} (\bibinfo{year}{2015}),
  \urlprefix\url{https://link.aps.org/doi/10.1103/PhysRevLett.115.170402}.

\bibitem[{\citenamefont{Serafini et~al.}(2017)\citenamefont{Serafini,
  Galantucci, Iseni, Bienaim\'e, Bisset, Barenghi, Dalfovo, Lamporesi, and
  Ferrari}}]{lampo2}
\bibinfo{author}{\bibfnamefont{S.}~\bibnamefont{Serafini}},
  \bibinfo{author}{\bibfnamefont{L.}~\bibnamefont{Galantucci}},
  \bibinfo{author}{\bibfnamefont{E.}~\bibnamefont{Iseni}},
  \bibinfo{author}{\bibfnamefont{T.}~\bibnamefont{Bienaim\'e}},
  \bibinfo{author}{\bibfnamefont{R.~N.} \bibnamefont{Bisset}},
  \bibinfo{author}{\bibfnamefont{C.~F.} \bibnamefont{Barenghi}},
  \bibinfo{author}{\bibfnamefont{F.}~\bibnamefont{Dalfovo}},
  \bibinfo{author}{\bibfnamefont{G.}~\bibnamefont{Lamporesi}},
  \bibnamefont{and} \bibinfo{author}{\bibfnamefont{G.}~\bibnamefont{Ferrari}},
  \bibinfo{journal}{Phys. Rev. X} \textbf{\bibinfo{volume}{7}},
  \bibinfo{pages}{021031} (\bibinfo{year}{2017}),
  \urlprefix\url{https://link.aps.org/doi/10.1103/PhysRevX.7.021031}.

\end{thebibliography}

\end{document}